\documentclass[a4paper,12pt,oneside]{report}
\usepackage[IL2]{fontenc}
\usepackage[utf8]{inputenc}
\usepackage[czech,british]{babel}
\usepackage{ifthen}
\newcommand{\draftoption}{nedraft} 
\usepackage[svgnames]{xcolor}
\newcommand{\FTTNameAndSurname}{
Aleš Flandera%
}
\newcommand{\FTTTitleOfThesis}{
Geometry of isolated horizons%
}
\newcommand{\FTTTitleOfThesisCzech}{
Geometrie izolovaných horizontů%
}
\newcommand{\FTTDepartment}{
Institute of Theoretical Physics%
}
\newcommand{\FTTDepartmentCzech}{
Ústav teoretické fyziky%
}
\newcommand{\FTTSupervisor}{
Mgr.\ Martin Scholtz, Ph.D.%
}
\newcommand{\FTTSupervisorDepartment}{
Institute of Theoretical Physics%
}
\newcommand{\FTTSupervisorDepartmentCzech}{
Ústav teoretické fyziky%
}
\newcommand{\FTTYear}{
2016%
}
\newcommand{\FTTBibliographySettings}{
numbered%
}
\newcommand{\FTTStudyProgramme}{
Physics%
}
\newcommand{\FTTSpecialization}{
Theoretical Physics%
}
\newcommand{\FTTAbstractCzech}{%
\textcolor{red}{Abstrakt}
}
\newcommand{\FTTAbstract}{%
While the formalism of isolated horizons is known for some time, only quite recently the near horizon solution of Einstein's equations has been found in the Bondi-like coordinates by Krishnan in 2012. In this framework, the space-time is regarded as the characteristic initial value problem with the initial data given on the horizon and another null hypersurface. It is not clear, however, what initial data reproduce the simplest physically relevant black hole solution, namely that of Kerr--Newman which describes stationary, axisymmetric black hole with charge. Moreover, Krishnan's construction employs the non-twisting null geodesic congruence and the tetrad which is parallelly propagated along this congruence. While the existence of such tetrad can be easily established in general, its explicit form can be very difficult to find and, in fact it has not been provided for the Kerr--Newman metric. The goal of this thesis was to fill this gap and provide a full description of the Kerr--Newman metric in the framework of isolated horizons. In the theoretical part of the thesis we review the spinor and Newman--Penrose formalism, basic geometry of isolated horizons and then present our results. Thesis is complemented by several appendices.
}
\newcommand{\FTTKeywordsCzech}{%
\textcolor{red}{Klíčová slova}
}
\newcommand{\FTTKeywords}{%
Newman--Penrose formalism, non-expanding horizons, isolated horizons, Kerr--Newman space-time, non-twisting congruences of geodesics
}
\usepackage[pdftex,unicode]{hyperref}   
\hypersetup{pdftitle=\FTTTitleOfThesis}
\hypersetup{pdfauthor=\FTTNameAndSurname}
\hypersetup{pdfkeywords=\FTTKeywords}
\ifthenelse{\equal{\draftoption}{nedraft}}%
{\hypersetup{colorlinks,citecolor=LimeGreen}}%
{\hypersetup{citecolor=LimeGreen}}%
\usepackage{./backend_settings/fales3}
\usepackage{amsfonts,amsmath,amssymb,amsthm}
\usepackage{bm}
\usepackage{IEEEtrantools}
\usepackage{longtable}
\usepackage[acronym,nomain,toc]{glossaries}
\usepackage{graphicx}
\graphicspath{{./chapters/appendix_3/epss/}}
\usepackage{epstopdf}
\usepackage{subcaption}
\usepackage{color}
\usepackage{url}
\usepackage{booktabs} 
\usepackage{array}
\usepackage{dcolumn} 
\usepackage{multirow}
\usepackage{nicefrac}
\usepackage{lastpage} 
\usepackage{calc}
\usepackage{etoolbox}
\usepackage[labelfont=bf]{caption}
\usepackage{tikz,tikz-3dplot}
\usetikzlibrary{shapes,fadings,positioning,intersections,through,calc,decorations.markings,decorations.pathreplacing,decorations.pathmorphing,plotmarks,snakes}
\tikzset{>=stealth}

\tikzset{
    set arrow inside/.code={\pgfqkeys{/tikz/arrow inside}{#1}},
    set arrow inside={end/.initial=>, opt/.initial=},
    /pgf/decoration/Mark/.style={
        mark/.expanded=at position #1 with
        {
            \noexpand\arrow[\pgfkeysvalueof{/tikz/arrow inside/opt}]{\pgfkeysvalueof{/tikz/arrow inside/end}}
        }
    },
    arrow inside/.style 2 args={
        set arrow inside={#1},
        postaction={
            decorate,decoration={
                markings,Mark/.list={#2}
            }
        }
    },
}

\newcommand\pgfmathsinandcos[3]{%
  \pgfmathsetmacro#1{sin(#3)}%
  \pgfmathsetmacro#2{cos(#3)}%
}
\newcommand\LongitudePlane[3][current plane]{%
  \pgfmathsinandcos\sinEl\cosEl{#2} 
  \pgfmathsinandcos\sint\cost{#3} 
  \tikzset{#1/.estyle={cm={\cost,\sint*\sinEl,0,\cosEl,(0,0)}}}
}
\newcommand\LatitudePlane[3][current plane]{%
  \pgfmathsinandcos\sinEl\cosEl{#2} 
  \pgfmathsinandcos\sint\cost{#3} 
  \pgfmathsetmacro\yshift{\cosEl*\sint}
  \tikzset{#1/.estyle={cm={\cost,0,0,\cost*\sinEl,(0,\yshift)}}} %
}

\newcommand\DrawLatitudeCircle[2][1]{
  \LatitudePlane{\angEl}{#2}
  \tikzset{current plane/.prefix style={scale=#1}}
  \pgfmathsetmacro\sinVis{sin(#2)/cos(#2)*sin(\angEl)/cos(\angEl)}
  \pgfmathsetmacro\angVis{asin(min(1,max(\sinVis,-1)))}
  \draw[current plane] (\angVis:1) arc (\angVis:-\angVis-180:1);
  \draw[current plane,dashed] (180-\angVis:1) arc (180-\angVis:\angVis:1);
}

\newcommand\DrawMyNewLongitudeCircle[2][1]{
  \LongitudePlane{\angEl}{#2}
  \tikzset{current plane/.prefix style={scale=#1}}
  \pgfmathsetmacro\angVis{atan(sin(#2)*cos(\angEl)/sin(\angEl))} %
  \draw[current plane] (0:1) arc (0:\angVis+180:1);
  \draw[current plane,dashed] (\angVis-180:1) arc (\angVis-180:0:1);
}
\newcommand\DrawMyNewColorLongitudeCircle[3][1]{
  \LongitudePlane{\angEl}{#2}
  \tikzset{current plane/.prefix style={scale=#1}}
  \pgfmathsetmacro\angVis{atan(sin(#2)*cos(\angEl)/sin(\angEl))} %
  \draw[current plane,#3] (0:1) arc (0:\angVis+180:1);
  \draw[current plane,dashed,#3] (\angVis-180:1) arc (\angVis-180:0:1);
}
\usepackage{tocbibind}
\usepackage{pdfpages}

\usepackage{tensor}

\usepackage{enumitem}

\input{./macros/widebar}
\newcommand{\qm}[1]{``#1''}
\renewcommand{\vect}[1]{\boldsymbol{#1}}
\newcommand{\levicivita}{\oldepsilon}

\newcommand{\cconj}[1]{\widebar{#1}}
\newcommand{\contran}[1]{#1^{+}}
\usepackage{mathrsfs}
\def\pd{\partial}
\def\scri{\mathscr{I}}
\newcommand{\MM}{\mathcal{M}}
\usepackage{marginnote}

\newcommand{\anote}[1]{\textcolor{blue}{#1}}
\newcommand{\ntext}[1]{\textcolor{purple}{#1}}

\newcommand{\snote}[1]{}

\newtheorem{theorem}{Theorem}[chapter]
\theoremstyle{definition}
\newtheorem{definition}{Definition}[chapter]
\newenvironment{myproof}{\par\medskip\noindent%
   \textit{Proof.} \small \rmfamily}{\qed \bigskip}

\newcommand{\tetcom}[1]{\mathbf{#1}}

\newcommand{\focc}[1]{\textbf{#1}}

\newcommand{\ProjectOnS}[1]{^{\sharp}\!\left(#1\right)}

\newcommand{\ricsp}{\Phi}
\def\ii{\mathrm{i}}
\def\Lie{\pounds}
\newcommand{\msp}{\oldphi}
\newcommand{\szero}{|_{\mathcal{S}_0}}
\newcommand{\hor}{^{(0)}}
\newcommand{\rone}{^{(1)}}
\newcommand{\scrho}{\varrho}
\newcommand{\abs}[1]{\lvert#1\rvert}
\newcommand{\zero}{_{(0)}}
\newcommand{\one}{_{(1)}}

\newcommand{\matcom}[1]{\texttt{#1}}

\newcommand{\sss}{\kern 0.0833em}

\newcommand{\szerot}{}
\usepackage{calc}
\usepackage{geometry}
\let\openright=\cleardoublepage

\ifthenelse{%
\equal{\FTTBibliographySettings}{harvard}%
}%
{%
\newcommand{\FTTBibliographyChoice}{%
\usepackage{csquotes}

\usepackage[
language=english, 
sortlocale=en_US, 
maxbibnames=30, 
minbibnames=4, 
maxcitenames=2, 
style=authoryear,
firstinits=true, 
backend=biber,
dashed=false, 
uniquelist=false,
uniquename=false,
]{biblatex}

\DeclareFieldFormat[article]{title}{#1} 
\DeclareFieldFormat[inproceedings]{title}{\textit{#1}} 
\DeclareNameAlias{sortname}{last-first} 
\AtBeginBibliography{\renewcommand*{\mkbibnamelast}[1]{\textsc{#1}}} 

\DeclareBibliographyDriver{lecture}{%
  \usebibmacro{bibindex}%
  \usebibmacro{begentry}%
  \usebibmacro{author/translator+others}%
  \setunit{\labelnamepunct}\newblock
  \usebibmacro{title}%
  \newunit 
  \printfield{howpublished}%
  \newunit 
  \printfield{month} \setunit{\addspace} \printfield{year}%
  \newunit 
  \printfield{note}%
  \setunit{\addspace}%
  \printfield{url}%
  \usebibmacro{finentry}
}

\addbibresource{./bibliography/bibliography.bib}

}%
}%
{%
\newcommand{\FTTBibliographyChoice}{%
\usepackage{csquotes}

\usepackage[
maxnames=5,
style=numeric,
firstinits=true, 
backend=bibtex,
sorting=none,
]{biblatex}

\DeclareFieldFormat[article]{title}{\textit{#1}}
\DeclareNameAlias{default}{last-first}
\AtBeginBibliography{\renewcommand*{\mkbibnamelast}[1]{\textsc{#1}}}

\DeclareBibliographyDriver{lecture}{%
  \usebibmacro{bibindex}%
  \usebibmacro{begentry}%
  \usebibmacro{author/translator+others}%
  \setunit{\labelnamepunct}\newblock
  \usebibmacro{title}%
  \newunit 
  \printfield{howpublished}%
  \newunit 
  \printfield{month} \setunit{\addspace} \printfield{year}%
  \newunit 
  \printfield{note}%
  \setunit{\addspace}%
  \printfield{url}%
  \usebibmacro{finentry}
}

\addbibresource{./bibliography/bibliography.bib}

}%
}%
\makeatletter
\def\@makechapterhead#1{
  {\parindent \z@ \raggedright \normalfont
   \Huge\bfseries \thechapter. #1
   \par\nobreak
   \vskip 20\p@
}}
\def\@makeschapterhead#1{
  {\parindent \z@ \raggedright \normalfont
   \Huge\bfseries #1
   \par\nobreak
   \vskip 20\p@
}}
\makeatother

\usepackage{csquotes}

\usepackage[
maxnames=5,
style=numeric,
firstinits=true, 
backend=bibtex,
sorting=none,
]{biblatex}

\DeclareFieldFormat[article]{title}{\textit{#1}}
\DeclareNameAlias{default}{last-first}
\AtBeginBibliography{}

\DeclareBibliographyDriver{lecture}{%
  \usebibmacro{bibindex}%
  \usebibmacro{begentry}%
  \usebibmacro{author/translator+others}%
  \setunit{\labelnamepunct}\newblock
  \usebibmacro{title}%
  \newunit 
  \printfield{howpublished}%
  \newunit 
  \printfield{month} \setunit{\addspace} \printfield{year}%
  \newunit 
  \printfield{note}%
  \setunit{\addspace}%
  \printfield{url}%
  \usebibmacro{finentry}
}

\def\HH{\mathcal{H}}
\def\DD{\mathcal{D}}
\def\NN{\mathcal{N}}
\def\surfkappa{\kappa_{(l)}}
\def\Real{\mathbb{R}}
\def\SS{\mathcal{S}}
\def\two{{}^{(2)}}
\def\dd{\mathrm{d}}
\def\RR{\mathcal{R}}
\def\PP{\mathcal{P}}

\makeglossaries

\begin{document}
\newacronym{ads}{AdS}{Anti-de Sitter}
\newacronym{lhs}{LHS}{left-hand side}
\newacronym{rhs}{RHS}{right-hand side}
\newacronym{lis}{LIS}{local inertial system}
\newacronym{eq}{eq.}{equation}
\newacronym{np}{NP}{Newman--Penrose}
\newacronym{coef}{coef.}{coefficient}
\newacronym{cc}{c.c.}{complex conjugate}
\newacronym{adm}{ADM}{Arnowitt--Deser--Misner}
\newacronym{ghp}{GHP}{Geroch--Held--Penrose}


\lefthyphenmin=2
\righthyphenmin=2

\ifthenelse{\equal{\draftoption}{draft}}
{
}{

\pagestyle{empty}
\hypersetup{pageanchor=false}
\begin{center}

\centerline{\mbox{\includegraphics[width=166mm]{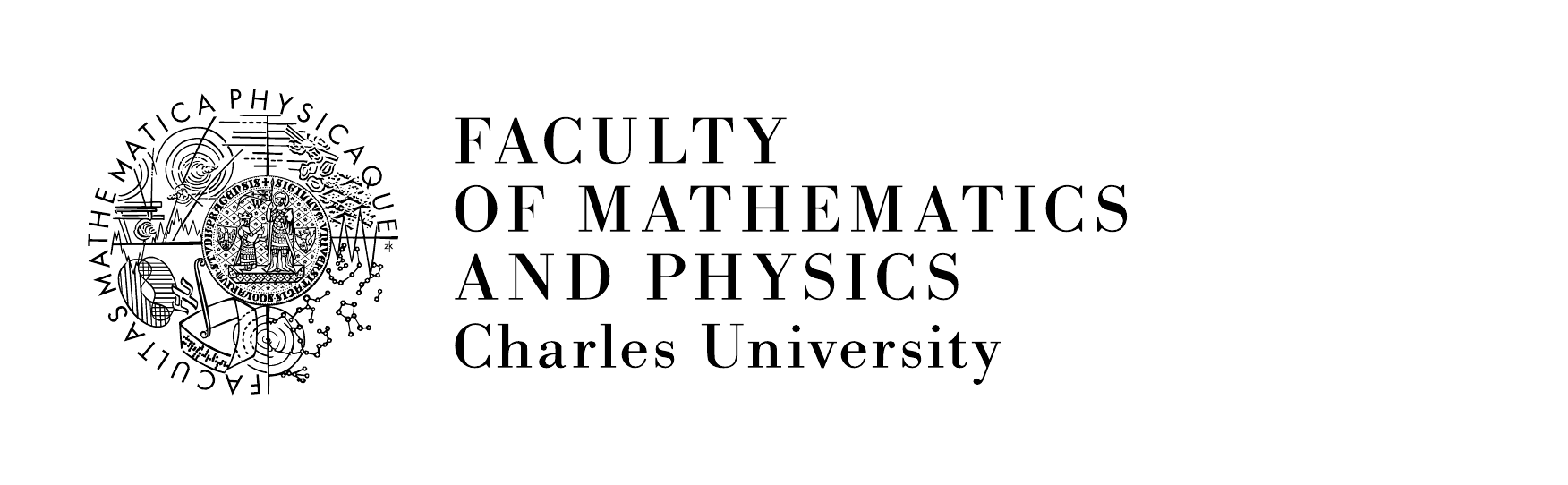}}}

\vspace{-8mm}
\vfill

{\bf\Large MASTER THESIS}

\vfill


{\LARGE \foreignlanguage{czech}{\FTTNameAndSurname}}

\vspace{15mm}

{\LARGE\bfseries \FTTTitleOfThesis}

\vfill

\FTTDepartment

\vfill

\begin{tabular}{rl}

Supervisor of the master thesis: & \FTTSupervisor \\
\noalign{\vspace{2mm}}
Study programme: & \FTTStudyProgramme \\
\noalign{\vspace{2mm}}
Study branch: & \FTTSpecialization \\
\end{tabular}

\vfill

Prague \FTTYear

\end{center}

\newpage

\openright
\hypersetup{pageanchor=true}


\pagestyle{plain}
\pagenumbering{roman}

\vglue 0pt plus 1fill

\noindent
I declare that I carried out this bachelor thesis independently, and only with the cited
sources, literature and other professional sources and with the help of my supervisor.

\medskip\noindent
I understand that my work relates to the rights and obligations under the Act No.
121/2000 Coll., the Copyright Act, as amended, in particular the fact that the Charles
University in Prague has the right to conclude a license agreement on the use of this
work as a school work pursuant to Section 60 paragraph 1 of the Copyright Act.

\vspace{10mm}

\hbox{\hbox to 0.5\hsize{%
In Prague on the 26th of July, 2016
\hss}\hbox to 0.5\hsize{%
signature of the author
\hss}}

\vspace{20mm}
\newpage

%
\openright

\vbox to 0.5\vsize{
\setlength\parindent{0mm}
\setlength\parskip{5mm}

Title:
\FTTTitleOfThesis

Author:
\foreignlanguage{czech}{\FTTNameAndSurname}

Department:
\FTTDepartment

Supervisor: 
\FTTSupervisor, \FTTSupervisorDepartment

Abstract: \FTTAbstract


\vss}

\newpage

\openright

\noindent
I would like to express my grateful thanks to my supervisor for his generous help. Not only he introduced me this interesting topic and in long talks explained me threads of it, he, most importantly, has been encouraging me to finish this thesis.

\newpage
}


\openright
\pagestyle{plain}
\pagenumbering{arabic}
\setcounter{page}{1}
\tableofcontents

\setlength\parskip{2mm}
\chapter*{Introduction}
\addcontentsline{toc}{chapter}{Introduction}

This thesis deals with the framework of isolated horizons in general relativity. As a research goal, we will construct a Newman--Penrose null tetrad with some specific properties in the Kerr--Newman space-time, namely a null tetrad such that vector $n^a$ is non-twisting, geodesic, and all other vectors of the tetrad are covariantly constant along $n^a$. This is, however, a very specific goal and the essential part of the thesis is devoted to the physical and mathematical background of the topic we study.
The result itself is to be presented in detail in a paper which is to be submitted to the journal \emph{Physical Review~D}.
Here we motivate the study of isolated horizons from a wider perspective and give an overall review of structure of the thesis.

\section*{Black hole horizons}

Black holes became one of the most interesting object of study in gravitational physics already with the discovery of the Schwarzschild solution, the very first exact solution of Einstein's equations \cite{Schwarzschild1916}. Although the singularity of the solution at the Schwarzschild radius turned out to be merely a coordinate singularity, rather than a singularity of the space-time, horizon is nevertheless a surface with very special and surprising properties. Moreover, there was initial scepticism that the singularity at the centre could represent anything physical. It was thought that the solution is valid only outside a spherical body and the interior of the solution should be disregarded. An important step to answer this question was the analysis of the spherical collapse by Oppenheimer and Snyder \cite{Oppenheimer1939}. They shown that during the collapse of spherically symmetric matter, a singularity and the event horizon will form. This is not yet conclusive because similar thing happens in the Newtonian theory, but only in the case of perfect spherical symmetry. Hence, the question was whether the formation of singularity is just an artefact of spherical symmetry or not. The most convincing argument that the formation of singularity is a real process in general relativity, is by Penrose and Hawking \cite{Hawking-Penrose-1970,Penrose1965,Hawking1973}. They have formulated a set of results known as the \emph{singularity theorems} according to which the formation of singularity is inevitable once a closed trapped surface exists in a space-time. Since the assumptions of the proof are very mild, requiring essentially only some sort of energy condition and global hyperbolicity \cite{Hawking1973,Wald1984}, the singularity theorems provide a powerful evidence that the formation of black holes is a generic feature of general relativity. An appropriate interpretation of the Schwarzschild solution as a black hole and clarifying its global causal structure is mainly due to Kruskal, who found maximal analytic extension of the Schwarzschild solution \cite{Kruskal1982}.

Today, black holes are considered as standard astrophysical objects and there are essentially no doubts about their existence. Black holes have been more or less directly observed in binary systems, they are located in the active galactic nuclei or powering quasars. For a review on observational evidence for black holes, see, e.g.\ \cite{Narayan2013} and references therein. The most direct and most striking confirmation of black holes is recent observation of gravitational waves where the wave pattern corresponding to a merger of two black holes has been observed \cite{Abbot2016}.

It was soon recognized that the event horizon exhibits many interesting and puzzling properties. Among the interesting ones, let us mention the uniqueness. In physics in general, exact solutions are usually just useful approximations to realistic situations. We expect that some simplified model will exhibit important properties of a studied system, but full system is supposed to be more complicated and complex and corresponding exact solution is infeasible.  Of course, this is also the case for black holes and we will return to that point later, but existing solutions describing black holes are, despite their high degree of symmetry, believed to be very realistic ones.

First, it was shown by Birkhoff \cite{Birkhoff1923} that \emph{any} spherically symmetric vacuum solution of Einstein's equations which is asymptotically flat must be isometric to the Schwarzschild solution. Notice that time dependence of the metric is allowed here. For example, we could have a pulsating or collapsing star which is spherically symmetric so that the exterior is always exactly given by the Schwarzschild metric. If, in addition, the space-time is required to be \emph{static}, the full geometry must coincide the the Schwarzschild solution. In this sense, Schwarzschild solution is the unique spherically symmetric black hole.

Similar uniqueness theorems apply to the Kerr--Newman metric which describes axially symmetric rotating black hole with charge \cite{Kerr2009,Adamo2014}, although here the results are not so strong. For a review on uniqueness theorems of black holes, see \cite{Chrusciel2012}. Essentially, it can be stated that any stationary axisymmetric solution which is asymptotically flat and regular everywhere except for the singularities below the horizon is necessarily the Kerr--Newman metric. 

These uniqueness results are summarized in the \emph{no-hair theorem} \cite{Israel1968,Carter1971,Robinson1975}. According to this theorem, black hole solution of Einstein--Maxwell equations is characterized by just three parameters, mass $M$, spin $a$ and (possibly magnetic) charge  $Q$. Taking different sources, e.g. the scalar field, one can elude the no hair theorem and produce a ``hairy black hole'', see, for example \cite{Herdeiro2014}.

To conclude, although the most important solutions representing black holes, the Schwarzschild and the Kerr--Newman solutions, exhibit high degree of symmetries, they actually represent highly astrophysically relevant solutions and we expect that a collapse of realistic matter will produce a Kerr--Newman black hole, provided there is no accretion disk surrounding the black hole. 

This class of black holes have very interesting properties, in particular, they satisfy the laws resembling the laws of classical thermodynamics. We will briefly review these laws in chapter \ref{chapter:isol}. This analogy shows that the surface gravity $\kappa$ of a black hole plays the role of temperature $T$, the relation between the two being $T = \kappa/2\pi$, and the area of the event horizon $A$ is related to the entropy $S$ by $S = A/4$. Originally it was thought that this analogy is purely formal, because in classical relativity, black hole is absolute black body with zero temperature, as it cannot emit anything. However, it was suggested by Bekenstein 
to interpret the entropy of black hole as a measure of information about the interior of black hole which is not accessible to the external observer \cite{Bekenstein1973}. Final justification for this hypothesis has been provided by Hawking who showed that, taking the quantum effects into account, a black hole must evaporate through the emission of the so-called \emph{Hawking radiation} which has a thermal spectrum \cite{Hawking1976}. Later, an interpretation of the Hawking particles as a tunnelling of particles through the horizon has been given in \cite{Wilczek2000}.

It turns out that the event horizon of the Kerr--Newman black hole can be characterized by several criteria. In general space-time, these criteria define different types of the horizon but in the Kerr--Newman case they all coincide. Geometrically, null geodesics in the space-time containing a black hole can be divided into two sets: those which are able to escape to future null infinity $\scri^+$ and those which end up in the singularity. All geodesics of the first family form the so-called \emph{causal past} of $\scri^+$, denoted by $J^-(\scri^+)$ \cite{Wald1984,Penrose1972}. In Minkowski space-time, all null geodesics can escape to infinity and hence $J^-(\scri^+)$ is in fact the entire space-time. If the space-time contains a black hole, $J^-(\scri^+)$ is not identical with the space-time $\MM$. Then, the black hole is defined as $B = \MM \setminus J^-(\scri^+)$ and the \emph{event horizon} is defined as the boundary $\HH = \pd B$ of the black hole region. 

We can see that the event horizon has a clear geometrical meaning but also suffers from some disadvantages. They have what is usually called a \emph{teleological nature} \cite{Frolov2012}. From the definition it is clear that one cannot identify the event horizon unless the whole space-time is known because one has to find the causal past of future null infinity; so, the knowledge of global solution is necessary. Moreover, during the collapse, the event horizon forms even before the actual black hole is formed. In a sense, the formation of a black hole is ``anticipated'' by the event horizon even before the entire mass of collapsing object falls under the event horizon which is the stage when the black hole is actually formed. This is also problem for numerical relativity, because the presence of the horizon cannot be detected locally. Only after the full evolution is obtained, one can integrate backwards from $\scri^+$ and find the event horizon \cite{Libson1996}.

For many other purposes and, in particular, in the context of singularity theorems, another definition of the horizon is important, namely that of apparent horizon. Imagine an observer above the horizon of black hole who is emitting signal oriented outwards with respect to a black hole, i.e. signals directed towards infinity. Next, suppose that this observer approaches the black hole. Then, at some distance, even the outward pointing light rays will eventually fall into black hole. The \emph{apparent horizon} is a boundary between region where outward pointing null geodesics will escape to infinity and region where even the light rays emitted in an outward direction fail to escape \cite{Hawking1973}. Compared to the event horizon, the advantage of the apparent horizon is its quasi-local nature: only a finite region must be known to locate the apparent horizon and this makes it more suitable for numerical computations \cite{Frauendiener2000}. 

Finally, an important notion is that of the Killing horizon. Kerr--Newman metric admits two Killing vectors: stationary Killing vector $\pd_v$ where $v$ is advanced time in the Kerr coordinates, and axial Killing vector $\pd_\phi$. Their linear combination is a ``helical'' Killing vector
\begin{align}
 K &= \pd_v + \Omega_{\mathrm{H}}\,\pd_\phi, \qquad \Omega_{\mathrm{H}} = \frac{a}{a^2+r_+^2},
\end{align}
where $a$ is the spin of a black hole, $r_+ = M+ \sqrt{M^2-a^2-Q^2}$ is the horizon of the black hole (with $Q$ being the charge) and $\Omega_{\mathrm{H}}$ is the angular frequency of the horizon. Since $\Omega_{\mathrm{H}}$ is constant, $K$ is also a Killing vector and it becomes null on the event horizon. In general, a \emph{Killing horizon} is a null hypersurface on which the norm $K_a K^a$ vanishes. In the case of the Kerr--Newman black hole it is easy to understand why Killing horizon is important. Above the horizon, orbits of the Killing vector are time-like curves and correspond to an observer which is stationary, i.e. hovering above the horizon at constant distance. Below the Killing horizon the norm $K_a K^a$ becomes negative and hence the orbits  of the Killing vector are space-like. Thus, they cannot correspond to any physical observer. This shows that under the Killing horizon it is impossible to have a stationary observer. 

In the case of the Kerr--Newman metric, all these three concepts of horizon, i.e. the event horizon, the apparent horizon and the Killing horizon, coincide, but they do not coincide in general. For example, during the Oppenheimer-Snyder collapse, the event horizon starts to form in the centre of spherical symmetry, its radius increases up to the Schwarzschild radius and then becomes constant. In contrast, the apparent horizon starts to form on the surface of collapsing matter and it approaches the event horizon only asymptotically. For an apparent horizon, the laws of thermodynamics cannot be satisfactorily formulated \cite{Ashtekar-LivRR}.

\section*{Isolated horizons and distorted black holes}

We have stressed that beside many simplifications exhibited by existing black hole solutions, they are in fact of great astrophysical importance thanks to uniqueness theorems. However, general as they are, these solutions do not encompass all important astrophysical situations. For the detection of black holes it is absolutely essential to have an accretion disk surrounding the black hole. Black holes admitting the accretion disk or arbitrary matter near the horizon are called \emph{distorted} or \emph{dirty} black holes. 

Natural approach to describing distorted black holes is to consider fixed Kerr--Newman geometry and study the test matter on given background or, to get better approximation, to study perturbations of the Kerr--Newman solution corresponding to the presence of the accretion disk. Usually, it is possible to neglect the mass of the accretion disk compared to the mass of black hole. However, recently, experiments were proposed in order to test the no-hair theorem mentioned above or possible deviations from this theorem using the Event Horizon Telescope with Sagittarius A* \cite{Psaltis2016,Johannsen2016}. These experiments will be sensitive to the presence of the accretion disk. It is important, because if some deviations from the no-hair theorem will be detected, one might tend to interpret it in favour of some alternative theory, but it might well be due to the presence of the accretion disk. Hence, it is necessary to study distorted black holes from a theoretical point of view and include the back-reaction of the disk on the geometry, not to treat the disk as a test matter on given background.

The presence of an accretion disk is also essential for measuring the properties of black holes. While the mass of a black hole can be calculated from the influence of a black hole on neighbouring stars, this cannot be done for measuring the spin, because the effect of frame dragging is far too small to be detected by observations of neighbouring stars. Instead, methods like continuum fitting or iron line method are employed \cite{Reis2014,Risaliti2013,McClintock2014,StrongGravity}. For example, the iron line method is based on the fact that X-ray photons emitted from the coronal region of a black hole will eventually hit the accretion disk. The matter forming the accretion disk usually comes from neighbouring star which forms a binary system with the black hole and contains light elements like hydrogen and helium, but also heavier elements, in particular the iron which was produced in the neighbouring star. Thanks to the Auger effect\footnote{Recall that the Auger effect is emission of an electron which accompanies the filling of a vacancy in an inner electron shell \cite{IUPAC}.}, lighter elements hit by X-ray photons will emit electrons on deexcitation,  while the iron atoms hit by X-rays will emit photons of energy $6.4\,\mathrm{keV}$. However, when observed, those photons are Doppler shifted because of the two reasons. First, different locations in the accretion disk have different projection of velocities on which the Doppler effect depends. Second, these photons are moving in the gravitational field of a black hole. Therefore, instead of single peak for energy $6.4\,\mathrm{keV}$, spectrum has characteristic extended shape from which the value of the spin can be inferred. Of course, if the back-reaction of the accretion disk on the space-time geometry is taken into account, expected profile of the spectrum of iron lines for a black hole of given spin will change.

For all these reasons, it is necessary to have a formalism which allows for distorted black holes, where the effect of surrounding material is taken into account. 

The framework of isolated horizons aims to provide a general definition of a black hole horizon which will:
\begin{itemize}
 \item be quasi-local, in order to circumvent difficulties related to the teleological nature of the event horizon and to make the notion of horizon convenient for numerical relativity;
 
 \item reproduce the laws of black hole thermodynamics;
 
 \item describe the black hole in the equilibrium with its neighbourhood.
\end{itemize}

Of course, these requirements are not all independent. In particular, one expects that the laws of (equilibrium) thermodynamics are related to the assumption that the black hole is in equilibrium with the neighbourhood. This means that there is no flux of matter or radiation through the horizon. The assumption of equilibrium seems plausible also from the numerical point of view, since the back-scattering effects for the late stage of the collapse become smaller than numerical errors. However, the notion of equilibrium here is much weaker than the assumption of stationarity for the Kerr--Newman black hole. It turns out that the intrinsic metric of the horizon \emph{is} stationary (i.e.\ it is Lie constant along the generators of the horizon), but the black hole itself can be embedded in otherwise dynamical space-time, e.g.\ in the expanding universe or as a component of a binary system. Since the isolated horizons are the main topic to be studied in this thesis, we relegate appropriate references and technical details to chapters \ref{chapter:non-exp} and \ref{chapter:isol}.

Isolated horizons play also an important role in loop quantum gravity. The fact that black holes possess entropy is very puzzling since the presence of entropy usually means the presence of some microscopic degrees of freedom. It is not clear, however, what do these degrees of freedom describe. According to the \emph{stretched horizon paradigm}, these degrees of freedom ``live'' on the space-like sphere with the radius $r_S + \ell_P$, where $r_S$ is the Schwarzschild radius and $\ell_P$ is the Planck length. This proposal was made by Susskind et al.\ in order to resolve the black hole information paradox \cite{Susskind1993}. This theory, however, is just phenomenological. One of the great achievements claimed by the string theory is the actual calculation of microscopic degrees of freedom for a black hole, and the result coincides with the usual Bekenstein--Hawking entropy \cite{Strominger1996} (however, see some objections to similar calculations, summarized in, e.g.\ \cite{Penrose2007}).

In loop quantum gravity, an attempt is made to introduce appropriate phase space for an isolated black hole and perform the quantization in the Hamiltonian framework. According to \cite{Ashtekar-LivRR}, there are obstacles if one considers the event horizon as a definition of a black hole. For example, the phase space of globally stationary solutions is too restricted in order to account for the quantum fluctuations. Formalism of isolated horizons was motivated also to circumvent these problems and allow for a black hole which is isolated but the space-time is not globally stationary.

\section*{Motivation of the thesis}

The formalism of isolated horizons turned out to be very fruitful and has applications which were not foreseen. This thesis is mainly motivated by work of G\"urlebeck and Scholtz \cite{Guerlebeck2016} where the Meissner effect for black holes was analysed. It describes the expulsion of the magnetic field from the horizon of extremal, axially symmetric black holes. This effect was known to exist for the Kerr--Newman black hole where the magnetic fields were treated as the test fields, although some exact results were known as well, see \cite{Guerlebeck2016} and the references therein for more complete discussion. In \cite{Guerlebeck2016}, the authors have employed the formalism of isolated horizons and generalized existing results for all types of distorted horizons provided they are axially symmetric. They also explain why the Meissner effect does not hold when the axial symmetry is violated.

We have mentioned that the no-hair theorem asserts that properties of black holes depend on the three parameters $M$, $a$ and $Q$ only. This is true for non-distorted black holes. An indication that the no-hair theorem can be extended to distorted black holes has been given by G\"urlebeck \cite{Guerlebeck2015}. He was able to prove in the static case that contributions to multipole moments of the space-time can be disentangled into those generated by the black hole and those which describe contribution from surrounding matter. The contribution from the black hole coincides with the multipole moments of the Schwarzschild space-time. In this sense, the no-hair theorem still holds.

This raises a natural question which properties of black holes are universal, like their ``baldness'', and which are special for the Kerr--Newman family of solutions. Scholtz and G\"urlebeck were able to show that the Meissner effect belongs to \emph{universal properties}. In their setting, based on the isolated horizons formalism, they considered arbitrary distorted black hole. For such general black hole, the Meissner effect does not occur, because the magnetic flux through the horizon is part of the free data which is not constrained with other geometrical quantities. In the axially symmetric case, however, it was proved that such constraints exist and, indeed, imply the Meissner effect. It is remarkable that this result is insensitive to distortions caused by external matter. Although there is an interesting result by Lewandowski and Pawlowski that the \emph{intrinsic} geometry of extremal isolated horizon is, in fact, isometric to the extremal Kerr--Newman case, the proof \cite{Guerlebeck2016} applies to full space-time geometry.

In practice, we expect that distorted black holes will be different from the Kerr--Newman black holes, but still similar to them. The main achievement of the formalism of isolated horizons, which was also used in \cite{Guerlebeck2016}, is that the back-reaction of the accretion disk is taken into account. Nevertheless, for small masses of accretion disks we expect that the geometry will not be too different from the Kerr--Newman one. In addition, since the Kerr--Newman geometry is well-understood, it is useful to have it as a reference point. For example, in \cite{Guerlebeck2016}, magnetic fields around distorted black holes have been visualised, where the deformed Kerr metric was assumed. In order to specify such a deformation, it is necessary first to translate the Kerr geometry into the language of isolated horizons and then consider specific deformations. In \cite{Guerlebeck2016}, this was partially done for the purposes of paper, i.e.\ the appropriate initial data reproducing Kerr geometry has been found. However, the deformations have been chosen \emph{ad hoc}, just to illustrate the transition from non-extremal case to the extremal one for a black hole different from pure Kerr. Nevertheless, physical interpretation of such deformations is highly desirable and for that a full analysis of the Kerr--Newman metric in the framework of isolated horizons is necessary.

In this thesis, the paper of central interest is by Krishnan \cite{Krishnan2012}. There, the author translates existing results on intrinsic geometry of isolated horizons into the Newman--Penrose formalism \cite{Penrose1962} and, in addition, provides perturbative expansion of the geometrical quantities (in terms of the Newman--Penrose formalism which, at the end, can be translated to the expansion of the metric) in the neighbourhood of the horizon. This is very similar to analogous expansion near null infinity $\scri^+$ \cite{NewmanUnti1962} by Newman and Unti. Such expansion are the Newman--Penrose version of asymptotic expansions obtained earlier by Bondi et al. \cite{Bondi1962,Sachs1962}. Historically, these works were fundamentally important because it was shown that despite the difficulties one encounters in order to define the notion of energy in general relativity, it is possible to define the global energy of a space-time. This mass is called \emph{Bondi mass} \cite{Bondi1960} and it is different from another concept, the \gls{adm} mass \cite{Arnowitt2008}, because it is not constant in time. Rather, it is measured at the null infinity $\scri$ and describes the mass of isolated system which decreases whenever the system emits gravitational or other type of radiation. Today, standard form of the Bondi mass is given in the Newman--Penrose formalism for electro-vacuum space-times \cite{SzabadosLRR} and it has been generalized recently to include scalar field sources, conformal scalar fields and interacting electromagnetic and scalar sources \cite{BicakScholtzTod2,Scholtz2014}. 

In the so-called \emph{Bondi coordinates} which are constructed on, say, past null infinity $\scri^-$ and in its neighbourhood, one employs a time coordinate $v$ along the generators of $\scri^-$ and two spherical coordinates on the cuts of constant $v$. Then, a congruence of non-twisting null geodesics is constructed in the neighbourhood of $\scri^-$ and the affine parameter $r$ along these geodesics is used as the fourth coordinate. The expansions are then performed in the coordinate $r$ and the asymptotic solution of the field equations is given in the form of series in $r$.

In \cite{Krishnan2012}, similar construction has been applied in the neighbourhood of isolated horizon $\HH$. Again, a non-twisting congruence is constructed and entire Newman--Penrose null tetrad is parallelly propagated along this congruence. While the topology and other properties of isolated horizons guarantee that this construction is always possible, the construction has not been performed explicitly even in the Kerr--Newman case.

With the Bondi-like coordinates and adapted null tetrad at hands, the space-time in the neighbourhood of isolated horizon is a solution to characteristic initial value problem with the initial data given on the horizon $\HH$ and some, arbitrarily chosen, null hypersurface transversal to $\HH$. In \cite{Krishnan2012} it was analysed which Newman--Penrose quantities are free, which part of the Einstein equations are constraints, and which are the evolution equations giving the solution in the region between the two null hypersurfaces. Construction has been given in general for arbitrary isolated horizon, but the author claims, without proof, that it has been checked numerically that the construction works for Kerr--Newman space-time. However, it is not discussed what initial data reproduce the Kerr--Newman metric, neither what is the explicit form of the tetrad and the metric in such Bondi-like coordinates.

On the other hand, kind of generalized Bondi coordinates have been found by Fletcher and Lun \cite{Fletcher2003a} for the Kerr metric (i.e.\ without charge). The authors employ the fact that the geodesic equation on the Kerr--Newman space-time is separable using the Hamilton--Jacobi equation, as was first demonstrated by Carter \cite{Carter1968}. This allows one to parametrize all geodesics of the Kerr--Newman metric by four constants of motion, namely the norm (equal to 1 for time-like, and 0 for null geodesics), the energy $E$, the angular momentum $L$ and the so-called (fourth) \emph{Carter constant}. The latter arises either from the separation of the Hamilton--Jacobi equation or from the projection of the Killing tensor \cite{Walker1970}. Authors of \cite{Fletcher2003a} choose the simplest choice of these constants which yields the non-twisting congruence, arguing that non-twisting congruence has vanishing angular momentum, $L=0$. This allows them to construct generalized Bondi coordinates in the Kerr space-time. The full null tetrad and description in the Newman--Penrose formalism is not discussed in \cite{Fletcher2003a}.

\section*{Goals and organization of the thesis}

In this thesis we aim to combine the approaches of \cite{Krishnan2012} and \cite{Fletcher2003a} in order to explicitly construct the tetrad in the Kerr--Newman space-time which meets the criteria imposed in \cite{Krishnan2012}, generalizing the technique employed in \cite{Fletcher2003a}. We proceed as follows:

\begin{itemize}
 \item we construct the null tetrad satisfying the criteria of \cite{Krishnan2012} \emph{on} the outer horizon of the Kerr--Newman metric;
 
 \item calculating relevant spin coefficients, we extract appropriate initial data given on the horizon;
 
 \item we construct general null congruence of geodesics emanating from the horizon and analyze the integrability conditions for the congruence be non-twisting;
 
 \item we adjust the Carter ``constants'' (they became space-time functions for a congruence) in order to reproduce the null tetrad we find on the horizon;
 
 \item using the Killing--Yano tensor of the Kerr--Newman metric, we construct full null tetrad \emph{in the neighbourhood} of the horizon such that all vectors of the tetrad are parallelly transported along vector $n^a$ of the null tetrad;
 
 \item we identify the congruence of \cite{Fletcher2003a} as a special case of our construction and compare both approaches.
 
\end{itemize}
%
These
results are not a part of the main text of the thesis, but they are to be published in the aforementioned paper. 
The thesis itself is organized in the following way.

Chapter \ref{chapter:tetrad form} contains a review of the tetrad formalism and its connection to the spinor formalism. We follow \cite{Kinnersley1969}, \cite{Chandrasekhar1983} and \cite{Newman2009}. While the tetrad formalism can be formulated in purely tensorial language, the actual motivation for particular definitions comes from the formalism of the two-component spinors on curved space-times. For this reason, we develop both Newman--Penrose formalism and the spinor formalism together.

In the chapters \ref{chapter:non-exp} and~\ref{chapter:isol}, we introduce, step by step, the formalism of non-expanding and isolated horizons, emphasizing the aspects important for our work. Ultimately, we describe the procedure of \cite{Krishnan2012} and clarify some details of the construction which are missing or even misleading in that article. In particular, we spend some effort on proving that the Lie dragging of the tetrad along the horizon is mathematically consistent which is not discussed in \cite{Krishnan2012}.

The last part of the work, mainly contained in chapter~\ref{chapter:kerr}, establishes a non-twisting null tetrad in the Kerr--Newman space-time. In this chapter, we explain the gauge freedom in the choice of the Newman--Penrose null tetrad and give explicit relations for all possible transformations, because not all of them are present in the literature and we had to employ all of them in our construction. 

In appendix \ref{app:np formalism}, we give basic definitions and relations of the Newman--Penrose formalism, in appendix \ref{app:series} we list perturbative expansions referenced earlier in the text. 

In order to achieve the aforementioned goals, we have used the computer algebra system \texttt{Mathematica}. For the work with the Newman--Penrose formalism, we have employed existing script made by the supervisor of this thesis and introduced in detail in \cite{Scholtz2012}. Moreover, in appendix \ref{appendix:mathematica} we list Mathematica code developed for the purposes of this thesis, so that the reader can follow the calculations that have been done.

\chapter{Tetrad formalism}
\label{chapter:tetrad form}
The metric tensor is the fundamental object of study in the standard formulation of general relativity. It is represented by a matrix the elements of which are components of the metric tensor with respect to a particular \emph{coordinate} system $x^\mu$, and its coordinates-induced basis vectors $\pd_\mu$ and dual one-forms $\d x^\mu$.\footnote{The term $\pd_\mu$ is the usual abbreviation of $\pd/\pd{x^\mu}$.} The metric tensor, together with its derivatives, is the only constituent of the Riemann tensor, and therefore is the solution of the Einstein field equations.

However, in some cases, it is more convenient to work in a \emph{non-coordinate} basis, called an \focc{$\boldsymbol{n}$-ade}. The entire process of introducing this approach was for the 4-dimensional space-time nicely described in \cite[sections 6 and 7]{Chandrasekhar1983}. However, there is not only vector motivated point of view, but also spinor motivated one, also described in \cite[section 102]{Chandrasekhar1983}, which was the one initiating a special choice of tetrad we are going to use. Therefore we shall describe these two sides of a coin side by side. We shall discuss the benefits of such a basis later. At this point, we want to construct the non-coordinate basis, and find a relation between the two frames (coordinate and non-coordinate). 

Before we set sail, let us start with a useful notation.

\section{Abstract index notation}
We will be using lower-case Latin letters as so called \focc{abstract indices} which were introduced by Roger Penrose. Such an index only tells us to which vector space the tensor belongs to. It says nothing about its components -- it is not an $n$-tuple. This notation is kind of a compromise between a physicist and a mathematician view. It uses the advantages of the index notation and is avoiding the drawback of having to refer to a particular basis, whether explicitly or not. To do so, we need to introduce basis-free operation which could be mirrored exactly to the ones used during computation with tensor components.

For example, $V^a$ is a vector from abstract vector space (module) $\mathfrak{S}^a$ and $\alpha_b$ a one-form from $\mathfrak{S}_b$. A tensor of type $(p,q)$ is then analogically 
$Q^{a_1\dots a_p}_{b_1\dots b_q}$.

To be still able to use the standard tensor notation with components, we shall use a hat to denote a set of components; hence, $V^{\hat{a}}$ stands for $(V^0,\dots,V^3)$. When using a number as an index, it is clear that it is a component, therefore there is no need for the hat, still it will sometimes be used to make obvious link between quantities. In the following section, we shall use also bold face upright indices with the same meaning (they will be indexing components), the only difference lies in the fact that we want to distinguish between an arbitrary tetrad (denoted by bold indices) and a particular one with which will have a great significance to us (we use the hat for this tetrad).

In the tensor notation, as we know, $V^{\hat{a}} \neq V^{\hat{b}}$. Therefore, we need two objects, $V^a$ and $V^b$, which both stand for the same vector $\vect{V}$ but are completely \emph{different}, to be able to rewrite terms including both $V^{\hat{a}}$ and $V^{\hat{b}}$ to the abstract index notation. Thus any vector $\vect{V}$ has to be associated with an infinite\footnote{Arbitrarily long expressions must be allowable.} collection of different copies $V^a,V^b,V^c,\dots$; each being an element of one copy of the module $\vect{V}$ belongs to. 

For the modules refer to \cite[page 76]{Penrose1987}. For more information on this topic, e.g.\ the axioms, one can take a look at \cite[chapter 2]{Penrose1987} or \cite[section 12.8]{Penrose2007}.

Now we are arriving back to the tetrads.
\section{Geometrical structures on manifolds}

We can assign a tangent space $T_P \MM$ to any point $P$ of the manifold $\MM$ which forms the space-time. Also cotangent space $T_P^\ast \MM$ can be constructed and is, by definition, dual to the tangent space. These two spaces have the same dimensionality $n$ as the manifold $\MM$. Consequently, the tensor algebra $\mathcal{T}(T_P \MM)$ exists. In this tangent space, we can choose an arbitrary basis, both coordinate and non-coordinate. We pick the non-coordinate one and denote it $e_{\tetcom{a}}$. The index ${\tetcom{a}}$ runs from $0$ to $n - 1$ and the basis is a set of $n$ vectors. The vectors $e_{\tetcom{a}}$ are thus said to constitute an $n$-ad (also called \focc{vielbein}\footnote{This word from German stands for \qm{many legs} and is pronounced as [feelbain].}). It has several special names reflecting the particular dimension of the space-time. For $3$-dimensional manifold it is referred to a \focc{triad} (or \focc{dreibein}). In the usual $(3\! + \! 1)$-dimensional physical case, which we are interested in, we talk about a \focc{tetrad} (\focc{vierbein}) and it can be represented by a set of 4 vectors $\{e_0, e_1, e_2, e_3\}$.

We could also be interested in the tangent spaces at all points of the manifold (or points from its connected subset) altogether, then a disjoint union of the tangent spaces assigned to these points is a \focc{tangent bundle}. The vielbeins within the tangent bundle become fields because the basis is prescribed separately for every tangent space and thus each component of the basis is a vector chosen at every point of the manifold.\footnote{We are restricting ourselves to smooth manifolds; hence, the vector fields have to be smooth too, which means that the choice is not completely arbitrary.} They are called \focc{frame fields}.

We can also generalize this procedure and project all tensors onto vectors (vielbeins) and rephrase all equations in terms of these projections. This is known as the \focc{Cartan formalism} or, equivalently, as the formalism of \focc{rep\`{e}res mobiles}.

We have been talking about coordinate and non-coordinate bases and we hope that the usage of the non-coordinate one will give us some simplification or new results; hence, we shall explain the difference between these two bases. Both of them live in the tangent space. The non-coordinate bases are those which cannot be derived from coordinate systems. The operators $\pd_\mu$ and $\pd_\nu$ commute for all $\mu,\nu$, but two arbitrary vector fields do not. The commutator of two vector fields is also a vector field whose components do not vanish in general \cite{Schutz1980}. If the two vector fields are two elements of a basis and the commutator is non-zero, then it is not possible to re-express the basis as derivatives with respect to any coordinates, therefore the basis is non-coordinate.

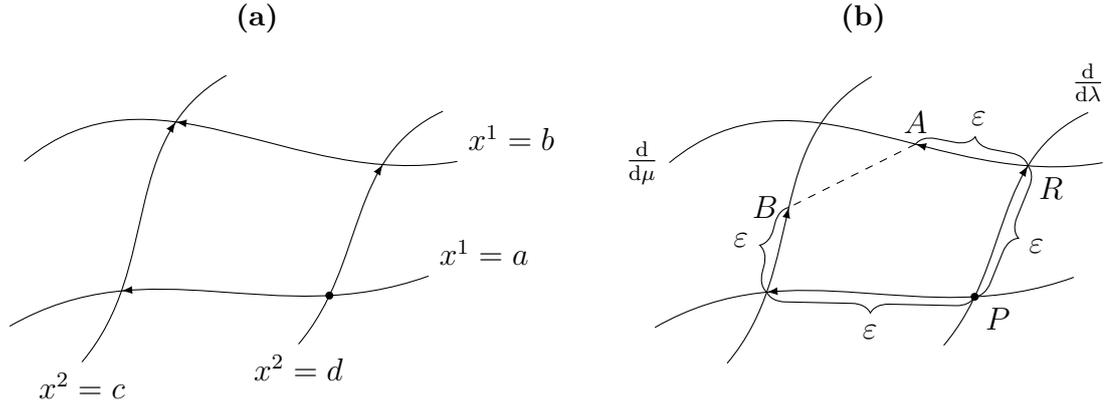
\begin{figure}
\begin{minipage}{\textwidth}
	\centering
		\begin{subfigure}[t]{0.45\linewidth}
			\centering
			\caption{}
			\begin{tikzpicture} [scale=0.95]
\path[draw, name path=top] (-0.8,2.8) to [out=40, in=190] (5.2,2.8) node[anchor=south west] {$x^1 = b$};
\path[draw, name path=bottom] (-1,0.5) to [out=30, in=200] (4.8,1.2) node[anchor=south west] {$x^1 = a$};
\path[draw, name path=right] (3,0.25) node[anchor=north] {$x^2 = d$} to [out=50, in=205] (5,3.5);
\path[draw, name path=left] (0,0) node[anchor=north] {$x^2 = c$} to [out=50, in=210] (2,4);


\path[name intersections={of = bottom and left}];
\coordinate (sec-lb)  at (intersection-1);
\path[name intersections={of = top and right}];
\coordinate (sec-rt)  at (intersection-1);
\path[name intersections={of = top and left}];
\coordinate (sec-lt)  at (intersection-1);
\path[name intersections={of = bottom and right}];
\coordinate (sec-rb)  at (intersection-1);

\node[circle,minimum size=2pt,inner sep=0pt,name path=circle-lb] at (sec-lb) {};
\node[circle,minimum size=2pt,inner sep=0pt,name path=circle-rt] at (sec-rt) {};
\node[circle,minimum size=2pt,inner sep=0pt,name path=circle-lt] at (sec-lt) {};

\path[name intersections={of = circle-lb and bottom}];
\coordinate (tan-lb)  at (intersection-1);

\path[name intersections={of = circle-rt and right}];
\coordinate (tan-rt)  at (intersection-2);

\path[name intersections={of = circle-lt and top}];
\coordinate (tan-A)  at (intersection-2);

\path[name intersections={of = circle-lt and left}];
\coordinate (tan-B)  at (intersection-2);

\draw[latex-] (sec-lb) -- (tan-lb);
\draw[latex-] (sec-rt) -- (tan-rt);
\draw[latex-] (sec-lt) -- (tan-A);
\draw[latex-] (sec-lt) -- (tan-B);

\node[circle,fill=black,minimum size=3pt,inner sep=0pt] at (sec-rb) {}; 

\end{tikzpicture} 
			\label{img > tetrad formalism > integral curves of coordinate basis}
		\end{subfigure}
	\hfill
		\begin{subfigure}[t]{0.45\linewidth}
			\catcode`-=12
			\centering
			\caption{}
			\begin{tikzpicture} [scale=0.95]
\path[draw, name path=left] (0,0) node[anchor=north,white] {$x^2 = c$} to [out=50, in=210] coordinate[pos=0.5] (B) (2,4);
\path[draw, name path=right] (3,0.25) to [out=50, in=205] (5,3.5) node[anchor=south] {$\der{}{\lambda}$};
\path[draw, name path=bottom] (-1,0.5) to [out=30, in=200] (4.8,1.2);
\path[draw, name path=top] (-0.8,2.8) node[anchor=east] {$\der{}{\mu}$} to [out=40, in=190] coordinate[pos=0.6] (A) (5.2,2.8);
\path[draw,name path=A--B] [dashed] (A) -- (B);
\draw [name intersections={of=bottom and right, by=p}] (p) node[anchor=north west] {$P$};
\draw [name intersections={of=top and right, by=r}] (r) node[anchor=north west] {$R$};
\draw (B) node[anchor=east] {$B$};
\draw (A) node[anchor=south] {$A$};
\draw [name intersections={of=bottom and left, by=LB}];
\draw [decorate,decoration={brace, mirror, amplitude=7pt}] (p) -- (r); 
\coordinate (base) at ($(p)!0.5!(r)$);
\coordinate (ortho) at ($ (0,0) ! 1 ! 90:($ (r) - (p) $) $);
\node[inner sep=2pt] at ($(base)!-15pt!($(base)+(ortho)$)$) {$\epsilon$};
\draw [decorate,decoration={brace, amplitude=7pt}] (A) -- (r); 
\coordinate (base) at ($(A)!0.5!(r)$);
\coordinate (ortho) at ($ (0,0) ! 1 ! 270:($ (r) - (A) $) $);
\node[inner sep=2pt] at ($(base)!-15pt!($(base)+(ortho)$)$) {$\epsilon$};
\draw [decorate,decoration={brace, amplitude=7pt}] (p) -- (LB); 
\coordinate (base) at ($(p)!0.5!(LB)$);
\coordinate (ortho) at ($ (0,0) ! 1 ! 270:($ (LB) - (p) $) $);
\node[inner sep=2pt] at ($(base)!-15pt!($(base)+(ortho)$)$) {$\epsilon$};
\draw [decorate,decoration={brace, amplitude=7pt}] (LB) -- (B); 
\coordinate (base) at ($(LB)!0.5!(B)$);
\coordinate (ortho) at ($ (0,0) ! 1 ! 270:($ (B) - (LB) $) $);
\node[inner sep=2pt] at ($(base)!-15pt!($(base)+(ortho)$)$) {$\epsilon$};
\path[name intersections={of = bottom and left}];
\coordinate (sec-lb)  at (intersection-1);
\path[name intersections={of = top and right}];
\coordinate (sec-rt)  at (intersection-1);

\node[circle,minimum size=2pt,inner sep=0pt,name path=circle-lb] at (sec-lb) {};
\node[circle,minimum size=2pt,inner sep=0pt,name path=circle-rt] at (sec-rt) {};
\node[circle,minimum size=2pt,inner sep=0pt,name path=circle-A] at (A) {};
\node[circle,minimum size=2pt,inner sep=0pt,name path=circle-B] at (B) {};

\path[name intersections={of = circle-lb and bottom}];
\coordinate (tan-lb)  at (intersection-1);

\path[name intersections={of = circle-rt and right}];
\coordinate (tan-rt)  at (intersection-2);

\path[name intersections={of = circle-A and top}];
\coordinate (tan-A)  at (intersection-2);

\path[name intersections={of = circle-B and left}];
\coordinate (tan-B)  at (intersection-2);

\draw[latex-] (sec-lb) -- (tan-lb);
\draw[latex-] (sec-rt) -- (tan-rt);
\draw[latex-] (A) -- (tan-A);
\draw[latex-] (B) -- (tan-B);

\node[circle,fill=black,minimum size=3pt,inner sep=0pt] at (p) {}; 


\end{tikzpicture}
			\label{img > tetrad formalism > integral curves of non-coordinate basis}
		\end{subfigure}
	\caption[A comparison of integral curves of coordinate and non-coordinate bases.]{A comparison of integral curves of coordinate and non-coordinate bases.\ \textbf{(a)}~Integral curves of a coordinate basis.\ \textbf{(b)}~Integral curves of a non-coordinate basis.%
	\footnote{The images were done according to \cite{Schutz1980}.}
	}
\end{minipage}
\end{figure}

The distinct character of the two types of bases cannot be seen at a single point. It depends on derivatives and accordingly is a matter of some neighbourhood in the manifold. The commutator $\left[X, Y\right]$ of the two vector fields $X$ and $Y$ is called the \focc{Lie bracket}. As an example, we can look at a coordinate grid on a 2-dimensional mani\-fold as it is discussed in \cite{Schutz1980}. The coordinates are constant alongside the integral curves of the remaining coordinate. This is the reason why they commute, an example is shown in the picture \ref{img > tetrad formalism > integral curves of coordinate basis}. However, the integral curves of two arbitrary vector fields are not necessarily curves of constant parameter of the other field. When not, the basis is non-coordinate. The Lie bracket has geometrical interpretation of the difference between paths on a parallelogram carried out alongside the integral curves in reversed order as it can be seen in picture \ref{img > tetrad formalism > integral curves of non-coordinate basis}.

A basis with vanishing commutators is called \focc{holonomic} (this is the coordinate one), otherwise it is \focc{non-holonomic}\footnote{\focc{Anholonomic} can also be seen.}.

The convenience
%
%
of usage of vielbeins arises from the fact that it can reflect important physical aspects of the space-time, e.g.\ the symmetries of the space-time. The symmetry should be taken in account when choosing the vector bases of the tangent bundle.%

Since both the coordinate basis $\pd_\mu$ and the tetrad $e_{\tetcom{a}}$ form the basis of the same tangent space, we can express the one in terms of the other. Thus, the general tetrad $e_{\tetcom{a}}$ is a linear combination of the basis vectors $\pd_\mu$,
\begin{equation}
e_{\tetcom{a}} = e_{\tetcom{a}}{}^\mu\,\pd_\mu \,, \footnote{This type of transformation would give us a new, possibly holonomic, basis in general. However, we will have mainly non-holonomic bases in mind.}
\end{equation}
where $e_{\tetcom{a}}{}^\mu$ is a regular matrix, $\det e_{\tetcom{a}}{}^\mu \neq 0$ (otherwise $e_{\tetcom{a}}$ would be linearly dependent and hence not constitute a basis). This set of vector fields is chosen in such a way that
\begin{equation}
\label{tf > vielbein definition}
g(e_{\tetcom{a}}, e_{\tetcom{b}}) = \eta_{\tetcom{a}\tetcom{b}} \,,
\end{equation}%
where $g$ is the space-time metric tensor and $\eta$ is a (habitually constant\snote{Chandrasekhar p. 35 pod rovnicí (243) objasňuje, že to můžou být funkce}) non-de\-ge\-ne\-ra\-te matrix, usually the Minkowski matrix, which is only an expression of the tangent space metric tensor in another frame. The equation \eqref{tf > vielbein definition} can be rewritten in terms of components of the Lorentz frame $e_{\tetcom{a}}{}^\mu$ with respect to the basis $\partial_\mu$ as
\begin{equation}
\label{tf > metric connection by vielbein}
g_{\mu\nu} e_{\tetcom{a}}{}^\mu e_{\tetcom{b}}{}^\nu = \eta_{\tetcom{a}\tetcom{b}} \,.
\end{equation}
We can define the dual tetrad in terms of the inverse matrix
\begin{equation}
\( e_{\tetcom{a}}{}^\mu \)^{-1} = e_\mu{}^{\tetcom{a}}
\end{equation}
and invert also \eqref{tf > metric connection by vielbein} to get the space-time metric in terms of inner product in the tetrad frame
\begin{equation}
g_{\mu\nu} = \eta_{\tetcom{a}\tetcom{b}} e_\mu{}^{\tetcom{a}} e_\nu{}^{\tetcom{b}} \,. 
\end{equation}
Note that the inverse tetrad vectors are denoted by the same character and look like
\begin{equation}
e^{\tetcom{a}} = e_\mu{}^{\tetcom{a}}\, \d x^\mu
\end{equation}
and the only thing distinguishing the two inverse matrices is the position of their indices.

Why do we actually use the coordinate bases? The reason is that a coordinate basis not only provides the dot product, as the Minkowski metric (or any other non-coordinate metric) but, moreover, it allows us to compute Christoffel symbols by means of derivatives of the metric, and it gives us the line element. None of these can be done with a non-coordinate basis.

The choice of the vielbein is a matter of convenience. One of the most fruitful choices is that of Newman and Penrose (1962), \cite{Newman2009}. In honour of the authors, it is called the \gls{np} formalism.

\section{Newman--Penrose tetrad}
The \gls{np} formalism is also known as the \focc{spin-coefficient formalism}. It is based on a \focc{null tetrad}, a tetrad composed of null vectors. This is convenient, for example, when treating gravitational waves, because then we are investigating \emph{null geodesics}, and when finding exact solutions. As is nicely summed up in \cite{Newman2009}, the main advantages are:
\begin{enumerate}
	\item All the equations are of the first order.
	\item They can be grouped into sets of linear equations.
	\item The number of equations is reduced by half as they are complex.
	\item The equations can be written out explicitly without index and summation conventions.
	\item One can focus on individual scalar equations and their geometrical and physical meaning.
	\item It allows to search for solutions with specific features.
\end{enumerate}

To introduce the null tetrad, we start with the Minkowski set of vectors $\left\{ t^a, x^a, y^a, z^a \right\}$. We know that in the four-dimensional space-time, we can find two independent null vectors -- these are, e.g., the ones which lie on the opposite sides of the light cone in $t^a$ and $z^a$ plane. 
\begin{figure}
\centering
\begin{tikzpicture}
\draw [latex-latex] (0,3) node [left] {$t$} -- (0,0) -- (3,0) node [below] {$z$};
\draw (0,0) -- (2,2);
\draw (0,0) -- (-2,2);
\draw (-2,2) to [out=30, in=150] (0,2);
\draw (-2,2) to [out=-30, in=-150] (0,2);
\draw (0,2) to [out=30, in=150] (2,2);
\draw [-latex, thick] (0,0) -- (1,1) node [below right] {$l^a$};
\draw [-latex, thick] (0,0) -- (-1,1) node [below left] {$n^a$};
\end{tikzpicture}
\caption{A construction of the null vectors from the Minkowski time coordinate and $z$-direction.}
\label{img:tetrad formalism:null vectors}
\end{figure}
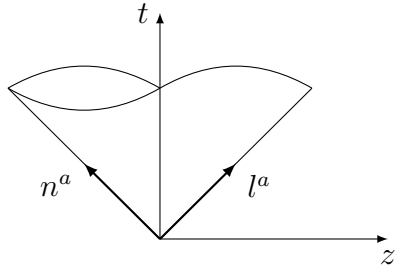
They are given by
\begin{subequations}
\begin{align}
l^a &= \frac{1}{\sqrt{2}} \!\( t^a + z^a \)\! \,,\\
n^a &= \frac{1}{\sqrt{2}} \!\( t^a - z^a \)\! \,.
\end{align}
The situation is covered by figure~\ref{img:tetrad formalism:null vectors}.

Any other two null vectors which are a linear combination of the Minkowski set of vectors with real coefficients would necessarily be linearly dependent (at least one of them).
\begin{myproof}
We would like to know how many vectors $v^a$ exist that they are of the form $v^a = a\, t^a + b\, x^a + c\, y^a + d\, z^a$ (with $a$, $b$, $c$, $d \in \mathbb{R}$), that they are null ($v^a\,v_a = 0$) and they are linearly independent. The condition of nullness gives us constraining equation for the coefficients: $a^2 = b^2 + c^2 + d^2$. This is an equation for a sphere with radius $a$. This sphere represent a 3-parametric space -- we can check this by using spherical coordinates. Hence, only three independent vectors can be found. At least the last one has to have complex coefficients.
\end{myproof}

For symmetry reasons and latter convenience, we introduce two other complex vectors to form a tetrad which could be such a basis in the tangential space that all the vectors are null. We choose them as
\begin{align}
{m}^a &= \frac{1}{\sqrt{2}} \!\( x^a - \im y^a \)\! \,,\\
\cconj{m}^a &= \frac{1}{\sqrt{2}} \!\( x^a + \im y^a \)\! \,.
\end{align}\label{eq:null to orthonormal}
\end{subequations}

Which contractions are non-zero? We can easily compute this with usage of the Minkowski vectors, and find out there are only two of them which are non-zero. They are
\begin{subequations}
\begin{align}
l^a n_a &= 1 \,, \label{eq > tetrad formalism > contractions of basis vectors > ln} \\ 
m^a \cconj{m}_a &= -1 \,. \label{eq > tetrad formalism > contractions of basis vectors > mm}
\end{align}
\label{eq > tetrad formalism > contractions of basis vectors}
\end{subequations}

When we have been discussing the~\eqref{tf > metric connection by vielbein}, we have pointed out that the matrix $\eta_{\tetcom{a}\tetcom{b}}$ is usually the Minkowski matrix. However, this is not the case in the \gls{np} formalism. Let's rewrite~\eqref{tf > vielbein definition} in the abstract index formalism and the components of the new tetrad as
\begin{equation}
\label{np > from minkowski to null metric}
\eta_{\hat{a}\hat{b}} = g_{ab} e_{\hat{a}}{}^a e_{\hat{b}}{}^b \,.
\end{equation}
Then we get an analogue of~\eqref{tf > metric connection by vielbein} where $\eta_{\hat{a}\hat{b}}$ is a matrix representing the metric in the \gls{np} tetrad frame and $g_{ab}$ is the metric tensor. We remind that indices which have a hat over them refer to the components of tensors with respect to the null tetrad (they are indexing the basis vectors). Undecorated Latin indices are the abstract indices. The tetrad itself is $e_{\hat{a}}{}^a = (l^a,n^a,m^a,\cconj{m}^a)$.

We would also like to have the dual basis to the $e_{\hat{a}}{}^a$. It has to meet
\begin{equation}
e_{\hat{a}}{}^a e^{\hat{b}}{}_a = \delta_{\hat{a}}^{\hat{b}} \,,
\end{equation}
where $\delta_{\hat{a}}^{\hat{b}}$ is the unit matrix -- $\text{diag}(1,1,1,1)$.\footnote{Do not confuse $\delta_{\hat{a}}^{\hat{b}}$ which is a matrix with $\delta_a^b$ which is a tensor, equivalently a map between two copies of the same abstract vector space.} It can be easily found, with usage of the contractions between basis vectors, that $e_a{}^{\hat{a}} = (n_a,l_a,-\cconj{m}_a,-m_a)$.

Why were there no coordinate indices in~\eqref{np > from minkowski to null metric}?\snote{je tento odstavec smysluplný?}
The reason is that in the first section we were basically explaining how to work locally in a flat Minkowski space instead of a general curved space-time. Now we do not want to work in Minkowski frame anymore, we want to work with the null tetrad. However, we defined its basis as the linear combination of Minkowski basis, therefore we are locally rewriting a curved space-time into the Minkowski one and then using the procedure again to work in a null frame. We have gone directly from the coordinate basis to the null tetrad to avoid the Minkowski metric just by plugging \eqref{tf > metric connection by vielbein} into \eqref{np > from minkowski to null metric} and joining $e_{\tetcom{a}}{}^\mu e_{\hat{a}}{}^{\tetcom{a}}$ into~$e_{\hat{a}}{}^\mu$ and then used the general, basis free, abstract notation. According to this, we wanted to emphasize that it is a similar process which differs in the notation of indices, only.

It is very important to distinguish between the Minkowski metric tensor $\eta_{ab}$, matrix $\eta_{\tetcom{a}\tetcom{b}}$ and the representation of the metric in the null basis $\eta_{\hat{a}\hat{b}}$. Though same notations are used for them, they are completely different. The matrix $\eta_{\hat{a}\hat{b}}$ is not even diagonal as $\eta_{\tetcom{a}\tetcom{b}}$ is, and $\eta_{ab}$ is not a matrix at all.

We only need the knowledge of the contractions between the null basis vectors to find out how $\eta_{\hat{a}\hat{b}}$ looks like. To compute an element of the matrix, we set the numerical value of its indices and then only compute inner product between two basis vectors of the tetrad, e.g., the element $\eta_{\hat{0}\hat{1}}$:
\begin{equation}
\eta_{\hat{0}\hat{1}} = g_{ab} e_{\hat{0}}{}^a e_{\hat{1}}{}^b = g_{ab} l^a n^b = l^a n_a = 1 \,.
\end{equation}
We do not need to know how the representation of the metric tensor $g_{ab}$ looks like, it is enough to know that it performs the inner product. The resulting matrix is
\begin{equation}
	\eta_{\hat{a}\hat{b}} =
		\begin{pmatrix}
			0 & 1 & 0 & 0 \\
			1 & 0 & 0 & 0 \\
			0 & 0 & 0 & -1 \\
			0 & 0 & -1 & 0
		\end{pmatrix}
\end{equation}

In relativity, usually, we solve the Einstein equations where the unknown variable is the metric. Now we know how it looks like for the \gls{np} tetrad. So, have all the work been done already? Of course not. We have \emph{chosen} how the matrix representation of the metric looks like. Usually, when using a coordinate tetrad, the metric gives us covariant derivative, Christoffel symbols and because the Riemann tensor is given only by metric and its derivatives, it is also granted. Now, they are \emph{not}. Not even the metric is really known, we know only its representation in the Newman--Penrose tetrad, but the tetrad is not given, we have to find it in order to get the metric tensor. It is clear from~\eqref{np > from minkowski to null metric} that only by contracting the equation with the tetrad, we arrive at an expression for the metric:
\begin{equation}\label{eq:metric from tetrad}
	g_{ab} = 2 l_{(a} n_{b)} - 2 m_{(a} \cconj{m}_{b)}
\end{equation}
where the parentheses denote symmetrization. We shall discuss this topic later on in section \ref{sec: unknown parameters}.

\section{Spin coefficients}
We start a discussion of the formalism with the computation of a covariant derivative in the null frame and its projection onto the null tetrad.

The following symbol $\nabla_{\hat{a}}$ is the covariant derivative projected onto the tetrad, this can be written as $\nabla_{\hat{a}} = e_{\hat{a}}{}^a \nabla_a$. The covariant derivative we shall be using, unless otherwise stated explicitly, is the metric compatible and torsion free \emph{Levi-Civita} covariant derivative.

The \focc{Ricci rotation coefficients} are defined, \cite{Newman2009}, by
\begin{equation}\label{eq:tetrad formalism:def spin coef}
	\nabla_{\hat{a}} e_{\hat{b}}{}^a = \gamma_{\hat{a}\hat{b}}{}^{\hat{c}}\, e_{\hat{c}}{}^a \, 
\end{equation}
and\snote{toto nemusí být součístí definice ale důsledkem vlastností $\eta$ - chandrasekhar p. 37}
\begin{equation}
	\gamma_{\hat{a}\hat{b}\hat{c}} = - \gamma_{\hat{b}\hat{a}\hat{c}} = \eta_{\hat{a}\hat{d}} \gamma^{\hat{d}}{}_{\hat{b}\hat{c}} \,.
\end{equation}
Some of them are real, some pure imaginary and the rest of them is complex.

Let $V^b$ be a vector field. Using the Ricci coefficients, we can calculate its covariant derivative as
\begin{equation}
	\nabla_{\hat{a}} V^b = 
		\nabla_{\hat{a}}\! \(V^{\hat{b}}\, e_{\hat{b}}{}^b \) = 
			\(e_{\hat{a}}\!\(V^{\hat{c}}\) + \gamma_{\hat{a}\hat{b}}{}^{\hat{c}}\, V^{\hat{b}} \)\! e_{\hat{c}}{}^b \,.
\end{equation}
As it is usual in mathematical notation, by the action of $e_{\hat{a}}$ on $V^{\hat{c}}$, $e_{\hat{a}}\!\(V^{\hat{a}}\)$, we mean $e_{\hat{a}}{}^\mu\,\partial_\mu V^{\hat{c}}$. 

\begin{table}
	\catcode`-=12
	\centering
	\begin{tabular}{*{4}{c}}
\toprule
$\nabla$ & $m^a\,\nabla l_a$ & $\frac{1}{2}\(n^a\, \nabla l_a - \cconj{m}^a\, \nabla m_a\)$ & $-\cconj{m}^a\, \nabla n_a$ \\
\midrule
$D$ & $\kappa$ & $\epsilon$ & $\pi$ \\
$\Delta$ & $\tau$ & $\gamma$ & $\nu$ \\
$\delta$ & $\sigma$ & $\beta$ & $\mu$ \\
$\cconj{\delta}$ & $\scrho$ & $\alpha$ & $\lambda$ \\
\midrule
$\nabla$ & $o^A\, \nabla o_A$ & $o^A\, \nabla \iota_A = \iota^A\, \nabla o_A$ & $\iota^A\, \nabla \iota_A$ \\
\bottomrule
\end{tabular} 
%
%
	\caption{A table of connection coefficients.}
	\label{tab > tetrad formalism > spin coef > spin coef}
\end{table}
We can find a list of \focc{spin coefficients} in table~\ref{tab > tetrad formalism > spin coef > spin coef}, \cite{Stewart1993}. As can be seen from the table, they are simply given as complex combinations of the Ricci spin coefficients. The~$\nabla$ symbol in the first and last lines is substituted by the corresponding letter in the first column. The letters in it denote the projections of covariant derivative operator\footnote{It is not a projection of the covariant derivative because that would be projected twice. The result of this operation is a vector, not a scalar as it would be in the case of projection of covariant derivative.} onto direction of the bases vectors as follows
\begin{equation}
	D = l^a\, \nabla_a \,, 
		\qquad 
	\Delta = n^a\, \nabla_a \,, 
		\qquad 
	\delta = m^a\, \nabla_a \,, 
		\qquad 
	\cconj{\delta} = \cconj{m}^a\, \nabla_a \,.
\end{equation}
We could also write it as $D = \nabla_{\hat{0}}\,,\dots\,, \cconj{\delta} = \nabla_{\hat{3}}$.

The last line in the table is the spinorial form of the first one. This way of description will be discussed in the section to follow. From this definition arises the complex combinations. This point of view is discussed also in \cite{Newman2009}.

As an example of how to read the table, we take the coefficient $\kappa$, it is determined as
\begin{equation}
\kappa = m^a D l_a = m^a l^b\, \nabla_b l_a \,.
\end{equation}

Let us compute the covariant derivative of the first basis vector in the direction of itself to demonstrate the usage of the spin coefficients. It surely has to be given in terms of the basis such as
\begin{equation}\label{eq > spin coefficients > example of covariant derivative}
D l^a = a l^a + b n^a + c m^a + \cconj{c} \cconj{m}^a \,.
\footnote{One might have wondered why is the last term proportional to the \gls{coef}~$\cconj{c}$ instead of some~$d$. The reason is that the \gls{lhs} is real and therefore the \gls{rhs} also has to be real.}
\end{equation}

In order to express $a,b$ and $c$ in terms of the spin coefficients, we project the equation successively onto the tetrad and use the contractions \eqref{eq > tetrad formalism > contractions of basis vectors}. Contracting with $l_a$, we get
\begin{equation}
l_a\, D l^a = b \,.
\end{equation}
And using that the basis vectors are null, it follows
\begin{align}
l_a\, l^a &= 0  \quad/\, D \,, \\
2 l_a\, D l^a &= 0 \,.
\end{align}
This gives us

\begin{equation}
b = 0 \,.
\end{equation}
To compute $a$, we just multiply~\eqref{eq > spin coefficients > example of covariant derivative} by $n_a$. For we get $n_a\, D l^a = a$, we are interested in $\epsilon$ spin coefficient from the table~\ref{tab > tetrad formalism > spin coef > spin coef}. Our situation is slightly complicated by the fact that this coefficient is not given only by the term we want, but also by $\cconj{m}^a \nabla m_a$. This term is complex in contrast to the real one we want to compute. On the other hand, considering both $\epsilon$ and its \gls{cc}, we find
\begin{align}
\epsilon + \cconj{\epsilon} 
&= n^a D l_a - \frac{1}{2} \!\( \cconj{m}^a D m_a + m^a D \cconj{m}_a \)  \nonumber\\
&= n^a D l_a - \frac{1}{2} \!\( \cconj{m}^a D m_a + D \!\(m^a \cconj{m}_a\) - \cconj{m}^a D m_a \)  \nonumber\\
&= n^a D l_a \nonumber\\
&= a \,.
\end{align}
We used the contraction~\eqref{eq > tetrad formalism > contractions of basis vectors > mm} and the fact that covariant derivative of a constant is zero. Note that complex conjugate of $D$ is $D$ itself because covariant derivative is real operator and so is the vector $l^a$. When treating the term $m^a D \cconj{m}_a$, we have commuted the two resulting terms using the symmetry of the scalar product and also used that we can lower and raise indices around the \emph{covariant} derivative (by the means of the metric, covariant derivative of which is zero, for we are using metric covariant derivative).

We are left with computation of $c$. As we can see, it is related to the spin \gls{coef}\ $\kappa$:
\begin{equation}
\cconj{c} = - m^a D l_a = - \kappa \,.
\end{equation}

The final expression for the covariant derivative of the first tetrad vector in its own direction is
\begin{equation}\label{eq:spin coef:example}
D l^a = \( \epsilon + \cconj{\epsilon} \) l^a - \cconj{\kappa}\, m^a - \kappa\, \cconj{m}^a \,.
\end{equation}
We have not chosen this certain example for no reason. The resulting equation has two important explications.

Firstly, it is an equation that tells us how is $l^a$ propagated to the space-time. For this reason, it is called \focc{transport equation}. For the complete set of transport equations, one can take a look into appendix~\ref{app:sec:transport eq}.

The second significance of this particular transport equation is even more important for us. If the spin \gls{coef}\ $\kappa = 0$, the equation~\eqref{eq:spin coef:example} becomes a \focc{geodesic equation}. As we can see, the real part of $\epsilon$ then measures deviation from the affine parametrization. It turns out that the case $\kappa = 0$ is the one we are typically interested in.

Before we describe in the next chapter why $\kappa$ should have such a specific value in our investigation, let us make a few digressions, we start with introduction of a complementary notation to the \gls{np} tetrad.

\section{Spinor formalism}
As mentioned earlier, the spinor formalism motivated our particular choice of tetrad and, in fact, is the underlying reason for the appearance of terms in the tetrad formalism. It is also more powerful for some kinds of computations, which is useful thanks to a straight link to the tetrad allowing us to change the point of view. Nevertheless, the tetrad is the main object of study for us, and, therefore, we establish the spinorial basis to meet the tetrad although it was originally done otherwise. One can take a look into the original paper by Penrose \cite{Penrose1962} or into some more recent works, e.g.\ \cite{Penrose1987}.

When using the spinor formalism, we think of the \gls{np} tetrad as constructed from a pair of spinors: $(o^A, \iota^A)$.\footnote{To read equations properly, please note that the first spinor is usually denoted by Greek letter omicron.}
The spin-frame basis is normalised by conditions
\begin{equation}\label{eq:spinform:basisnorm}
	\levicivita_{AB} o^{A} o^{B} = 0 \,, 
		\qquad 
	\levicivita_{AB} o^A \iota^B = 1 \,,
\end{equation}
where $\levicivita_{AB}$ is a skew-symmetric tensor in 2-dimensions (the Levi-Civita tensor). The first normalization condition is necessarily fulfilled due to (anti-)symmetry. The Levi-Civita symbol behaves similarly to the metric tensor on the spinors -- it is performing the inner 
product
\begin{equation}
\levicivita_{AB} \xi^A \eta^B = \xi_B \eta^B = - \xi^A \eta_A \,.
\end{equation}
Alternatively, we could write
\begin{equation}\label{eq:spinform:lowerraise}
	\xi^A = \levicivita^{AB} \xi_B \,, 
		\qquad 
	\xi_A = \levicivita_{BA} \xi^B \,,
\end{equation}
where $\levicivita_{AB}$ and $\levicivita^{AB}$ are inverse to each other in the sense of
\begin{equation}
	\levicivita^{AC} \levicivita_{BC} = \delta^A_B \,.
\end{equation}

For~\eqref{eq:spinform:basisnorm} holds true, we can easily determine the dual basis to be $(-\iota_A, o_A)$.

Also the Levi-Civita symbol can be expressed via the spinor basis. It has two indices and is skew-symmetric, therefore, it has to have the form
\begin{equation}
	\levicivita_{AB} = a o_A \iota_B + b o_B \iota_A \,.
\end{equation}
The coefficients $a$ and $b$ can be determined from the fact that it raises and lowers indices, see~\eqref{eq:spinform:lowerraise}.
\begin{align}
	\levicivita_{AB} o^A 	&= a o_A \iota_B o^A + b o_B \iota_A o^A  \nonumber \\
							&= 0 + b o_B \levicivita_{CA} \iota^C o^A \nonumber \\
							&\overset{!}{=} o_B \,.
\end{align}
Hence, $b = -1$. Analogously we would get $a = 1$. The Levi-Civita symbol in the spinor formalism is given as
\begin{equation}\label{eq:spinor levicivita}
	\levicivita_{AB} = o_A \iota_B - o_B \iota_A \,.
\end{equation}
The convention is $\levicivita_{01} = 1$, which leads to the matrix representation of $\levicivita_{AB}$ in this basis:
\begin{equation}
	\levicivita_{AB} = 
		\begin{pmatrix}
			0 & 1 \\
			-1 & 0
		\end{pmatrix}\! \,.
\end{equation}

Using the spinor formalism, we can also perform scalar product of 4-vectors thanks to their equivalence. Let $V^a = o^A \cconj{o}^{A'}$, where the \emph{primed} spin-space is \gls{cc}\ of the unprimed. Hence, the conjugation is
\begin{equation}
\cconj{\xi^A} = \cconj{\xi}{}^{A'} \,.
\end{equation}

As indicated, there is a correspondence between the spin-frame and the tetrad, therefore, there is also a relation between the indices -- one lower case abstract index is in accordance to a pair of upper case spinor indices -- $a \sim A A'$. The vectors of the dyad have two components, in comparison with the tetrad indices
\begin{equation}
	\hat{a} \in \left\{0,1,2,3\right\} , 
		\qquad 
	\hat{A} \in \left\{0,1\right\} .
\end{equation}

The normalization of the vector $V^a$ is
\begin{equation}
V^a V_a = o^A \cconj{o}^{A'} o_A \cconj{o}_{A'} = o^A \cconj{o}^{A'} \levicivita_{BA} \levicivita_{B'A'} o^B \cconj{o}^{B'} = 0 \,.
\end{equation}
The dot product $V^a V_a$ is zero because of the anti-symmetry of $\levicivita_{BA}$ and the symmetry of $o^A o^B$. The pair $o^B \cconj{o}^{B'}$ can be regarded as $V^b$ from which we have that
\begin{equation}
V^a V_a = \levicivita_{BA} \levicivita_{B'A'} V^a V^b \,,
\end{equation}
so the pair of metrics on the two \gls{cc} spin-spaces is the metric in the 4-dimensional space
\begin{equation}
\levicivita_{BA} \levicivita_{B'A'} = g_{ab} \,.
\end{equation}
As we can see, the anti-symmetry of the spinor metric is in the relation to the \gls{np} metric (which is symmetric) compensated by multiplicity of the $\levicivita_{AB}$. Similarly, one can show that the spinorial equivalent of the volume form $\levicivita_{abcd}$ is
\begin{align}\label{eq:levi-civita symbol}
 \levicivita_{abcd} = \im \left( \levicivita_{AB} \levicivita_{CD}  \levicivita_{A'C'} \levicivita_{B'D'} - \levicivita_{A'B'} \levicivita_{C'D'} \levicivita_{AC} \levicivita_{BD}\right)\! \,.
\end{align}

We have chosen one particular vector's spinor form, what other choices could we have done? One can easily see that for another choice $W^a = \iota^A \cconj{\iota}^{A'}$, the situation is the same -- it is null. We have found two null vectors, what is their scalar product?
\begin{equation}
V^a W_a = o^A \cconj{o}^{A'} \levicivita_{AB} \levicivita_{A'B'} \iota^B \cconj{\iota}^{B'} = 1 \,.
\end{equation}
These two vectors are therefore good candidates for spinor form of the first two \gls{np} tetrad vectors. We would also like to find the remaining two of them.

We have been combining $o^A$ and $\iota^A$ with themselves, we are left with their combinations
\begin{equation}
	U^a = o^A \cconj{\iota}^{A'} \,, 
		\qquad 
	\cconj{U}^a = \iota^A \cconj{o}^{A'} \,.
\end{equation}
It can be easily checked that they are null, and meet the desired contractions without any factor.

We arrive at simple relations between the tetrad and the spin-frame
\begin{align}\label{eq:tf:spinors:tetrad}
	l^a = o^A \cconj{o}^{A'}\,, 
		\qquad 
	n^a = \iota^A \cconj{\iota}^{A'} \,, 
		\qquad 
	m^a = o^A \cconj{\iota}^{A'} \,, 
		\qquad 
	\cconj{m}^a = \iota^A \cconj{o}^{A'} \,.
\end{align}
These relations can be also taken as the definition of the spin-frame.

\section{Geometrical introduction of spinors}
\def\conu{\mathfrak{z}}
The spinors can also be established from geometrical considerations. We start with a \emph{complex plane} described by a complex number $\conu$. The number can be expressed as a quotient of two numbers $\zeta$ and $\eta$: $\conu = \zeta / \eta$. The plane is stereographically projected onto a sphere (zero is projected to one pole a and the complex infinity to the other), the situation is shown in the figure~\ref{img:geospin:stereoprojection}.
\begin{figure}
\begin{minipage}{\textwidth}
  \centering
  \begin{tikzpicture} [mark coordinate/.style={inner sep=0pt,outer sep=0pt,minimum size=3pt,
    fill=black,circle}] 


\def\R{2.5} 
\def\angEl{25} 
\def\angAz{-105} 
\def\angPhi{-48} 
\def\angBeta{30} 


\pgfmathsetmacro\H{\R*cos(\angEl)} 
\tikzset{xyplane/.estyle={cm={cos(\angAz),sin(\angAz)*sin(\angEl),-sin(\angAz),
                              cos(\angAz)*sin(\angEl),(0,0)}}}
\LongitudePlane[xzplane]{\angEl}{\angAz}
\LongitudePlane[pzplane]{\angEl}{\angPhi}
\LatitudePlane[equator]{\angEl}{0}

\draw[name path=Circle1] (\R,0) arc (0:180:\R);
\draw[name path=Circle2,dashed] (\R,0) arc (0:-180:\R);


\coordinate (O) at (0,0,0);
\coordinate[mark coordinate] (N) at (0,\H);
\coordinate[mark coordinate] (S) at (0,-\H);
\path[pzplane] (\angBeta:\R) coordinate[mark coordinate] (P);
\path[pzplane] (\R,0) coordinate (PE);
\path[xzplane] (\R,0) coordinate (XE);
\coordinate[mark coordinate] (Phat) at (intersection cs: first line={(N)--(P)},
                                        second line={(O)--(PE)});


\DrawLatitudeCircle[\R]{0} 
\DrawMyNewLongitudeCircle[\R]{\angAz} 
\DrawMyNewLongitudeCircle[\R]{\angAz+90} 
\DrawMyNewColorLongitudeCircle[\R]{\angPhi}{red} 


\draw[xyplane,latex-latex,blue] (2*\R,0) node[below] {$X$} -- (0,0) -- (0,2*\R)
    node[right] {$Y$};
\draw[-latex,blue] (0,0) -- (0,1.6*\R) node[above] {$Z$};

\draw[dashed, thick,red] (P) -- (N) +(0.3ex,0.6ex) node[above left] {$\mathbf{N}$};
\draw[thick,red] (P) -- (Phat) node[right] {$\conu$};
\path (S) +(0.4ex,-0.4ex) node[below] {$\mathbf{S}$};
\draw[dashed,red] (O) -- (P) node[above right] {$\hat{\conu}$};
\draw[dashed,red] (O) -- (PE);
\draw[red] (PE) -- (Phat);
\draw[pzplane,-latex,thin] (90:0.5*\R) to[bend left=15]
    node[pos=0.4,right] {$\theta$} (\angBeta:0.5*\R);
\draw[equator,-latex,thin] (\angAz:0.4*\R) to[bend right=30]
    node[pos=0.4,below] {$\phi$} (\angPhi:0.4*\R);

\draw[xyplane] (-2*\R,-2*\R) -- (+3*\R,-2*\R) -- (3*\R,3*\R) -- (-2*\R,3*\R);

\LatitudePlane[equator]{\angEl}{0}

\pgfmathsetmacro\cosaz{ {cos(\angAz)} }
\pgfmathsetmacro\sinaz{ {sin(\angAz)} }
\pgfmathsetmacro\cosel{ {cos(\angEl)} }
\pgfmathsetmacro\sinel{ {sin(\angEl)} }

\pgfmathsetmacro\oxl{-2*\R}
\pgfmathsetmacro\oyl{-2*\R}
\pgfmathsetmacro\ozl{0}
\pgfmathsetmacro\nxl{\cosaz*\oxl - \sinaz*\oyl}
\pgfmathsetmacro\nyl{\sinaz*\oxl + \cosaz*\oyl}
\pgfmathsetmacro\nzl{0}
\pgfmathsetmacro\nnxl{\nxl}
\pgfmathsetmacro\nnyl{\cosel*\nyl - \sinel*\nzl}
\pgfmathsetmacro\nnzl{\sinel*\nyl + \cosel*\nzl}
\pgfmathsetmacro\oxr{-2*\R}
\pgfmathsetmacro\oyr{3*\R}
\pgfmathsetmacro\ozr{0}
\pgfmathsetmacro\nxr{\cosaz*\oxr - \sinaz*\oyr}
\pgfmathsetmacro\nyr{\sinaz*\oxr + \cosaz*\oyr}
\pgfmathsetmacro\nzr{0}
\pgfmathsetmacro\nnxr{\nxr}
\pgfmathsetmacro\nnyr{\cosel*\nyr - \sinel*\nzr}
\pgfmathsetmacro\nnzr{\sinel*\nyr + \cosel*\nzr}

\coordinate (LB) at (\nnxl,\nnzl);
\coordinate (RB) at (\nnxr,\nnzr);

\path[name path=mypath] (LB) -- (RB);

\path[name intersections={of=mypath and Circle1, total=\tot, sort by = curve 1}]
	\foreach \i in {1,...,\tot} {(intersection-\i) coordinate (Inter-\i)};

\draw (LB) -- (Inter-2);
\draw[dashed] (Inter-1) -- (Inter-2);
\draw (Inter-1) -- (RB);

\end{tikzpicture}
  \caption[Stereographical projection onto the complex plane.]{Stereographical projection onto the complex plane.%
  \footnote{A part of the code for this figure was adopted and modified from an example of usage of the Tikz system written by \textsc{Trzeciak, Tomasz M.} which was on the 8th of February, 2016, available at \url{http://www.texample.net/tikz/examples/map-projections/}.}
  }
  \label{img:geospin:stereoprojection}
\end{minipage}
\end{figure}

\begin{figure}
  \centering
\pgfmathsetmacro\th{80}
\pgfmathsetmacro\az{100}

\tdplotsetmaincoords{\th}{\az}

\pgfmathsetmacro\R{1}
\pgfmathsetmacro\v{1}
\pgfmathsetmacro\dif{0.2}
\pgfmathsetmacro\RR{(\R*(\v+\dif))/\v}

\begin{tikzpicture} [scale=3, tdplot_main_coords, axis/.style={-latex,blue,thick},
vector/.style={-stealth,black,very thick},
vector guide/.style={dashed,black,thick}]

\coordinate (O) at (0,0,0);

    \draw[axis] (0,0,0) -- (2,0,0) node[anchor=north east]{$X$};
    \draw[axis] (0,0,0) -- (0,2,0) node[anchor=north west]{$Y$};
    \draw[blue,thick,dashed] (0,0,0) -- (0,0,\v);
    \draw[axis] (0,0,\v) -- (0,0,2) node[anchor=south]{$t$};

\pgfmathsetmacro\cott{{cot(\th)}}
\pgfmathsetmacro\fraction{\R*\cott/\v}
\pgfmathsetmacro\angle{{acos(\fraction)}}

\pgfmathsetmacro\PhiOne{(\az-90)+\angle}
\pgfmathsetmacro\PhiTwo{(\az-90)-\angle}

\pgfmathsetmacro\sinPhiOne{{sin(\PhiOne)}}
\pgfmathsetmacro\cosPhiOne{{cos(\PhiOne)}}
\pgfmathsetmacro\sinPhiTwo{{sin(\PhiTwo)}}
\pgfmathsetmacro\cosPhiTwo{{cos(\PhiTwo)}}

\pgfmathsetmacro\sinazp{{sin(\az-90)}}
\pgfmathsetmacro\cosazp{{cos(\az-90)}}
\pgfmathsetmacro\sinazm{{sin(90-\az)}}
\pgfmathsetmacro\cosazm{{cos(90-\az)}}

\tdplotdrawarc[tdplot_main_coords,dashed,red]{(0,0,\v)}{\R}{\PhiOne}{360+\PhiTwo}{anchor=north}{}
\tdplotdrawarc[tdplot_main_coords,red]{(0,0,\v)}{\R}{\PhiTwo}{\PhiOne}{anchor=north}{}

\tdplotdrawarc[tdplot_main_coords,name path=Circle1]{(0,0,\v+\dif)}{\RR}{0}{360}{anchor=north}{}

\path[name path=left] (O) -- (\RR*\cosPhiOne,\RR*\sinPhiOne,\v+\dif);
\path[name path=right] (O) -- (\RR*\cosPhiTwo,\RR*\sinPhiTwo,\v+\dif);
\path[name path=front] (2,2,1) -- (2,-1.5,1);
\path[name intersections={of=left and front, total=\tot, sort by = curve 1}]
	\foreach \i in {1,...,\tot} {(intersection-\i) coordinate (L-\i)};
\path[name intersections={of=right and front, total=\tot, sort by = curve 1}]
	\foreach \i in {1,...,\tot} {(intersection-\i) coordinate (R-\i)};
\draw (O) -- (L-1);
\draw (O) -- (R-1);
\draw[dashed] (L-1) -- (\R*\cosPhiOne,\R*\sinPhiOne,\v);
\draw[dashed] (R-1) -- (\R*\cosPhiTwo,\R*\sinPhiTwo,\v);
\draw (\R*\cosPhiOne,\R*\sinPhiOne,\v) -- (\RR*\cosPhiOne,\RR*\sinPhiOne,\v+\dif);
\draw (\R*\cosPhiTwo,\R*\sinPhiTwo,\v) -- (\RR*\cosPhiTwo,\RR*\sinPhiTwo,\v+\dif);


\draw[red] (-2,2,1) -- (2,2,1) -- (2,-1.5,1) -- (-2,-1.5,1);
\path[name path=mypath] (-2,-1.5,1) -- (-2,2,1);

\path[name intersections={of=mypath and Circle1, total=\tot, sort by = curve 1}]
	\foreach \i in {1,...,\tot} {(intersection-\i) coordinate (Inter-\i)};

\draw[red] (-2,-1.5,1) -- (Inter-2);
\draw[dashed,red] (Inter-1) -- (Inter-2);
\draw[red] (Inter-1) -- (-2,2,1);

\node[red] at (1,2.3,\v) {$t = 1$};


\end{tikzpicture}
  \caption[Section of the null cone which creates a sphere.]{Section of the null cone which creates a sphere. The direction of the coordinate $Z$ is not displayed.}
  \label{img:geospin:hyperplane}
\end{figure}
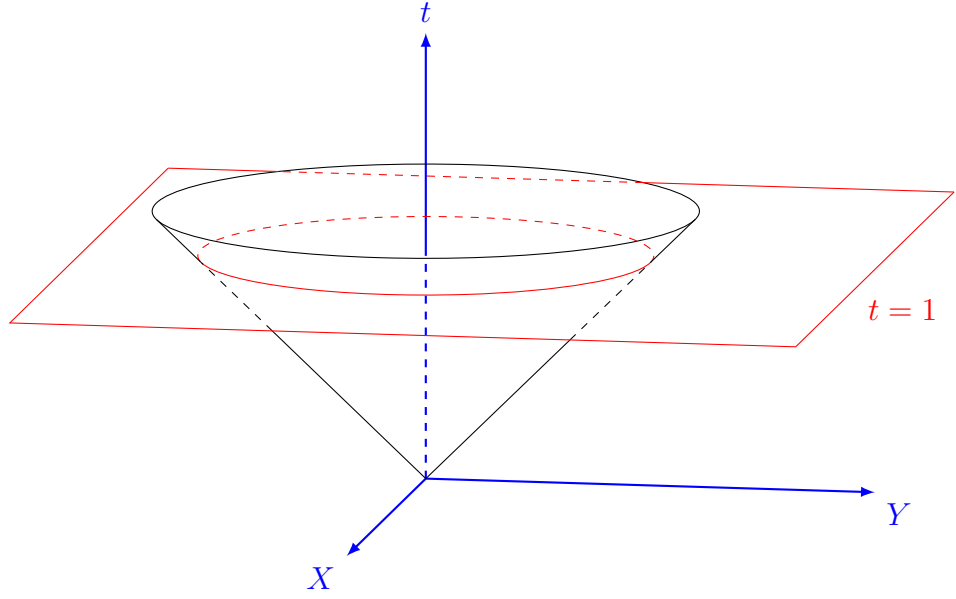

We can introduce a Cartesian coordinate system on the sphere, where the coordinates are
\begin{align}
X = \frac{\zeta \cconj{\eta} + \cconj{\zeta}\eta}{\left| \zeta \right|^2 + \left| \eta \right |^2} \,, 
\qquad
Y = \frac{1}{\im} \frac{\zeta \cconj{\eta} - \cconj{\zeta}\eta}{\left| \zeta \right|^2 + \left| \eta \right |^2} \,, 
\qquad
Z = \frac{\zeta \cconj{\zeta} - \eta \cconj{\eta}}{\left| \zeta \right|^2 + \left| \eta \right |^2} \,.
\end{align}
This sphere is a space-like hyperplane. Let us add the time-like direction to complete the space-time. We can start with a choice $t = 1$ as our first simple pick to get the sphere as a section of the space-time. The situation can be seen in the figure~\ref{img:geospin:hyperplane}.
From the picture it is clear that any other plane for different time is also a sphere, with a different radius. We can choose 
\begin{equation}\label{eq:geospin:tchoice}
	t = \zeta \cconj{\zeta} + \eta \cconj{\eta}
\end{equation}
in our convenience. It is easily seen that this term is exactly the denominator of $X$, $Y$ and $Z$. It was chosen so we can adopt their nominators as lower case version of these coordinates:
\begin{equation}
  x = \zeta \cconj{\eta} + \cconj{\zeta} \eta \,, \qquad y = - \im \( \zeta \cconj{\eta} - \cconj{\zeta} \eta \)\! \,, \qquad z = \zeta \cconj{\zeta} - \eta \cconj{\eta} \,.
\end{equation}

We introduce a spinor $\xi^A$ as a pair 
\begin{equation}
	\xi^A =
    	\begin{pmatrix}
      		\zeta \\ 
      		\eta 
    	\end{pmatrix} 
	\! \,.
\end{equation}
The point is that we are able to express the new coordinates using the spinor and its Hermitian conjugate which is to be denoted with $+$ as superscript. Therefore, e.g.\
\begin{equation}
	t = \contran{\xi}\! A_t \xi = 
		\begin{pmatrix}
			\cconj{\zeta} & \cconj{\eta}
		\end{pmatrix}
	A_t 
		\begin{pmatrix} 
			\zeta \\ 
			\eta 
		\end{pmatrix} 
	\! \,,
\end{equation}
where $A_t$ is a matrix. This matrix has to be
\begin{equation}
	A_t =
		\begin{pmatrix}
			1 & 0 \\
			0 & 1
		\end{pmatrix}
\end{equation}
in order to meet the choice~\eqref{eq:geospin:tchoice}. In analogy we get
\begin{equation}
	x = \contran{\xi}\! A_x \xi = 
		\begin{pmatrix}
			\cconj{\zeta} & \cconj{\eta}
		\end{pmatrix}
	A_x 
		\begin{pmatrix} 
			\zeta \\ 
			\eta 
		\end{pmatrix} 
\end{equation}
with
\begin{equation}
	A_t =
		\begin{pmatrix}
			0 & 1 \\
			1 & 0
		\end{pmatrix}
\end{equation}
and so on.

The coordinates $t$, $x$, $y$, $z$ are components of a null vector $k^a$ which lies in the light cone. The covariant form of expressing its coordinates using spinors is then given by equation
\begin{equation}
	k^{\hat{a}} = \cconj{\xi^A} \sigma^{\hat{a}}_{AA'} \xi^A \,,
\end{equation}
where $\sigma^{\hat{a}}_{AA'}$ is a vector of matrices $A_i$ with $i = t$, $x$, $y$ and $z$. It is a soldering form -- an isomorphism between two tangent spaces \cite{Esposito}.
As was broached, it is a vector of matrices and in our case they are the Pauli matrices. This particular one takes tensors into spinors and vice versa.

\section{Set of unknown parameters and their field equations} 
\label{sec: unknown parameters}
We have already indicated that the situation is different from the most usual case when we are solving Einstein equations to find out the proper metric. We shall arrange our variables and field equations for our case. 

However, let us firstly discuss a set of transformations of the tetrad. The topic is included at this place because at the end of the discussion, we will be able to see that the neat notation for our variables is connected with two of them. Transformations which we proceed to introduce will be later used to discuss the gauge freedom in the choice of the null tetrad (its fixation). We start with two similar transformations important for the notion: the \focc{spin} and the \focc{boost}. The transformations can be found in~\cite{Penrose1987}.


\subsection{Spin weight}\label{sec:spin weights}
The first one is a \emph{rotation} in the plane spanned by $m^a, \cconj{m}^a$ and is given by 
\begin{equation}
	m^a \mapsto \eu^{\im\! \chi} m^a \,, \qquad
	\cconj{m}^a \mapsto \eu^{-\im\! \chi} \cconj{m}^a \,,
\label{eq:tf:spinweight:transformation}
\end{equation}
where $\chi$ is a real function. Therefore a function $\eta$ is said to have \focc{spin weight} $s$ when it transforms under the rotation as
\begin{equation}
	\eta \mapsto \eu^{\im\! s \chi} \eta \,.
\end{equation}

The origin of the transformation rules~\eqref{eq:tf:spinweight:transformation} can be easily seen on the spinorial level. Clearly, a transformation
\begin{equation}
	o^A \mapsto \eu^{\im\!\frac{\chi}{2}} o^A \,, \qquad
	\iota^A \mapsto \eu^{-\im\!\frac{\chi}{2}} \iota^A \,.
\end{equation}
preserves the normalization condition~\eqref{eq:spinform:basisnorm} and by means of~\eqref{eq:tf:spinors:tetrad} it leads to \eqref{eq:tf:spinweight:transformation}.


In the next subsection we will introduce a similar concept, the boost weight. A quantity can have none of the weights, one of the weights or even both of them defined. Note that undefined weight is different state from having zero weight. An example of a quantity which does not have defined the spin weight is the product of the Newman--Penrose operator $\delta$ (and of course also the $\cconj{\delta}$) and a spin-weighted function. Let us proceed to show how $\delta$ acting on an arbitrary function with spin weight $s$ transforms under the spin:
\begin{align}
	\delta \eta = m^a\, \nabla_a \eta \mapsto 
		{}&\eu^{\im\!\chi} m^a\, \nabla_a \! \left( \eu^{\im\!s\chi} \eta \right) 
		= \eu^{\im\!\chi} m^a  \!\left( \eu^{\im\!s\chi}\, \nabla_a\eta +  \im\! s \eta \eu^{\im\!s\chi}\, \nabla_a \chi \right) \nonumber \\
		&\qquad = \eu^{\im \(s+1\) \chi} \!\left( \delta \eta + \im\! s \eta\, \delta \chi \right)\! \,.
\end{align}
At first, it might look like it has the spin weight $s + 1$, but there is an additional inhomogeneous term $\im\! s \eta\, \delta \chi$. Therefore, $\delta$ and $\cconj{\delta}$ do not preserve the spin weight. However, we can add a compensation term to $\delta$ to define a \emph{new operator} $\eth$, 
\begin{equation}
	\eth \eta = \delta \eta + s \!\( \cconj{\alpha} - \beta \)\! \eta \,. \label{eq:eth op}
\end{equation}
This operator transforms homogeneously with spin weight $s + 1$:
\begin{equation}
	\eth \eta \mapsto \eu^{\im \(s + 1\) \chi}\, \eth \eta \,.
\end{equation}
There is also a conjugate operator by which $\cconj{\delta}$ can be replaced:
\begin{equation}
	\cconj{\eth} \eta = \cconj{\delta} \eta - s \!\(\alpha - \cconj{\beta} \)\! \eta \,. \label{eq:eth op cc}
\end{equation}
These two operators will be with convenience used later.

\subsection{Boost weight}
The boost is, on the other hand, a transformation in a plane spanned by $l^a$ and $n^a$. Therefore, we want to scale the spinors which give these directions. If we impose the change
\begin{equation}
	o^A \mapsto A^\frac{1}{2} o^A \,, \qquad \iota^A \mapsto A^{-\frac{1}{2}} \iota^A \,,
\end{equation}
the equations of the transformation in the tetrad formalism are
\begin{equation}
	l^a \mapsto A l^a \,, \qquad n^a \mapsto A^{-1} n^a \,. \label{eq:boost}
\end{equation}
Function $A$ is real and positive. Vectors $m^a$ and $\cconj{m}^a$ and the normalization $l_a n^a = 1$ alike are unchanged. Similarly to the spin weight, a function $\eta$ has \focc{boost weight} $w$ if it transforms according to the relation
\begin{equation}
	\eta \mapsto A^w \eta \,.
\end{equation} 

Unlike in the case of the rotation, it is not immediately obvious that the transformation is the boost. Let us demonstrate this. The transformation of vector $t^a$ is
\begin{align}
	t^a = \frac{1}{\sqrt{2}} \!\( l^a + n^a \) \mapsto 
		{}&\frac{1}{\sqrt{2}} \!\( A l^a + A^{-1} n^a \) = \frac{1}{2} \bigl(A \!\(t^a + z^a \) + A^{-1} \!\( t^a - z^a \) \bigr) \nonumber \\
		&\qquad = a t^a - \tilde{a} z^a 
\end{align}
where
\begin{equation}
	a = \frac{1}{2} \!\( A^{-1} + A \)\! \,, \qquad \tilde{a} = \frac{1}{2} \!\( A^{-1} - A \)\! \,.
\end{equation}
It is straightforward to check that, independently of $A$, equation
\begin{equation}
	a^2 - \tilde{a}^2 = 1
\end{equation}
holds true. For this reason, we can write $a$ and $\tilde{a}$ as
\begin{equation}
	a = \cosh\oldtheta \,, \qquad \tilde{a} = \sinh\oldtheta \,.
\end{equation}
Situation is analogical for $z^a$, which gives us the usual formulation of the boost transformation
\begin{align}
	t^a &\mapsto \cosh \oldtheta \, t^a - \sinh \oldtheta \, z^a \,, \\
	z^a &\mapsto \cosh \oldtheta \, z^a - \sinh \oldtheta \, t^a \,.
\end{align}

It is a well known fact, \cite{Penrose1987}, that the \focc{Lorentz group}, which consists of rotations and boosts, has $6$ parameters. So far, we have introduced only two of them. This should lead us to a search of other, more complicated, transformations which would complete the set. It turns out that they can be represented by two more general rotations: rotation about $l^a$ and about $n^a$. Both of them give us two more parameters. We shall show the one around $l^a$, the second one is analogous. 

As before, its form can be better understood from the spinorial frame. The vector $l^a$, about which we want to rotate, is in this formalism given only by $o^A$, therefore, we want to leave $o^A$ as it is. The transformation is given by a change of $\iota^A$. We have already scaled it and changed its phase, we are left with addition of $o^A$ to it. Together the transformation is
\begin{equation}
	o^A \mapsto o^A \,, \qquad
	\iota^A \mapsto \iota^A + c o^A \,.
\end{equation}
This gives us the Newman--Penrose form of the transformation
\begin{subequations}\label{eq:rotation about l}
\begin{align}
	l^a &{}\mapsto l^a \,, \\
	m^a &{}\mapsto m^a + \cconj{c} l^a \,, \\
	n^a &{}\mapsto n^a + c m^a + \cconj{c} \cconj{m}^a + \abs{c}^2 l^a \,.
\end{align}
\end{subequations}
The function $c$ is complex; hence, gives us two parameters.

Beside the Newman--Penrose formalism, there exists so called \gls{ghp} formalism in which equations are written gauge invariantly, \cite{Geroch1973a}.

It is about time to get back to what is different from Einstein's formulation of general relativity.

\subsection{The usual approach}
The usual approach is pretty well-known, and we very briefly review it only to be able to point out the differences we are experiencing with the Newman--Penrose formulation. One can refer, e.g., to \cite{Wald1984} for more information on the topic of the standard formulation.

Our unknown is the metric tensor $g_{\mu\nu}$ which is found by means of the Einstein equations\footnote{We use the geometrized units; $G = c = 1$.}
\begin{equation}
G_{\mu\nu}\!\(g,\partial g,\partial^2 g\) = 8\pi T_{\mu\nu} \,.
\end{equation}
We have omitted the term with the cosmological constant for we shall not consider it later on. As we have indicated, the Einstein tensor $G_{\mu\nu}$ depends on the metric tensor up to its second derivatives (through the Ricci curvature tensor). The Einstein equations are therefore partial differential equations of the second-order.

\subsection{Spin coefficient approach}\label{sec:field equations}
We would like to simplify the situation, namely we do not want to solve second-order differential equations, instead, we can have only first-order differential equations whose solution would yield the metric. To achieve this, we have to increase the number of equations and variables. This is somewhat analogous to passing from the Lagrangian formulation to the Hamiltonian one, where the reduction of the order of equations of motion is compensated by an increase of the number of equations.

Identities which are usually trivially satisfied, namely the Ricci and Bianchi identities, are now our field equations. These equations are projected onto the tetrad, and expressed in terms of the spin coefficients and components of the Riemann tensor -- these are the variables of ours. The reason is that the Riemann tensor actually cannot be calculated by the well-known definition because we do not know how the connection looks like, its components are the desirable spin coefficients. On the other hand, Einstein's equations are now only an algebraic relation between the energy-momentum tensor and the Ricci tensor. This section is in correspondence with \cite{Chandrasekhar1983} and \cite{Stewart1993}, be aware that \cite{Chandrasekhar1983} uses a different sign convention, we use the one referred to as \qm{$-\,-\,-$}\footnote{Not to be confused with the signature, though, its sign is included in the first minus which shows we are using the mostly negative one.} in \cite{Misner1973} and used by Roger Penrose.

As mentioned, components of the Riemann tensor are included into our variables. It turns out that they can be represented by $5$ non-zero complex components of the Weyl tensor and $9$ components of the Ricci tensor, for both of which a neat labelling arises from the spinor formalism, completed with the scalar curvature~$R$ (or $\Lambda$, to be explained) which gives us $16$ unknowns. Therefore, we proceed with a decomposition of the Riemann tensor in the spinor formalism. It is well known that it can be divided into parts according to their trace as (in four dimensions)
\begin{equation}
	R_{abcd} = C_{abcd} + \( g_{a[c}\mathrm{Ric}_{d]b} - g_{b[c}\mathrm{Ric}_{d]a} \) - \frac{1}{3}R g_{a[c} g_{d]b} \,.
\end{equation}
The square brackets denote antisymmetrization.

The Weyl tensor $C_{abcd}$ is in the spinor formalism given by the \focc{Weyl spinor} and the metric-like acting Levi-Civita tensor as
\begin{equation}
\label{eq:weyl in spinor form}
	C_{abcd} = C_{AA'BB'CC'DD'} = \Psi_{ABCD} \levicivita_{A'B'} \levicivita_{C'D'} + \cconj{\Psi}_{A'B'C'D'} \levicivita_{AB} \levicivita_{CD} \,.
\end{equation}
Similarly, the Ricci tensor in the spinor formalism 
can be divided into trace-free part consisting of the \focc{Ricci spinor} and the scalar curvature, while the scalar curvature in the spinor formalism is written as $\Lambda = R/24$:
\begin{equation}
	\mathrm{Ric}_{ab} = \mathrm{Ric}_{AA'BB'} = -2\ricsp_{ABA'B'} + 6 \Lambda \levicivita_{AB} \levicivita_{A'B'} \,.
\end{equation} 

Finally, the Riemann tensor is in spinor terms decomposed as follows:
\begin{align}
	R_{abcd} &= \Psi_{ABCD} \levicivita_{A'B'} \levicivita_{C'D'} + \cconj{\Psi}_{A'B'C'D'} \levicivita_{AB} \levicivita_{CD} \nonumber \\
		&\qquad{}+\ricsp_{ABC'D'} \levicivita_{A'B'} \levicivita_{CD} + \ricsp_{CDA'B'} \levicivita_{AB} \levicivita_{C'D'} \nonumber \\
		&\qquad{}-2 \Lambda \levicivita_{AB} \levicivita_{CD} \levicivita_{A'(C'} \levicivita_{D')B'} -2 \Lambda \levicivita_{A'B'} \levicivita_{C'D'} \levicivita_{A(C} \levicivita_{D)B} \,.
\end{align}

The five non-zero components of the Weyl spinor are\snote{použít nějak spin a boost?}
\begin{subequations}
\begin{align}
	\Psi_0 & = \Psi_{0000} = \Psi_{ABCD} o^A o^B o^C o^D = C_{abcd}\, l^a m^b l^c m^d \,, \\
	\Psi_1 & = \Psi_{0001} = \Psi_{ABCD} o^A o^B o^C \iota^D = C_{abcd}\, l^a n^b l^c m^d \,, \\
	\Psi_2 & = \Psi_{0011} = \Psi_{ABCD} o^A o^B \iota^C \iota^D = C_{abcd}\, l^a m^b \cconj{m}^c n^d \,, \\
	\Psi_3 & = \Psi_{0111} = \Psi_{ABCD} o^A \iota^B \iota^C \iota^D = C_{abcd}\, l^a n^b \cconj{m}^c n^d \,, \\
	\Psi_4 & = \Psi_{1111} = \Psi_{ABCD} \iota^A \iota^B \iota^C \iota^D = C_{abcd}\, \cconj{m}^a n^b \cconj{m}^c n^d \,,
\end{align} \label{eq:weyl scalars}
\end{subequations}
where the tensorial form is, for completeness, also shown. The components of the Ricci spinor are
\begin{subequations}
\begin{align}
	\ricsp_{00} &= \ricsp_{ABA'B'} o^A o^B \cconj{o}^{A'} \cconj{o}^{B'} = -\tfrac{1}{2} \mathrm{Ric}_{ab}\, l^a l^b \,, \label{eq:definition of ricsp_00} \\
	\ricsp_{01} &= \ricsp_{ABA'B'} o^A o^B \cconj{o}^{A'} \cconj{\iota}^{B'} = -\tfrac{1}{2} \mathrm{Ric}_{ab}\, l^a m^b \,, \\
	\ricsp_{02} &= \ricsp_{ABA'B'} o^A o^B \cconj{\iota}^{A'} \cconj{\iota}^{B'} = -\tfrac{1}{2} \mathrm{Ric}_{ab}\, m^a m^b \,, \\
	\ricsp_{10} &= \ricsp_{ABA'B'} o^A \iota^B \cconj{o}^{A'} \cconj{o}^{B'} = -\tfrac{1}{2} \mathrm{Ric}_{ab}\, l^a \cconj{m}^b\,, \\
	\ricsp_{11} &= \ricsp_{ABA'B'} o^A \iota^B \cconj{o}^{A'} \cconj{\iota}^{B'} = -\tfrac{1}{4} \mathrm{Ric}_{ab} \(l^a n^b + m^a \cconj{m}^b\)\!  \,, \\
	\ricsp_{12} &= \ricsp_{ABA'B'} o^A \iota^B \cconj{\iota}^{A'} \cconj{\iota}^{B'} = -\tfrac{1}{2} \mathrm{Ric}_{ab}\, n^a m^b \,, \\
	\ricsp_{20} &= \ricsp_{ABA'B'} \iota^A \iota^B \cconj{o}^{A'} \cconj{o}^{B'} = -\tfrac{1}{2} \mathrm{Ric}_{ab}\, \cconj{m}^a \cconj{m}^b \,, \\
	\ricsp_{21} &= \ricsp_{ABA'B'} \iota^A \iota^B \cconj{o}^{A'} \cconj{\iota}^{B'} = -\tfrac{1}{2} \mathrm{Ric}_{ab}\, n^a \cconj{m}^b \,, \\
	\ricsp_{22} &= \ricsp_{ABA'B'} \iota^A \iota^B \cconj{\iota}^{A'} \cconj{\iota}^{B'} = -\tfrac{1}{2} \mathrm{Ric}_{ab}\, n^a n^b \,.
\end{align}\label{eq:ricci scalars}
\end{subequations}
Notice that some of them are complex conjugates of others and we only need $\ricsp_{\hat{a}\hat{b}}$ with $\hat{a} \leq \hat{b}$.

The field equations for the spin coefficients (and components of the Weyl tensor) are broken into three sets according to their meaning. The equations from the latter two groups shall be covered only by examples -- to reveal their structure -- while the full sets of them can be found in the appendix~\ref{app:np formalism}. At this place, we only describe the purpose and origin of these equations. 
\begin{enumerate}
	\item \textbf{The commutation relations} are obtained by writing down commutations between the Newman--Penrose derivative operators explicitly using the spin coefficients. They are
	\begin{subequations}
	\begin{align}
		\Delta D - D \Delta &=
			\(\gamma + \cconj{\gamma}\) D + \(\epsilon + \cconj{\epsilon}\) \Delta - \(\cconj{\tau} + \pi\) \delta - \(\tau + \cconj{\pi}\) \cconj{\delta} \,, \label{eq:commutator Delta D}\\
		\delta D - D \delta &=
			\(\cconj{\alpha} + \beta - \cconj{\pi}\) D + \kappa\, \Delta - \(\cconj{\scrho} + \epsilon - \cconj{\epsilon}\) \delta - \sigma\, \cconj{\delta} \,, \\
		\delta \Delta - \Delta \delta &=
			-\cconj{\nu}\, D + \(\tau - \cconj{\alpha} - \beta\) \Delta + \(\mu - \gamma + \cconj{\gamma}\) \delta + \cconj{\lambda}\, \cconj{\delta} \,, \\
		\cconj{\delta} \delta - \delta \cconj{\delta} &=
			\(\cconj{\mu} - \mu\) D + \(\cconj{\scrho} - \scrho\) \Delta + \(\alpha - \cconj{\beta}\) \delta - \(\cconj{\alpha} - \beta\) \cconj{\delta} \,, \label{eq:commutator delta deltacc}
	\end{align}\label{eq:commutation relations}%
	\end{subequations}%
	and they are valid when acting on a scalar. They can also be seen in an alternative form which arises from a natural choice of the scalar -- coordinates. Then we have
	\begin{subequations}
	\begin{align}
		\Delta l^a - D n^a &=
			\(\gamma + \cconj{\gamma}\)\! l^a + \(\epsilon + \cconj{\epsilon}\)\! n^a - \(\cconj{\tau} + \pi\)\! m^a - \(\tau + \cconj{\pi}\)\! \cconj{m}^a \,, \\
		\delta l^a - D m^a &=
			\(\cconj{\alpha} + \beta - \cconj{\pi}\)\! l^a + \kappa n^a - \(\cconj{\scrho} + \epsilon - \cconj{\epsilon}\)\! m^a - \sigma \cconj{m}^a \,, \\
		\delta n^a - \Delta m^a &=
			-\cconj{\nu} l^a + \(\tau - \cconj{\alpha} - \beta\)\! n^a + \(\mu - \gamma + \cconj{\gamma}\)\! m^a + \cconj{\lambda} \cconj{m}^a \,, \\
		\cconj{\delta} m^a - \delta \cconj{m}^a &=
			\(\cconj{\mu} - \mu\)\! l^a + \(\cconj{\scrho} - \scrho\)\! n^a + \(\alpha - \cconj{\beta}\)\! m^a - \(\cconj{\alpha} - \beta\)\! \cconj{m}^a \,.
	\end{align}
	\end{subequations}
	
	\item \textbf{The spin-coefficient equations} which are in fact the tetrad version of the \textbf{Ricci identities}. They express derivatives of the spin coefficients in terms of themselves and components of the Riemann tensor. The list of them, other forms of them as well as procedure of getting the tetrad version is in section~\ref{sec:ricci identities}. We have 36 equations of this type. As an example we mention one of them:
	\begin{align}
		D\tau-\Delta\kappa &= 
		(\tau+\cconj{\pi})\scrho+(\cconj{\tau}+\pi)\sigma+(\epsilon-\cconj{\epsilon})\tau 
		\nonumber \\
		&\qquad{}-(3\gamma+\cconj{\gamma})\kappa+\Psi_1+\Phi_{01} \,.
	\end{align}
	
	\item \textbf{The Bianchi identities} are equations for components of the Riemann tensor, i.e.\ for the Weyl scalars, Ricci components and scalar curvature $\Lambda$. Details can be found in section~\ref{sec:bianchi identities}. There are 8 complex equations (a--h) and 4 real of which two are concatenated into one complex (i--k). The equations are of form
	\begin{align}
		&{}D\Psi_1-\cconj{\delta}\Psi_0-D\Phi_{01}+\delta\Phi_{00} = 
					(\pi - 4 \alpha) \Psi_0 + 2(2\scrho+\varepsilon)\Psi_1 - 3\kappa\Psi_2 + 2\kappa\Phi_{11} \nonumber\\ 
						&\qquad - (\cconj{\pi} - 2\cconj{\alpha} - 2\beta) \Phi_{00} - 2\sigma\Phi_{10} - 2(\cconj{\scrho}+\varepsilon)\Phi_{01} + \cconj{\kappa}\Phi_{02} \,
	\end{align}
	and can be found in section~\ref{sec:bianchi identities}.
\end{enumerate}

As we could already notice, the procedure of projecting the field equations onto the tetrad (and abandoning of the covariant derivative in favour of the Newman--Penrose operators) convert tensor equations into scalar ones. We have to give up on summations in order to use the spin coefficients explicitly, this leads to large sets of equations, however, space-time symmetries can help to reduce the number of equations, and, even though, that we have still not a few of them, they are in particular cases simple to use and solve.




As we have seen, there are 12  complex spin \gls{coef} carrying information equivalent to 24 independent Christoffel symbols. Next, the curvature is encoded in 6 complex components of the Ricci spinor $\ricsp_{mn}$ and the scalar curvature $\Lambda$, which describe the matter through Einstein equations. Finally, there are 5 Weyl scalars depicting degrees of freedom of the gravitation field itself.

\subsection{Maxwell equations}

Maxwell field tensor can be represented by a symmetric 2-spinor usually denoted $\oldphi_{AB}$. It is associated to the \focc{Maxwell tensor} via \cite{Stewart1993}
\begin{equation}
	\oldphi_{AB} = \frac{1}{2} F_{ABC'}{}^{C'} = \oldphi_{BA} \,.
\end{equation}
Equivalently,
\begin{equation}
	F_{ab} = F_{ABA'B'} = \msp_{AB} \levicivita_{A'B'} + \cconj{\msp}_{A'B'} \levicivita_{AB} \,.
\end{equation}
Using a relation $\levicivita^{AB} = o^A \iota^B - \iota^A o^B$ (which can be obtained analogously to~\eqref{eq:spinor levicivita}) the identity $\oldphi_{AB} = \levicivita_A{}^C \levicivita_B{}^D \oldphi_{CD}$ goes to
\begin{equation}
	\oldphi_{AB} = \oldphi_2 o_A o_B - 2 \oldphi_1 o_{(A} \iota_{B)} + \oldphi_0 \iota_A \iota_B \,
\end{equation}
where 3 complex scalars were identified
\begin{equation}
	\oldphi_0 = \oldphi_{AB} o^A o^B \,, \qquad \oldphi_1 = \oldphi_{AB} o^A \iota^B \,, \qquad \oldphi_{AB} = \oldphi_2 \iota^A \iota^B \,.
\end{equation}

Then we have four complex equations
\begin{equation}
	\nabla^{AA'} \oldphi_{AB} = 0 \,
\end{equation}
which stand for the eight real \focc{Maxwell equations}. The introduced spinor version of covariant derivative is an analogue to the vector one with one additional requirement -- any linear map obeying the Leibniz rule acting on spinors/tensors can be written as a contraction of tangential space-time vector $\xi^{AA'}$ and the covariant differential when acting on a spinor field. The complete set of axioms for the spinor version of covariant differential can be found in \cite[p. 81]{Stewart1993}.

This equations can be rewritten analogically to the Bianchi and the Ricci identities by means of a projection onto the basis where we employ the 3 complex scalars. The projections are
\begin{align}
	D \msp_1 - \cconj{\delta} \msp_0 &= \(\pi - 2\alpha\)\! \msp_0 + 2\scrho\msp_1 - \kappa\msp_2 \,, \label{eq:maxwell D1} \\
	D\msp_2 - \cconj{\delta}\msp_1 &= - \lambda \msp_0 + 2 \pi \msp_1 + \(\scrho - 2 \epsilon\)\! \msp_2 \,, \label{eq:maxwell D2} \\
	\Delta\msp_0 - \delta \msp_1 &= \(2\gamma - \mu\) - 2 \tau \msp_1 + \sigma \msp_2 \,, \label{eq:maxwell Delta0} \\
	\Delta\msp_1 - \delta\msp_2 &= \nu \msp_0 - 2 \mu \msp_1 + \(2\beta - \tau\)\! \msp_2 \,. \label{eq:maxwell Delta1}
\end{align}

In an electrovacuum space-time we can express the traceless Ricci tensor (no scalar curvature is, therefore, present) as
\begin{equation}
	\ricsp_{mn} = \msp_m \cconj{\msp_n} \,. \label{eq:electrovac einstein}
\end{equation}
In the electrovacuum, these are the Einstein equations.

In table~\ref{tab:spin weights}, spin weights for the scalars and the spin coefficients are listed. Note that the convenient spinor notation is directly connected to spin weights of the scalars. Only the spin coefficients possessing a spin weight are present. The boost possesses analogical property.

\begin{table}
	\centering
	\begin{tabular}{*{5}{c}}
\toprule
-2 & -1 & 0 & 1 & 2 \\
\midrule
$\lambda$ & $\nu$ & $\mu$ & $\kappa$ & $\sigma$ \\
 & $\pi$ & $\scrho$ & $\tau$ & \\
 & $\msp_2$ & $\msp_1$ & $\msp_0$ & \\
$\Psi_4$ & $\Psi_3$ & $\Psi_2$ & $\Psi_1$ & $\Psi_0$ \\
\bottomrule
\end{tabular} 
%
%
	\caption{A table of spin weights of important scalars in the Newman--Penrose formalism.}
	\label{tab:spin weights}
\end{table}

\chapter{Non-expanding horizons}
\label{chapter:non-exp}
Having established the convenient formalism for the null hypersurfaces, we are prepared to introduce a special class of horizons, so-called \focc{isolated horizons}, and analyse them in the Newman--Penrose formalism. Ultimately, we are interested in the intrinsic geometry of these horizons and in the geometry of the space-time near the horizon. A general analysis has been done in~\cite{Krishnan2012}, and the main goal of the thesis is to perform analogous analysis for the special case of the Kerr--Newman metric, as explained in the introduction. We warn reader that we employ different conventions than that of \cite{Krishnan2012}. Differences result in different signs in several equations.

In order to introduce isolated horizons properly, it is necessary to address the issue of kinematics of null geodesic congruences. This is most easily done in the Newman--Penrose formalism where certain spin coefficients are to be identified with the so-called \focc{optical scalars}. The definition of an isolated horizon imposes several restrictions on the Newman--Penrose quantities, which will reduce a huge gauge freedom in the formalism, as we developed it in chapter \ref{chapter:tetrad form}, and will allow us to define a null tetrad adapted to a space-times with isolated horizons in an invariant, geometrical way.

\section{Expansion, shear and twist}

One of the most powerful tool for understanding the geometry of a curved space-time is based on an analysis and visualization of geodesics of the given space-time. Among all geodesics, the null ones are clearly privileged, as they define the causal structure. However, the Schwarzschild space-time shows that the analysis of the geodesics must be done carefully: a freely falling observer will never cross the horizon from the point of view of an external observer, while the same observer penetrates the horizon in a finite proper time in his own frame of reference. Thus, an invariant, coordinate independent description is necessary. For this reason, it is more useful to characterize the behaviour of congruences of null geodesics by scalar quantities, in particular, by the Newman--Penrose spin coefficients. For a brief review of optical scalars in the \gls{np} formalism, see \cite{Krishnan2014}, for a general and comprehensive treatment of time-like and null congruences, see \cite{Poisson2004}.

We consider a congruence of null geodesics with the tangent vector field $l^a$. At each point of the congruence, we complete $l^a$ to a Newman--Penrose null tetrad and define a projector to the space orthogonal both to $l^a$ and $n^a$ by
\begin{equation}
q_{ab} = - m_a \cconj{m}_b - \cconj{m}_a m_b \,. 
\end{equation}
In fact, $q_{ab}$ plays the role of the metric on the 2-dimensional subspace, cf.\ \eqref{eq:metric on S0}. This is analogous to a similar construction for time-like congruences where a projector $h_{ab}$ to a 3-dimensional subspace orthogonal to a time-like vector $u^a$ is defined. However, such a construction fails for null congruences since the induced 3-dimensional metric is degenerate \cite{Hawking1973,Wald1984}.

The \focc{expansion} of $l^a$ is defined by the relation
\begin{equation}
	\Theta_{\(l\)} = q^{ab}\, \nabla_a l_b \,,
\end{equation}
i.e., the expansion is a 2-dimensional trace of $\nabla_a l_b$. The so-called \focc{shear tensor} is defined by
\begin{equation}
	\sigma_{ab} = \ProjectOnS{\nabla_{(a} l_{b)}} - \frac{1}{2} \Theta_{(l)} q_{ab} \,, \label{eq:shear tensor}
\end{equation}
where the parentheses denote symmetrization and $\ProjectOnS{\ldots}$ is a projection onto the 2-dimensional orthogonal space-like plane spanned by $m^a$ and $\cconj{m}^a$. The shear tensor $\sigma_{ab}$ is a rank two tensor but, for many purposes, information encoded in the scalar
\begin{equation}
\sigma_{ab}\sigma^{ab} = \hat{\sigma}^2
\end{equation}
is sufficient. In that case, $\hat{\sigma}$ is referred to simply as the \focc{shear}. We decorated the shear $\hat{\sigma}$ by a hat in order to distinguish it from the spin coefficient for which we reserve the symbol $\sigma$.

 The complementary antisymmetric part to the shear tensor~\eqref{eq:shear tensor} gives us analogically the \focc{twist tensor}
\begin{equation}
	\omega_{ab} = \ProjectOnS{\nabla_{[a} \xi_{b]}}
\end{equation}
and the \focc{twist scalar}
\begin{equation}
\omega_{ab}\omega^{ab} = \hat{\omega}^2 \,.
\end{equation}

\begin{figure}
	\centering
	\begin{subfigure}[t]{0.45\linewidth}
		\centering
		\caption{}
		\includegraphics[width=0.7\textwidth]{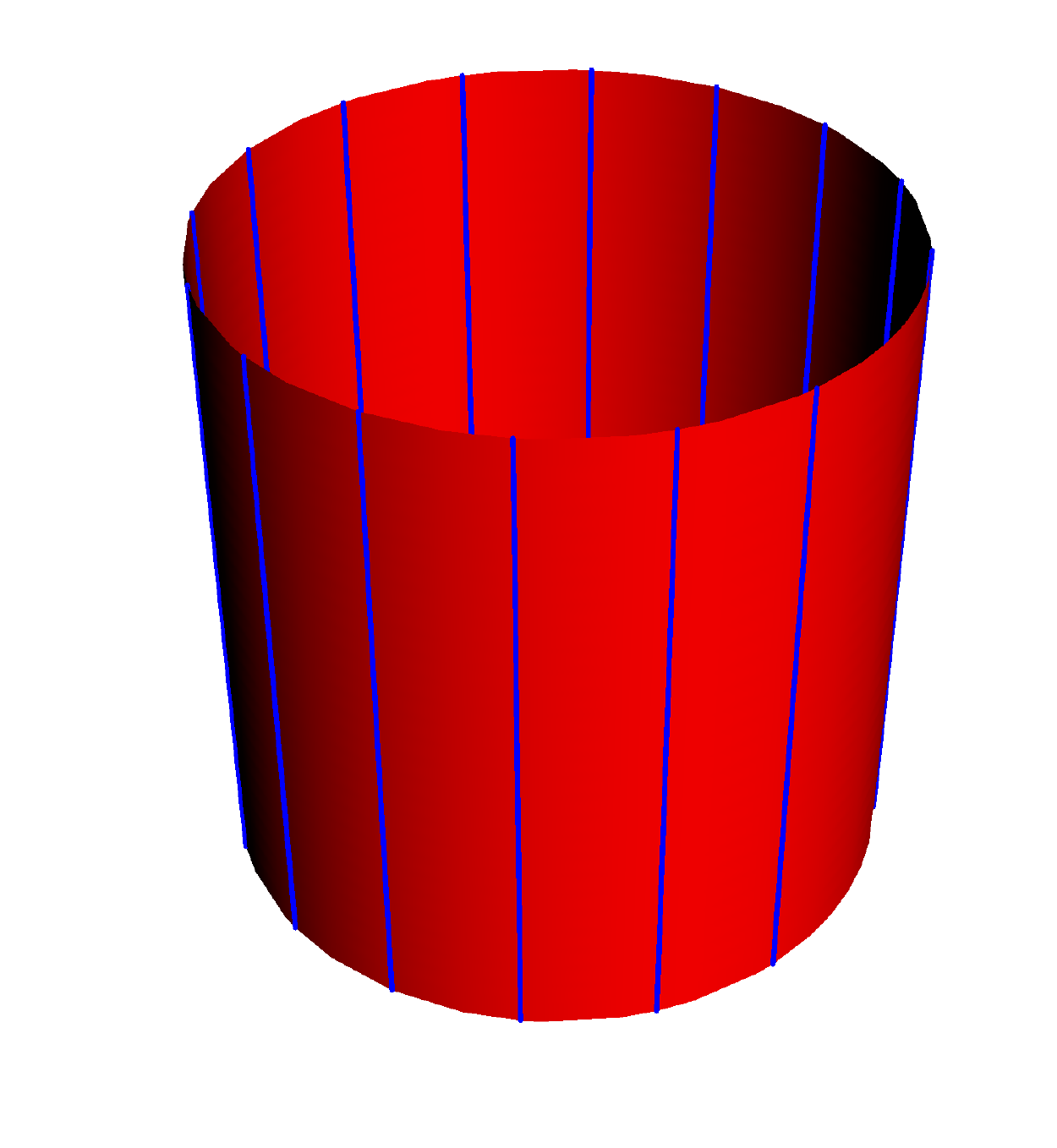}
	\end{subfigure}
	\hfill
	\begin{subfigure}[t]{0.45\linewidth}
		\centering
		\caption{}
		\includegraphics[width=0.7\textwidth]{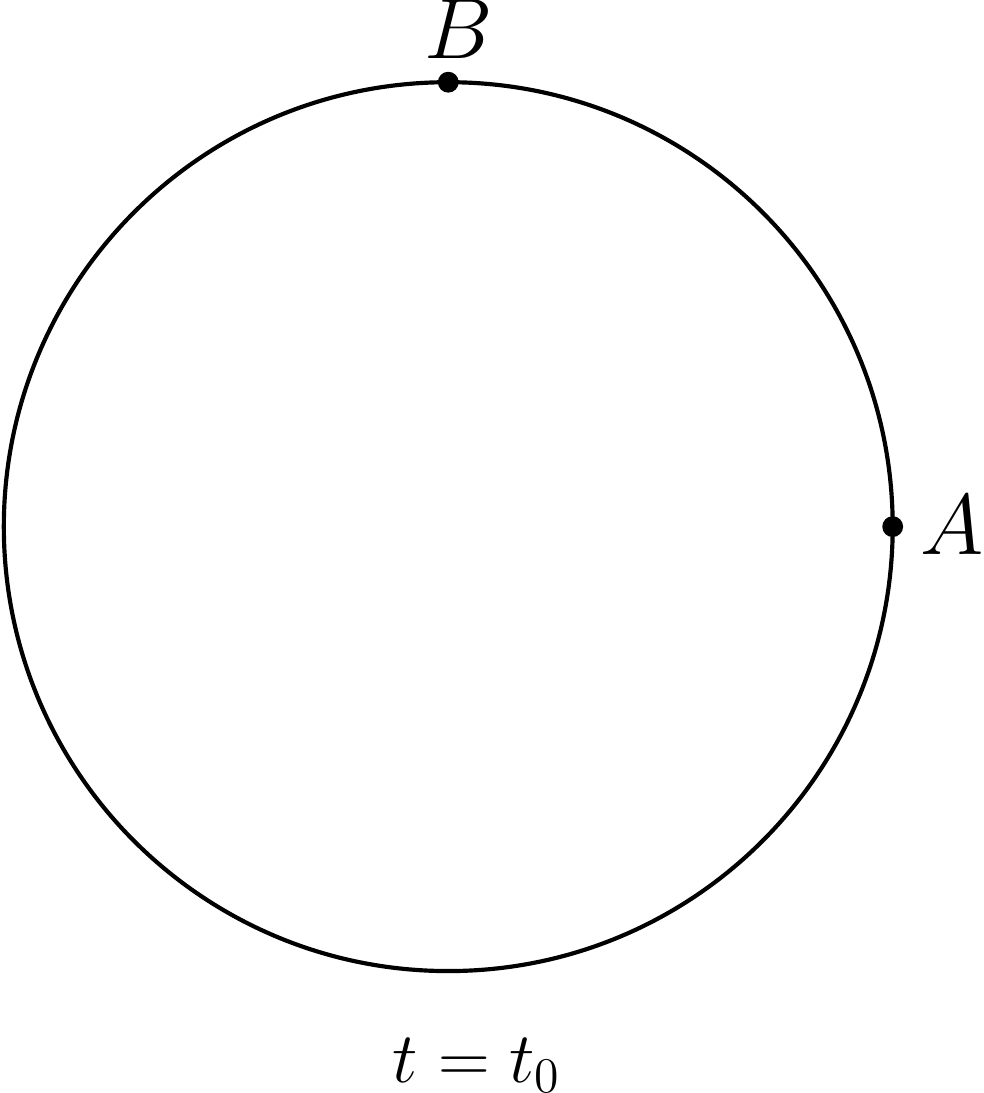}
		\label{fig:flat geodesics}
	\end{subfigure}
	\\ \hfill\\
	\begin{subfigure}[t]{0.45\linewidth}
		\centering
		\caption{}
		\includegraphics[width=0.7\textwidth]{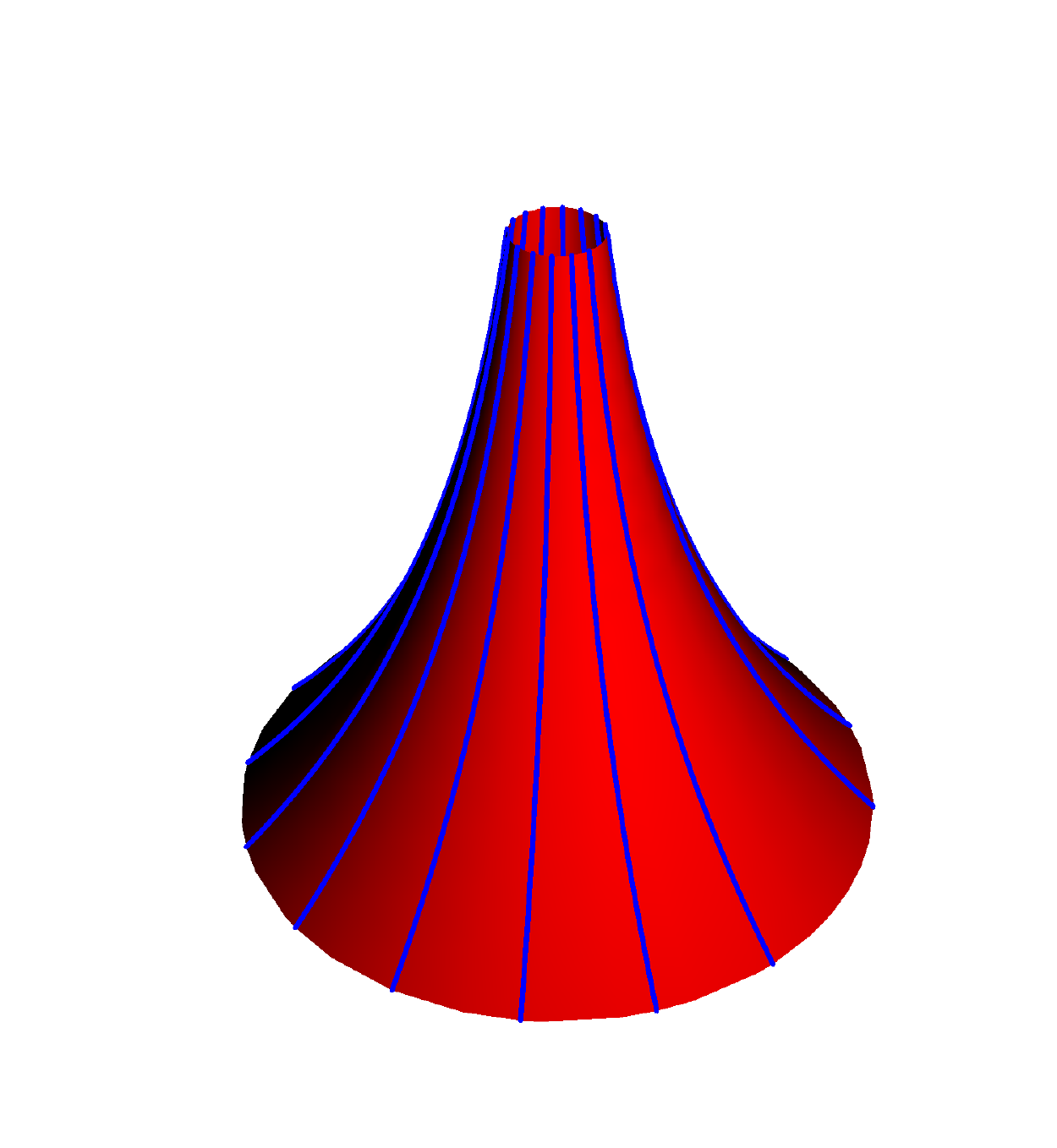}
		\label{fig:congruence expansion}
	\end{subfigure}
	\hfill
	\begin{subfigure}[t]{0.45\linewidth}
		\centering
		\caption{}
		\includegraphics[width=0.7\textwidth]{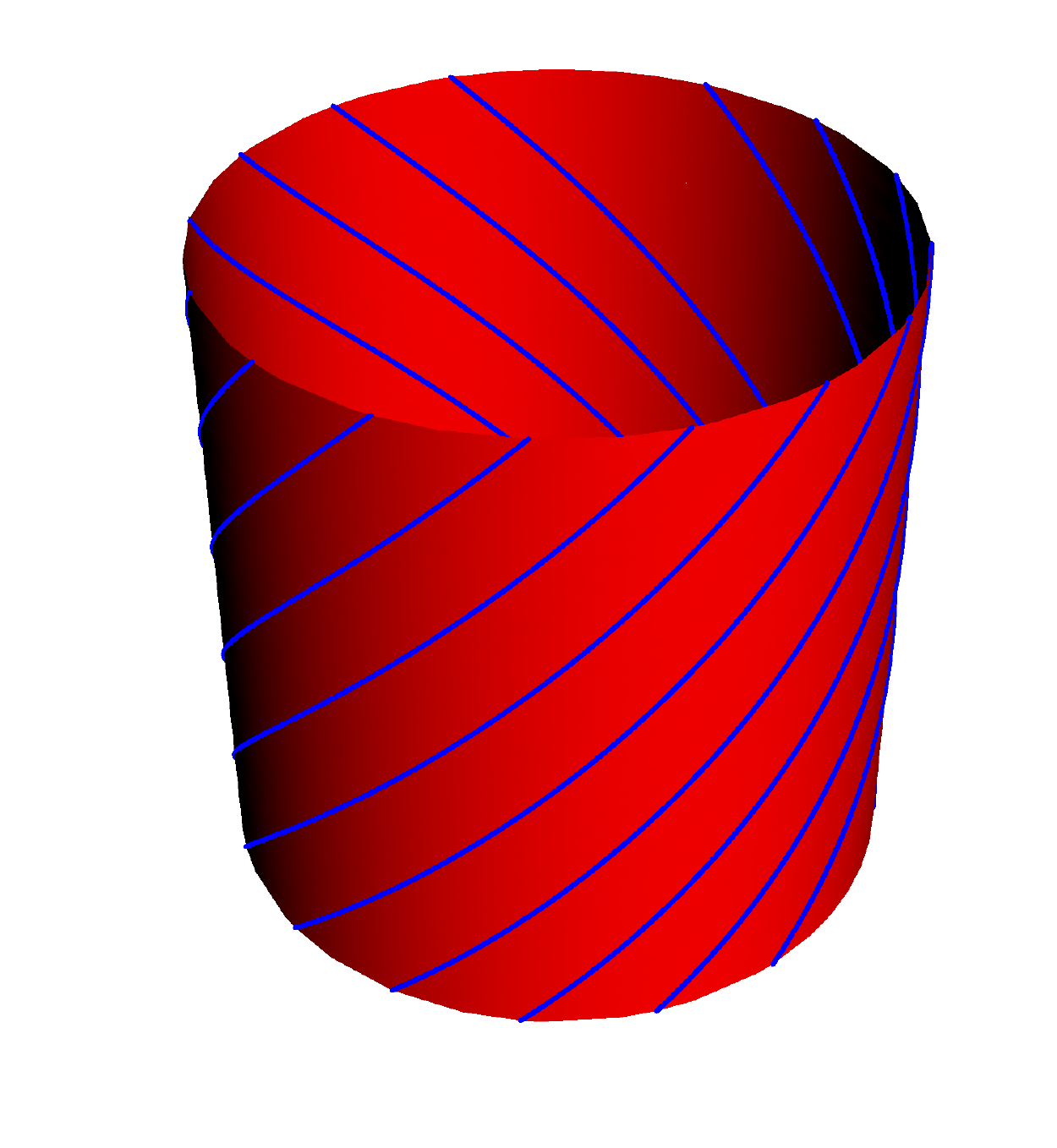}
		\label{fig:congruence twist}
	\end{subfigure}
	\\ \hfill\\
	\begin{subfigure}[t]{0.45\linewidth}
		\centering
		\caption{}
		\includegraphics[width=0.7\textwidth]{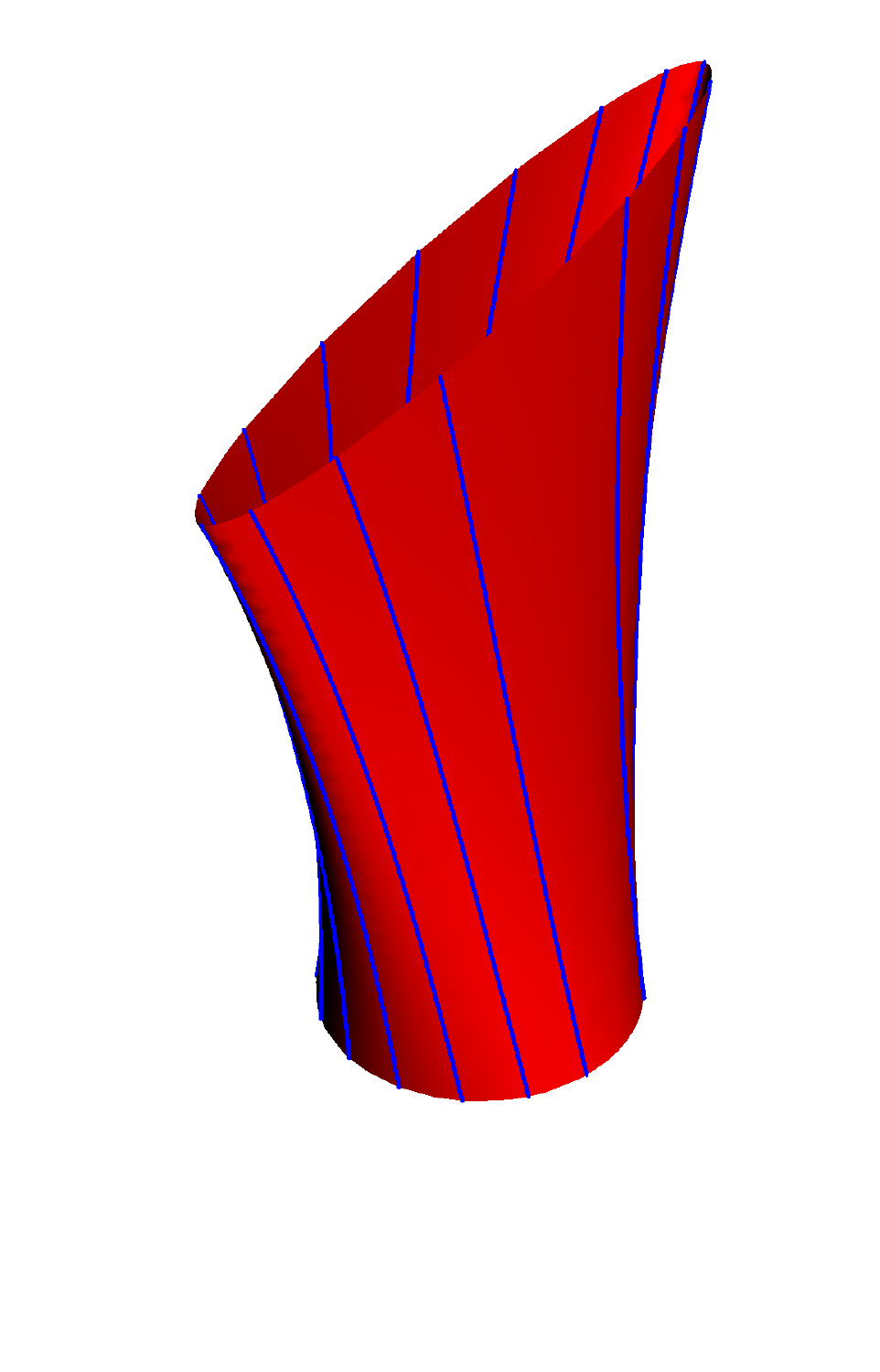}
		\label{fig:congruence shear}
	\end{subfigure}
	\hfill
	\begin{subfigure}[t]{0.45\linewidth}
		\centering
		\caption{}
		\includegraphics[width=0.7\textwidth]{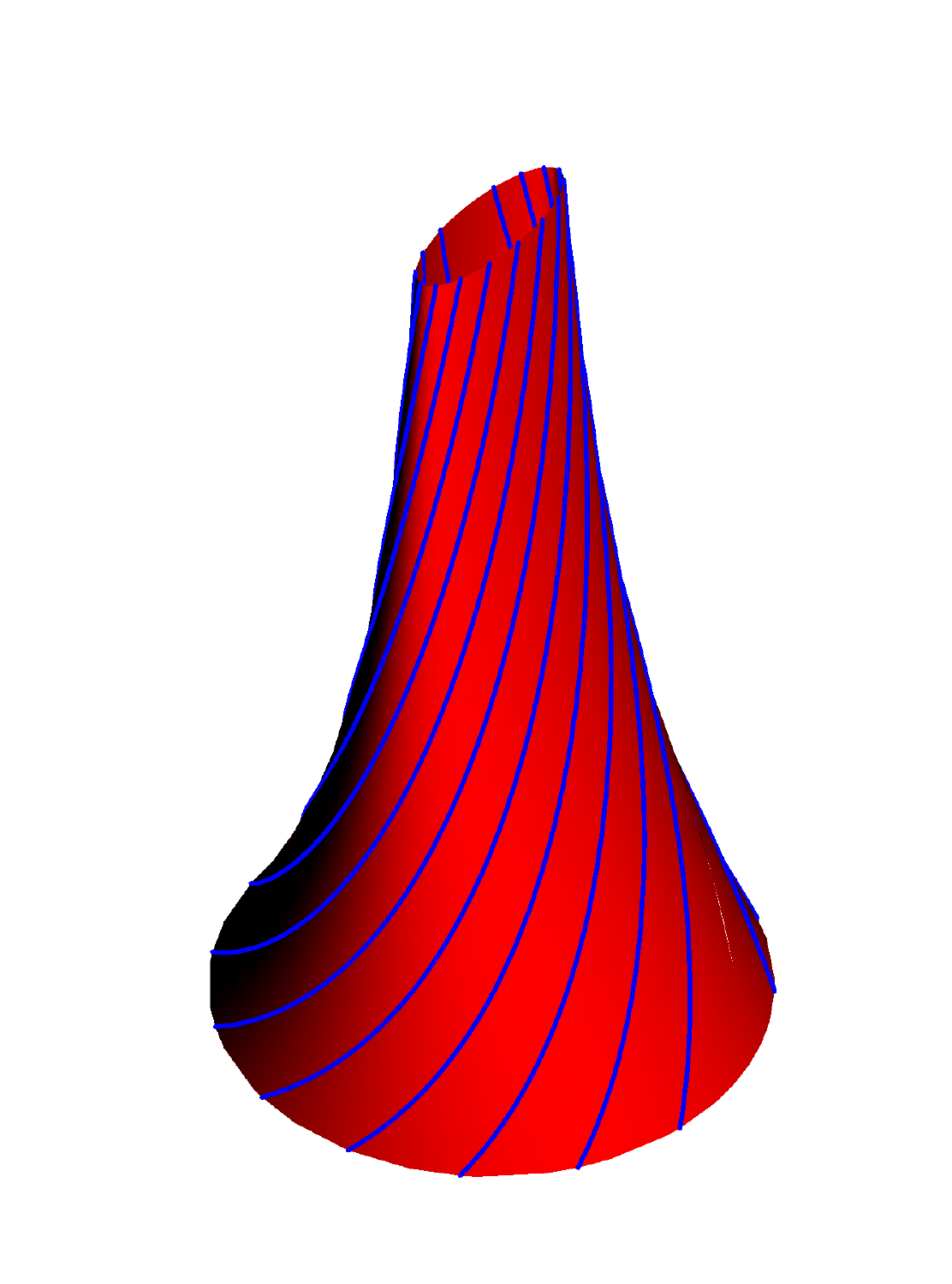}
		\label{fig:congruence combination}
	\end{subfigure}
	\caption[A family of null geodesics as it is affected by optical scalars.]{A family of null geodesics as it is affected by optical scalars: \textbf{(a)}~The family of null geodesics in the flat space-time forming a surface of a cylinder. \textbf{(b)}~The cross-section of the family is circular. \textbf{(c)}~Focusing of the geodesics in a region of negative expansion $\Theta_{l}$. \textbf{(d)}~Helical shape of the geodesics in the presence of twist. \textbf{(e)}~Shearing of geodesics preserves the area of a cross-section. \textbf{(f)}~A combination of negative expansion, shear, and twist.}
	\label{fig:shear combination}
\end{figure}

Optical scalars have intuitive geometrical meaning \cite{Stewart1993}. Consider the flat Min\-kow\-ski space-time and a family of null geodesics (which are straight lines) forming a surface of a cylinder. That is, any cross-section of the family is a circle, see figure~\ref{fig:flat geodesics}. Now, suppose that the family of geodesics enters a region with negative expansion, while remaining optical scalars vanish. The effect of negative expansion is an exponential \emph{focusing} of the geodesics at a rate given by $\Theta_{(l)}$, figure~\ref{fig:congruence expansion}. If, on the other hand, the geodesics enter a region with non-vanishing twist, with remaining optical scalars being zero, the cross-section of the family will remain circular but the geodesics are helices lying on the surface of the cylinder, see figure~\ref{fig:congruence twist}. Finally, in the region with just non-vanishing shear, the circular cross-section is deformed to an ellipse. One semi-axis of the ellipse tends exponentially to zero at the rate given by the shear, while the other semi-axis is increasing at the same rate, so that the area of the cross-section remains constant, see figure~\ref{fig:congruence shear}. In general, all the three effects are present, as illustrated in figure~\ref{fig:congruence combination}. Behaviour of the 2-dimensional cross-section in all the cases is also plotted in figure~\ref{img:optical-scalars}.

\section{Non-expanding horizon}
We have already planted all the seeds needed to understand the definition of the special horizon we are going to discuss in depth. The concept of an isolated horizon representing a black hole in equilibrium with its neighbourhood has been proposed by Ashtekar and collaborators \cite{Ashtekar1999}. Subsequently, a complete formalism has been developed in works~\cite{Ashtekar2000a,Ashtekar2000b,Ashtekar2001,Ashtekar2002}, and it was shown that isolated horizons satisfy usual laws of black-hole thermodynamics. In addition, multipole moments can be assigned to isolated horizons \cite{Ashtekar2004} which are different from usual Hansen--Geroch multipoles defined at the spatial infinity \cite{Hansen1974}.
\begin{definition}\label{def:non-expand}
A \focc{non-expanding horizon} $\HH$ is a null hypersurface with topology $\mathbb{R} \times \mathcal{S}^2$, where $\mathcal{S}^2$ is a two-sphere, which meets the following properties:
	\begin{enumerate}
		\item Any null normal to the horizon $l^a$ has vanishing expansion.
		\item Einstein's equations hold on $\HH$.
		\item For any future pointing null normal $l^a$, the vector $T_{ab}\sss l^b$ is also future pointing where $T_{ab}$ is the energy momentum tensor.
	\end{enumerate}
\end{definition}

Definition~\ref{def:non-expand} refers to an arbitrary null normal $l^a$. It is a specific feature of null hypersurfaces that any vector orthogonal to a null hypersurface is also tangent to it. Although this is a bit counter-intuitive, it is clear that any null vector is orthogonal to itself. Conversely, the only vector which is orthogonal to \emph{any} vector tangent to a null hypersurface, is itself a tangent. This is related to the fact that a three-dimensional metric induced on a null hypersurface is degenerate \cite{Hawking1973}. 

However, not only is the normal $l^a$ tangent to a null hypersurface, but its orbits are also necessarily null geodesics. These geodesics are called \emph{generators} of the null hypersurface. Different normals $l^a$ and $l'^a$ can differ, at most, by a scaling, i.e.\ $l'^a = c\sss l^a$, where $c$ is an arbitrary function. Both such normals have the same orbits (as sets), but they can be parametrized differently. In particular, any null normal $l^a$ is tangent to a null geodesic, but this is not necessarily affinely parametrized. In the following, we prove aforementioned properties of $l^a$.

\begin{myproof}
	A hypersurface can be given, at least locally, by a condition $u = 0$, where $u$ depends on all coordinates. Then a 1-form $l_a$ annihilating all tangent vectors (which means that $l^a$ is normal to the hypersurface) is therefore of the form $l_a |_{u=0} = f\, \nabla_a u$ where $f$ is an unspecified function, for it holds that any tangent vector $t^a$ satisfies $t^a\, \nabla_a u |_{u=0} = 0$, since $u = \mathrm{constant}$ on the hypersurface. Then
	\begin{equation}
		l^a f\, \nabla_a u |_{u=0} = l^a l_a = 0
	\end{equation}
	shows that $l^a$ is tangent and
	\begin{align}
		l^a\, \nabla_a l_b |_{u=0} &= l^a\, \nabla_a f\, \nabla_b u = f l^a\, \nabla_a \nabla_b u + l^a (\nabla_a f) (\nabla_b u) = f l^a\, \nabla_b \nabla_a u + l^a (\nabla_a f) (\nabla_b u) \nonumber \\
		&= f l^a\, \nabla_b l_a + l^a (\nabla_a f) (\nabla_b u) = \tfrac{1}{2} \nabla_b \(l_a l^a \) + l^a (\nabla_a f) (\nabla_b u) \nonumber \\
		&= 0 + l^a (\nabla_a f) (\nabla_b u) = \left( D \log f \right) l_b
	\end{align}
	is an equation of a non-affinely parametrised geodesic for $l^a$, where the \qm{acceleration}, i.e.\ deviation from the affine parametrization, is given by $D \log f$, where $D = l^a\, \nabla_a$ is the usual \gls{np} operator.
\end{myproof}

Let us get back to the fact that $l^a$ is a normal to the hypersurface $\HH$. It is well-known that not \emph{any} congruence of curves is hypersurface orthogonal, which means that, in general, for a given congruence, it is impossible to construct a foliation of a manifold $\mathcal{M}$ by hypersurfaces such that each vector of the congruence is orthogonal to the hypersurface containing that point. In more geometrical language, any 1-form $l_a$ defines a distribution on a manifold, that is, at each point $P \in \mathcal{M}$, it selects a subspace $V_P \subset T_P \mathcal{M}$ of the tangent space by the condition $l_a X^a = 0$. If these subspaces $V_P$ define a foliation of the manifold, distribution is said to be \focc{integrable}. 

A practical tool to decide whether a given distribution is integrable is provided by the Frobenius theorem~\cite{Wald1984,Choquet-Bruhat1982}: a vector field $l^a$ is hypersurface orthogonal, if relation
\begin{equation}
	l_{[a} \nabla_b l_{c]} = 0
\end{equation}
is satisfied. Therefore, the congruence of $l^a$ is also twist-free. In other words, the obstacle for a congruence being hypersurface orthogonal is the presence of twist. Without a rigorous proof, the Frobenius criterion can be justified as follows: If $l_a$ arises as a gradient of a scalar function, say, $l_a = \nabla_a u$, its twist is automatically zero, because $\nabla_{[a}l_{b]} = \nabla_{[a}\nabla_{b]}u = 0$ by the absence of a torsion. Moreover, a congruence will be twist free also if $l_a$ is just \emph{proportional} to a gradient, i.e. $l_a = f\, \nabla_a u$. However, the exterior derivative now does not automatically vanish, for we have
\begin{align}
 \nabla_{[b}l_{a]} &= -f^{-1} l_{[a}\nabla_{b]}f \,.
\end{align}
Multiplying with $l_c$ and antisymmetrizing in $[abc]$ we find
\begin{align}
 l_{[c} \nabla_b l_{a]} &= -f^{-1} l_{[c} l_{a} \nabla_{b]}f = 0 \,,
\end{align}
because of a symmetry of the expression in $(ca)$. Thus, the Frobenius criterion actually checks if a given 1-form $l_a$ is a gradient or a \emph{multiple} of a gradient. 

We shall show that the absence of expansion implies also zero shear. Thanks to that, we are treating horizons which do not expand, neither shear nor twist.

\section{Structure of the horizon}
%
We would like to be able to picture the structure of the horizon in terms of the Newman--Penrose tetrad adapted to the horizon in an appropriate sense. Since a null tetrad is a basis of the tangent space, we can always imagine a \qm{Minkowski} orthonormal tetrad $(t^a, x^a, y^a, z^a)$ induced by the null tetrad via relations \eqref{eq:null to orthonormal}. As early as in the definition, we have wordlessly separated the basis vectors of a null tetrad into two pairs. The elements of the first of them, $l^a$ and $n^a$, contain the Minkowski time direction $t^a$. When studying gravitational waves, we usually interpret $l^a$ as an outgoing wave and $n^a$ as an ingoing one (or vice versa). In other words, directions given by $l^a$ and $n^a$ contain the time evolution, and, typically, we choose (one of) them to be geodesics. On the other hand, vectors $m^a$ and $\cconj{m}^a$ are orthogonal to both ingoing and outgoing null directions and, hence, they do not contain the time direction. In fact, relations \eqref{eq:null to orthonormal} show that they arise from a complex rotation of vectors $x^a$ and $y^a$ of the Minkowskian frame. In this sense, they form a basis of space-like subspace of the tangent space at each point. Let us now turn to the question how to choose the null tetrad in the presence of an isolated horizon. Such a choice should respect the geometry of the horizon in some sense; let us clarify this point. 

\begin{figure}
  \centering
  \begin{tikzpicture} [scale=2,mark coordinate/.style={inner sep=0pt,outer sep=0pt,minimum size=4pt,fill=blue,circle}]

\coordinate (ld) at (-0.1,-0.5);
\coordinate (pd) at (2,-1);
\coordinate (lh) at (1.9,2.3);
\coordinate (ph) at (3.4,1.9);

\path (ld) to [bend left=20] coordinate[pos=0.1] (A1) (lh);
\path (ld) to [bend left=20] coordinate[pos=0.9] (A2) (lh);
\path (pd) to [bend left=20] coordinate[pos=0.1] (B1) (ph); 
\path (pd) to [bend left=20] coordinate[pos=0.9] (B2) (ph);

\draw[name path=dole] (A1) .. controls (0.1,-0.7) and (1.9,0.2) .. (B1);
\draw[name path=nahore] (A2) .. controls (1.6,1.5) and (2.9,2) .. (B2);

\pgfmathsetmacro\N{8}
\pgfmathsetmacro\Nn{(\N-1)}

\foreach \number [count=\x from 1] in {1,...,\N}{
	\pgfmathsetmacro\pos{(\number-1)/(\N-1)}
	
	\path (ld) to [bend left=20] coordinate[pos=\pos] (A\x)(pd);
	\path (lh) to [bend left=20] coordinate[pos=\pos] (B\x)(ph);
	\path[name path=linie\x] (A\x) to [bend left=20] (B\x);
	
	\path[name intersections={of = dole and linie\x}];
	\coordinate (D\x)  at (intersection-1);
	\path[name intersections={of = nahore and linie\x}];
	\coordinate (N\x)  at (intersection-1);
}

\path[draw,name path=leva] (D1) to [bend left=17] coordinate[pos=0.2] (L) (N1);

\path[draw,name path=prava] (D\N) to [bend left=17] coordinate[pos=0.2] (P) (N\N);

\foreach \number in {2,...,\Nn}{
	\draw[very thin] (D\number) to [bend left=17] (N\number);
}

\draw[name path=sfera] (L) to [bend left=20] (P);

\node[anchor=west] at (P) {$\mathcal{S}^2$};

\path[name path=linieted,
	decoration={
    markings,
    mark=at position 0.5 with {\arrow[thick]{latex}}},postaction={decorate}
	] 
    	(D4) to [bend left=17] coordinate[pos=0.5] (sipka) (N4);
\path[name intersections={of = sfera and linieted}];
\coordinate (bod)  at (intersection-1);
\begin{scope}
	\clip (bod) rectangle (sipka);
	\draw[thick,-latex] (D4) to [bend left=17] (N4);
\end{scope}

\node[anchor=west] at (sipka) {$l^a$};

\draw[thick,-latex] (bod) to +(-0.6,0.7) node[anchor=south east] {$n^a$};

\node[anchor=west] at (N\N) {$\mathcal{H}$};

\end{tikzpicture}
  \caption{Local topology of the space-time.}
  \label{img:spacetime:topology:1}
\end{figure}
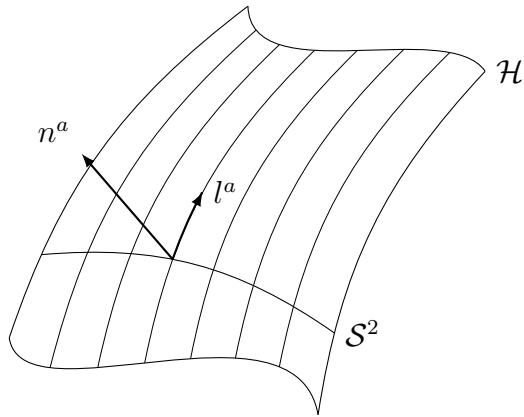

From the definition~\ref{def:non-expand} we know that the horizon has topology $\mathbb{R} \times \mathcal{S}^2$. The generators $l^a$ are topologically isomorphic  to $\mathbb{R}$. We already know that there is a 2-dimensional space-like hypersurface orthogonal to $l^a$, and it can be now easily concluded that we can choose vectors $m^a$ and $\cconj{m}^a$ as a basis of the (space tangent to the) sphere $\mathcal{S}^2$, and we will do so shortly. The structure is schematically illustrated in figure \ref{img:spacetime:topology:1}.
It is assumed that the whole space-time in a neighbourhood of the horizon can be foliated by such null hypersurfaces so that the topology of the space-time is, at least locally\footnote{In principle, global topology $\mathcal{S}^1 \times \mathbb{R}\times \mathcal{S}^2$ is not excluded, although we consider it as non-physical situation. However, even such topology is locally, in the neighbourhood of the horizon, of desired type.}, $\mathbb{R} \times \mathbb{R} \times \mathcal{S}^2$.

Now we would like to construct convenient coordinates on the horizon which reflect the structure we have found. Let us pick one particular 2-sphere, and denote it by $\mathcal{S}_0$.  We define a coordinate $v$ by putting it equal to zero on $\mathcal{S}_0$ and propagating it along the generators $l^a$:
\begin{equation}\label{eq:intro of v}
	v = 0 \quad \text{on $\mathcal{S}_0$}\,,\qquad D v \doteq 1 \,.
\end{equation}
The equality with a dot means that two quantities are equal on the horizon, however, they do not necessarily have to be equal elsewhere. Recall that $D$ is the Newman--Penrose covariant derivative operator. From the fact that the topology of $\mathbb{R}$ is given by $l^a$, it follows that the spheres are labelled by the coordinate $v$.

The spheres are 2-dimensional spaces, and we need to introduce two other coordinates $x^2$ and $x^3$ -- let them be the spherical coordinates $x^2 = \theta$ and $x^3 = \phi$ which are given on the sphere $\mathcal{S}_0$. They are propagated on the entire horizon $\HH$ as
\begin{equation}
	D x^I \doteq 0 \,,\quad I = 2,3 \,.
\end{equation}
Thus, we have established coordinates $(v, x^I)$ on $\HH$.

Next, we wish to complete the generator $l^a$ to a full null tetrad on the horizon. On $\mathcal{S}_0$, consistently with the discussion of the horizon topology above, we choose vectors $m^a$ and $\cconj{m}^a$ in an arbitrary way. With these vectors, we associate a projector on $\mathcal{S}_0$
\begin{equation}\label{eq:metric on S0}
	q_a^b = - m_a \cconj{m}^b - \cconj{m}_a m^b \quad \text{on $\mathcal{S}_0$}\,.
\end{equation}
It acts as identity on $T\mathcal{S}_0$ and annihilates $l^a$. It also gives us degenerate induced metric on the horizon $\HH$ and \emph{non-degenerate} metric on $\mathcal{S}_0$. 

Naturally, we want to extend vectors $m^a$ and $\cconj{m}^a$ \emph{off} the initial sphere $\mathcal{S}_0$ to the entire horizon. One natural choice would be to parallelly propagate $m^a$ along $l^a$, and such a choice is often done in the context of Bondi tetrad at the null infinity. However, an isolated horizon is supposed to describe a black hole in an equilibrium with its neighbourhood. This does not mean that the \emph{space-time} metric is stationary, i.e., the presence of a Killing vector is not required. In this sense, isolated horizons are generalizations of stationary Killing horizons. Nevertheless, the condition of being in an equilibrium imposes restrictions on the \emph{intrinsic} metric of the horizon. Namely, it turns out that for a weakly isolated horizon, the induced intrinsic metric \emph{is} stationary and the pull-back of $l_a$ is the Killing vector of the induced metric. For this reason, it is more convenient to propagate vectors $m^a$ off the sphere $\mathcal{S}_0$ in terms of the Lie derivative, rather than using the covariant derivative. That is, we would like to propagate $m^a$ requiring that the Lie derivative of $m^a$ vanishes along $l^a$. Then, the projector \eqref{eq:metric on S0} can be interpreted as a Lie constant metric on $\HH$. 

There is an issue, however. In general, a Lie transport does not preserve the scalar product. Hence, if we have a null tetrad at one point and we Lie drag it, in general, we end up with differently normalized set of vectors. In what follows, we carefully show that a non-expanding horizon is non-shearing and non-twisting as well. This implies that the Lie dragging preserves the normalization of vectors tangent to the horizon and, consequently, the Lie dragging preserves the induced metric, and we can conclude that the intrinsic metric is stationary.


\section{Geometrical interpretation of optical scalars and their relation to spin coefficients}
\label{sec:optical scalars interpretation}
To elucidate physical interpretation of equations from section~\ref{sec:field equations}, it is convenient to get some intuition for the geometrical interpretation of at least some of the spin coefficients. This topic is in detail covered in~\cite{Chandrasekhar1983}. We briefly review the most important parts.

The physical meaning of the optical scalars is not seen at a point, we need to investigate how the horizon change when the tetrad is being parallelly propagated along the congruence. The term $\nabla_b l_a$ is therefore in the centre of our interest. We recall the discussion of the geodesic equation~\eqref{eq:spin coef:example}. We have seen that the spin coefficient $\kappa = 0$. Moreover, it is possible to make the geodesic affinely parametrized, i.e.\ set $\epsilon = 0$,\footnote{It would be enough to zero the real part of $\epsilon$.} by a transformation of type~\eqref{eq:rotation about l}.

\begin{figure}
\begin{minipage}{\textwidth}
	\input{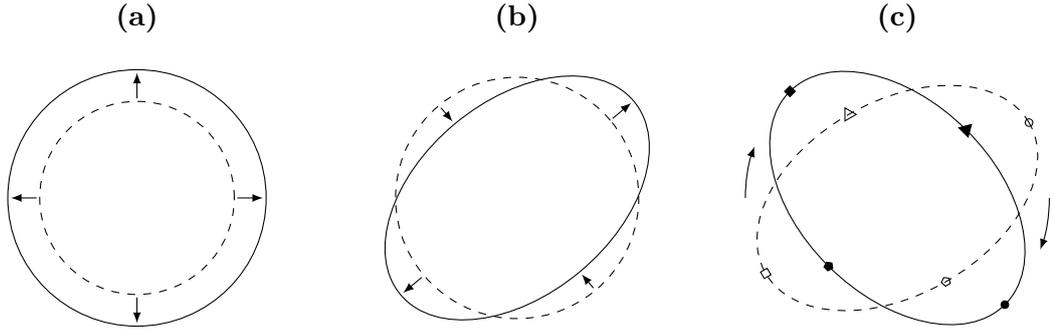}
	\begin{center}
		\begin{subfigure}[t]{\osw\linewidth}
			\begin{center}
				\caption{} 
				\begin{tikzpicture} [scale=\osscale,shorten <>/.style={ shorten >=\osshorten, shorten <=\osshorten }]

\pgfmathsetmacro{\osbc}{(0.9*\osbig+\osmiddle)/2}

\draw[dashed] (0,0) circle (\ossmall);
\draw (0,0) circle (\osbc);

\draw[-latex,shorten <>] (0,\ossmall) -- (0,\osbc);
\draw[-latex,shorten <>] (0,-\ossmall) -- (0,-\osbc);
\draw[-latex,shorten <>] (\ossmall,0) -- (\osbc,0);
\draw[-latex,shorten <>] (-\ossmall,0) -- (-\osbc,0);

\path (0,-\os) -- (0,\os);

\end{tikzpicture}%
			\end{center}%
		\end{subfigure}%
	\hfill
		\begin{subfigure}[t]{\osw\linewidth}
			\begin{center}
				\caption{} 
				\begin{tikzpicture} [rotate=40,scale=\osscale,shorten <>/.style={ shorten >=\osshorten, shorten <=\osshorten }]


\draw (0,0) ellipse ({\osbig} and \ossmall);

\draw[dashed] (0,0) circle (\osmiddle);

\draw[-latex,shorten <>] (0,\osmiddle) -- (0,\ossmall);
\draw[-latex,shorten <>] (0,-\osmiddle) -- (0,-\ossmall);
\draw[-latex,shorten <>] (\osmiddle,0) -- (\osbig,0);
\draw[-latex,shorten <>] (-\osmiddle,0) -- (-\osbig,0);

\path[rotate=-40] (0,-\os) -- (0,\os);

\end{tikzpicture}%
			\end{center}%
		\end{subfigure}%
	\hfill
		\begin{subfigure}[t]{\osw\linewidth}
			\begin{center}
				\caption{} 
				\begin{tikzpicture} [scale=\osscale]%

\draw[rotate=-45] (0,0) ellipse ({\osbig} and \ossmall);
\draw[dashed,rotate=30] (0,0) ellipse ({\osbig} and \ossmall);

\draw[-latex] ([shift=(0:{\osbig})]0,0) arc (0:-20:{\osbig});
\draw[-latex] ([shift=(180:{\osbig})]0,0) arc (180:160:{\osbig});

\draw[rotate=30] node[draw, circle ,inner sep=1pt] at ($(0,0)+(0:{\osbig} and {\ossmall})$) {};
\draw[rotate=30] node[rotate=30, draw, regular polygon, regular polygon sides=3,inner sep=1pt] at ($(0,0)+(90:{\osbig} and {\ossmall})$) {};
\draw[rotate=30] node[rotate=30, draw, regular polygon, regular polygon sides=4,inner sep=1pt] at ($(0,0)+(180:{\osbig} and {\ossmall})$) {};
\draw[rotate=30] node[rotate=30, draw, regular polygon, regular polygon sides=5,inner sep=1pt] at ($(0,0)+(270:{\osbig} and {\ossmall})$) {};

\draw[rotate=-45] node[draw, fill=black, circle ,inner sep=1pt] at ($(0,0)+(0:{\osbig} and {\ossmall})$) {};
\draw[rotate=-45] node[rotate=-45, draw, fill=black, regular polygon, regular polygon sides=3,inner sep=1pt] at ($(0,0)+(90:{\osbig} and {\ossmall})$) {};
\draw[rotate=-45] node[rotate=-45, draw, fill=black, regular polygon, regular polygon sides=4,inner sep=1pt] at ($(0,0)+(180:{\osbig} and {\ossmall})$) {};
\draw[rotate=-45] node[rotate=-45, draw, fill=black, regular polygon, regular polygon sides=5,inner sep=1pt] at ($(0,0)+(270:{\osbig} and {\ossmall})$) {};

\path (0,-\os) -- (0,\os);

\end{tikzpicture}%
			\end{center}%
		\end{subfigure}%
	\caption[A graphical illustration of the meaning of expansion, shear, and twist.]{A graphical illustration of the meaning of the \textbf{(a)}~expansion, \textbf{(b)}~shear, and \textbf{(c)}~twist. The circle (ellipse) is around a point in the null congruence and is in the 2-plane spanned by $l^a$ and $m^a$, which is one of the two space-like vectors orthogonal to $l^a$. The dashed lines represent the original horizon while the full lines are the final states of the horizon after the effect of the corresponding operation and the arrows symbolize the processes, which takes place in the future null direction of the congruence. The markings in the case of the twist show how points have moved to make it clear it is not the case of two shears with different polarization. Deeper discussion and especially pictures really reflecting the evolution in 3-dimensions can be found in~\cite{Scholtz2012}.%
	\footnote{The figure was inspired by the one in \cite[p. 5]{Krishnan2014}.}
	}
	\label{img:optical-scalars}
	\end{center}
\end{minipage}
\end{figure}

From the definitions of the optical scalars, we then get
\begin{align}
	\frac{1}{2} \nabla_a l^a &= - \frac{1}{2} \!\( \scrho + \cconj{\scrho} \) = \Theta_{\(l\)} \,, \\
	\frac{1}{2} \nabla_{[b} l_{a]} \nabla^b l^a &= - \frac{1}{4} \!\(\scrho - \cconj{\scrho}\)^2 = \hat{\omega}^2 \,, \\
	\frac{1}{2} \nabla_{(b} l_{a)} \nabla^b l^a &= \Theta_{\(l\)}{}^2 + \abs{\hat{\sigma}}^2\! \,.
\end{align}
We see that $\Theta_{\(l\)} = - \mathrm{Re} \scrho$ and $\hat{\omega} = \mathrm{Im} \cconj{\scrho}$. In the last equation, we have to consider that it is composed from the expansion and the shear. We already know that the expansion is given by $\scrho$, and, therefore, the rest gives us the shear, it turns out $\abs{\hat{\sigma}} = \sigma$.\footnote{The apparent chaos in naming the spin coefficients stemmed from the fact that some of the spin coefficients were already named quantities.} The phase of the shear correspond to its polarization -- see figure~\ref{img:optical-scalars}.

Figure~\ref{img:optical-scalars} shows how the three types of propagation of the horizon look like. The source of the shear is the Weyl tensor while the expansion is determined by matter, and the twist is merely an initial condition \cite{Hawking1973}.

\begin{theorem}\label{theorem:zero rho and sigma}
A non-expanding horizon has no twist and no shear, which means that the spin coefficients $\scrho \doteq \sigma \doteq 0$. Moreover, the Ricci spinor $\ricsp_{00} \doteq 0$.
\end{theorem}
\begin{myproof}
We have already found that vanishing expansion ensures zero real part of $\scrho$, and orthogonality to hypersurfaces from which it follows that the twist and therefore the imaginary part of $\scrho$ are zero. To be able to show the other two zeros, we can use the Sach's equation, which is only other name for one of the Ricci identities, namely~\eqref{np:RI:Drho}, when we employ that $l^a$ is a geodesic ($\kappa = 0$) it reads
\begin{equation}
	D \scrho = \scrho^2 + \(\epsilon + \cconj{\epsilon} \)\! \scrho + \sigma \cconj{\sigma} + \ricsp_{00} \,.
\end{equation}
Together with the zero expansion and twist on the horizon we have
\begin{equation}
	\left|\sigma\right|^2 + \ricsp_{00} \doteq 0 \,.
\end{equation}
The first term is manifestly greater or equal to $0$. To see that the second also is, we have to use the energy condition from definition~\ref{def:non-expand}. The Ricci spinor is defined to be~\eqref{eq:definition of ricsp_00}
\begin{equation}
	\ricsp_{00} = - \frac{1}{2} \mathrm{Ric}_{ab}\sss l^a l^b = 4\pi T_{ab}\sss l^a l^b \,,
\end{equation}
we have used also the Einstein's equations. The energy condition says that for $l^a$ is future pointing, $T_{ab}\sss l^b$ also is, therefore, $T_{ab}\sss l^a l^b \geq 0$. Both terms are non-negative and sum to zero, so both terms must in fact vanish separately.
\end{myproof}

\section{Adapted coordinates}\label{sec:coordinates}
To have the horizon completely described, we are left to introduce $m^a$ and $\cconj{m}^a$ on the entire horizon, recall they were chosen on $\mathcal{S}_0$. We have already proposed we can use the Lie-dragging as it is done in~\cite{Krishnan2012}, however, we should discuss why it is an appropriate choice which is not done in~\cite{Krishnan2012}. The fourth vector ($n^a$) is then completely determined by the triad $(l^a, m^a, \cconj{m}^a)$ and the conditions~\eqref{eq > tetrad formalism > contractions of basis vectors}.

We Lie-drag along the only vector we have in the entire $\HH$. The problem to discuss is that we need relations~\eqref{eq > tetrad formalism > contractions of basis vectors} to be preserved despite the fact the Lie-dragging does not conserve scalar products in general. It turns out, however, that vanishing of the expansion and, consequently, shear and twist, is sufficient for the Lie dragging to preserve the scalar products we need.

\begin{theorem}\label{theorem:horizon lie transport}
Let $(l^a, n^a, m^a, \cconj{m}^a)$ be a tetrad satisfying conditions~\eqref{eq > tetrad formalism > contractions of basis vectors}, defined on the spherical cut $\mathcal{S}_0$ of a non-expanding horizon $\HH$. The vector field $m^a$ on $\HH$ obtained by the Lie dragging
\begin{equation}\label{eq:horizon lie transport}
	\Lie_l m^a \doteq 0 \,
\end{equation}
preserves the normalization conditions
\begin{equation}
	l^a m_a = 0 \,, \qquad m^a m_a = 0 \,, \qquad m^a \cconj{m}_a = -1
\end{equation}
everywhere on $\HH$. Moreover, the spin coefficient $\epsilon$ is real on $\HH$:
\begin{equation}
	\epsilon - \cconj{\epsilon} \doteq 0 \,.
\end{equation}
\end{theorem}
\begin{myproof}
The Lie derivative \cite{Fecko2006} of vector and covector fields, respectively, is given by
\begin{align}
	\Lie_l X^a &= l^b\, \nabla_b X^a - X^b\, \nabla_b l^a \,, \\
	\Lie_l \alpha_a &= l^b\, \nabla_b \alpha_a + \alpha_b\, \nabla_a l^b \,. \label{eq:lie derivative of covector}
\end{align}
When applying the Lie derivative on the three scalar products, we get
\begin{align}
	\Lie_l \!\(m_a m^a \) &\doteq m^a \Lie_l m_a = m^a \!\bigl(l^b\, \nabla_b m_a + m_b\, \nabla_a l^b\bigr) \nonumber \\
	&\phantom{\doteq m^a \Lie_l m_a }\qquad\equiv m^a D m_a + m_b\, \delta l^b = \sigma = 0 \,, \\
	\Lie_l \!\(m_a l^a \) &\doteq l^a \Lie_l m_a = l^a D m_a + m_b\, D l^b = D \!\(l^a m_a\) - m_a\, D l^a + m_b\, D l^b = 0 \,, \\
	\Lie_l \!\(m_a \cconj{m}^a \) &\doteq  \cconj{m}^a \Lie_l m_a = \cconj{m}^a D m_a + m_b\, \cconj{\delta} l^b = \scrho + \cconj{\epsilon} - \epsilon = \cconj{\epsilon} - \epsilon \,.
\end{align}
We see that if the Lie transport preserves the scalar product, the term $\cconj{\epsilon} - \epsilon$ has to be zero. To prove that this is indeed the case, let us introduce another vector field $\widehat{m}^a$ which is identical with $m^a$ on $\mathcal{S}_0$ and is propagated by condition
\begin{equation}
	D \widehat{m}^a \doteq 0 \,.
\end{equation}
In this new basis, the term $\cconj{\hat{\epsilon}} - \hat{\epsilon}$ vanishes, as we wanted, for it is defined to be $\cconj{\epsilon} - \epsilon = \cconj{\widehat{m}}_a\, D \widehat{m}^a \doteq 0$. The original vector $m^a$ can be expressed in the new basis
\begin{equation}
	m^a \doteq A l^a + B \widehat{m}^a \,.
\end{equation}
Condition~\eqref{eq:horizon lie transport} then gives us requirements onto the unknown coefficients $A$ and $B$.
\begin{equation}
	DA \doteq B\big(\cconj{\hat{\alpha}} + \hat{\beta} \big) \,, \qquad DB \doteq B \!\(\cconj{\hat{\epsilon}} - \hat{\epsilon} \) \doteq 0 \,.
\end{equation}
The difference of the two spin coefficients in the original basis is then
\begin{equation}
	\epsilon - \cconj{\epsilon} \doteq \cconj{B} \!\( DB + B \!\( \hat{\epsilon} - \cconj{\hat{\epsilon}} \) \) \doteq 0 \,.
\end{equation}
Also other properties are conserved.
\end{myproof}

So far, we have constructed the tetrad on the horizon where the tangent triad $(l^a, m^a, \cconj{m}^a)$ is given in the introduced coordinates as
\begin{align}
	l^a &\doteq \(\pder{}{v} \)^a \! \,, \label{eq:l on H}\\
	m^a &\doteq \xi^I \! \(\pder{}{x^I}\)^a \! \,. \label{eq:m on H}
\end{align}

The vector $n^a$ is the only one which is not tangent to the horizon, and, therefore, we can conveniently use it to get the tetrad also off the horizon. An advantageous way is to construct a geodesic in the direction of $n^a$ at every point of the horizon. This means that we require
\begin{equation}\label{eq:parallel transport n}
	\Delta n^a = 0 \,.
\end{equation}
Then the easiest way to get $l^a$ and $m^a$ everywhere is to parallelly transport them because that preserves the normalizations:
\begin{equation}\label{eq:parallel transport l,m}
	\Delta l^a = \Delta m^a = 0 \,.
\end{equation}
Since the coordinates are closely related to the vectors on the horizon, we propagate them alike:
\begin{equation}
	\Delta v = \Delta x^I = 0 \,.
\end{equation}

We have the Bondi-like \cite{Krishnan2012} coordinates in the entire neighbourhood of the horizon; hence, we can write down how are the vectors of the tetrad given in terms of the coordinates. The simplest is the vector $n^a$ along which only the coordinate $r$ varies:
\begin{equation}
	n^a = \(\pder{}{r} \)^a \! \,. \label{eq:tetrad gen coor n}
\end{equation}
For the vector $l^a = l^\mu (\pd_\mu)^a$, the normalization condition $l^a n_a = 1$ determines the component $l^v$ to be $1$ because $n_a = (\d v)_a$. This statement follows from the fact that $n_a$ is normal to hypersurfaces $\NN_v$ labelled by $v$ which are generated by the geodesics along $n^a$ arising from the foliation $\mathcal{S}_v$ of the horizon. The situation can be seen in figure~\ref{img:spacetime structure}. The functions for the other three coordinates are outside of the horizon general, and we denote them in the following way:
\begin{equation}
	l^a = \(\pder{}{v} \)^a + U \!\(\pder{}{r}\)^a + X^I \!\(\pder{}{x^I} \)^a \label{eq:tetrad gen coor l}
\end{equation}
while the $m^a$ vector analogically is
\begin{equation}
	m^a = \Omega \!\(\pder{}{r} \)^a + \xi^I \!\(\pder{}{x^I}\)^a \! \,. \label{eq:tetrad gen coor m}
\end{equation}
The functions have to obey
\begin{equation}
	U \doteq X^I \doteq \Omega \doteq 0
\end{equation}
to meet the conditions for $l^a$ \eqref{eq:l on H} and $m^a$ \eqref{eq:m on H} on the horizon. However, the functions are not arbitrary even off the horizon. We know that the metric tensor is, in the Newman--Penrose formalism, related to the tetrad, cf.~\eqref{eq:metric from tetrad}. Therefore, the coordinate components of the tetrad vectors constitute the coordinate components of the metric tensor; for this reason we refer to them as the \focc{metric functions}. The so-called \focc{frame equations} \cite[section 3.10]{Stewart1993} for the metric functions then follow from the commutation relations \eqref{eq:commutation relations} by applying them onto the coordinates. Before writing them down, let us make a digression which will simplify them.

The construction of the tetrad gives us a further simplification by making some of the spin coefficients vanish. The conditions of the parallel transport are in fact directional derivatives of the tetrad, which are called transport equations. We have already seen one of them -- \eqref{eq:spin coef:example}, however, we are interested in the \qm{$\Delta$} set of them, namely: \eqref{np:transport eqs-Delta la}, \eqref{np:transport eqs-Delta na}, and \eqref{np:transport eqs-Delta ma}. Together with the conditions \eqref{eq:parallel transport n} and \eqref{eq:parallel transport l,m}, we have
\begin{equation}\label{eq:par tr zero spins}
	\gamma = \nu = \tau = 0 \,.
\end{equation}
Firstly, we use commutators~\eqref{eq:commutator Delta D} and~\eqref{eq:commutator delta deltacc} (together with the coordinate $v$) to get relations
\begin{equation}\label{eq:pi alpha beta real mu}
	\pi = \alpha + \cconj{\beta} \,, \qquad \cconj{\mu} = \mu \,.  
\end{equation}

The other two commutators applied on $v$ give identities while all the commutators~\eqref{eq:commutation relations} and the coordinates $r$ and $x^I$ result in the frame equations.

Now we can write the frame equations explicitly, taking into account the simplifications implied by \eqref{eq:par tr zero spins} and \eqref{eq:pi alpha beta real mu}. The first set of the frame equations consists of tangential derivatives and reads
\begin{subequations}
\begin{align}
D \Omega - \delta U &= -\kappa + \left( \cconj{\scrho} - \cconj{\epsilon} + \epsilon \right)\! \Omega + \sigma \cconj{\Omega} \,, \label{eq:DOmega-deltaU}\\
D\xi^I - \delta X^I &= \left( \cconj{\scrho}-\cconj{\epsilon} + \epsilon \right)\! \xi^I + \cconj{\sigma} \cconj{\xi}^I \,,
\label{eq:DxiI-deltaXI}\\
\cconj{\delta}\Omega - \delta\cconj{\Omega} &= \cconj{\scrho} - \scrho + \left(\alpha - \cconj{\beta} \right)\! \Omega - \left( \cconj{\alpha} - \beta \right)\!\cconj{\Omega} \,,\label{eq:deltabarOmega} \\
\delta \cconj{\xi}^I - \cconj{\delta}\xi^I &= \left(\cconj{\alpha} - \beta \right)\! \cconj{\xi}^I - \left( \alpha - \cconj{\beta} \right)\! \xi^I \,,\label{eq:deltaXi}
\end{align}
while the second set of the frame equations represents the evolution of the metric functions along $n^a$:
\begin{align}
 \Delta U &= \epsilon + \cconj{\epsilon} - \pi\Omega - \cconj{\pi}\cconj{\Omega} \,, \label{eq:DeltaU} \\
\Delta X^I &= - \pi \xi^I - \cconj{\pi} \cconj{\xi}^I \,, \label{eq:DeltaXI} \\
\Delta \Omega &= \cconj{\pi} - \mu \Omega - \cconj{\lambda} \cconj{\Omega} \,,\label{eq:DeltaOmega} \\
\Delta \xi^I &= -\mu \xi^I - \cconj{\lambda} \cconj{\xi}^I \,.\label{eq:DeltaxiI}
\end{align}\label{eq:eqs for metric functions}
\end{subequations}

\section{Geometry of the horizon}
\label{sec:geomtry of the horizon}
As we have seen, some of the spin coefficients can be eliminated by an appropriate choice of the tetrad, but these are not the only quantities which can be eliminated in this way. We have already found that the Ricci component $\ricsp_{00}$ vanishes on the horizon, and we shall investigate if there are others which do. It turns out there are.

Thanks to $\ricsp_{00} \doteq 0$, the electrovacuum Einstein equations~\eqref{eq:electrovac einstein} give
\begin{equation}
	\oldphi_0 \doteq 0 \,. \label{eq:msp0 0}
\end{equation}
Using the equation one more time, we find
\begin{equation}
	\ricsp_{0m} \doteq 0 \,, \quad m = 0,1,2 \,.
\end{equation}

The remaining equations for components of the Riemann tensor we have not used yet are the Ricci identities~\eqref{eq:ricci id}. Two of them are useful at this stage. From \eqref{np:RI:Dsigma}, we realize that
\begin{equation}
	\Psi_0 \doteq 0
\end{equation}
and \eqref{np:RI:deltarho} gives
\begin{equation}
	\Psi_1 \doteq 0 \,.
\end{equation}
Moreover, $\Psi_2$ is invariant on $\HH$ for null tetrads adapted to the horizon and together with $\oldphi_1$ is time independent, \cite{Krishnan2012}.

Next we turn our attention to the intrinsic covariant derivative of the horizon $\HH$. Given a full space-time connection $\nabla_a$, it is \emph{not} possible to introduce a \emph{unique} induced covariant derivative on a general \emph{null} hypersurface. This is different when the hypersurface is a space-like hypersurface, since we can use the projector that gives induced metric, and get the compatible covariant derivative by a projection. On the other hand, null hypersurfaces exhibit an additional complication. The induced metric is degenerate and, therefore, does not have a unique inverse. Consequently, usual Christoffel symbols cannot be defined. Thus, there is no preferred way of inducing the connection on a general null hypersurface from the ambient space-time. Nevertheless, the non-expanding horizons \emph{do have} a preferred one which we introduce following~\cite{Ashtekar2000b}.

\begin{theorem}\label{theorem:induced covder}
	Let $(\MM, g_{ab})$ be a space-time in which is embedded a non-ex\-pan\-ding horizon $\HH$ and let $\nabla_a$ be the standard Levi-Civita connection  compatible with metric $g_{ab}$. Then the \textbf{induced covariant derivative} $\DD_a$ on $\HH$ defined by
	\begin{equation}
		X^a\, \DD_a Y^b \doteq X^a\, \nabla_a Y^b \qquad \text{for any tangent vectors}~X^a, Y^a,
	\end{equation}
	is a well-defined covariant derivative on $T\HH$.
\end{theorem}
\begin{myproof}
In order to appreciate the theorem, it is necessary to understand potential obstacles. By the definition, derivative $\DD_a$ is just a restriction of the ambient connection to the horizon. The connection $\nabla_a$ operates on the tangent bundle $T \MM$ of the space-time, while a well-defined induced connection should operate on $T\HH$ only. The domain of $\DD_a$ is, by definition, just $T \HH$ because we restrict $\nabla_a$ to $\DD_a$. However, in general, it is not guaranteed that the \emph{image} of $\DD_a$ will be in $T\HH$, even if the pre-image is. In other words, a restriction of $\nabla_a$ to $T\HH$ can act only on tangential vectors $X,Y$, but the result may contain also non-tangential components. The point of the theorem is that this problem does not occur if $\DD_a$ is induced on a non-expanding horizon.

The tangent vectors can be expanded in terms of the tangent basis:
\begin{equation}
	X^a \doteq x_l l^a + x_m m^a + x_{\cconj{m}} \cconj{m}^a \,, \qquad Y^b \doteq y_l l^a + y_m m^a + y_{\cconj{m}} \cconj{m}^a \,.
\end{equation}
The covariant derivative in the direction of $X^a$ then is
\begin{equation}\label{eq:covder decomposition with X}
	X^a\, \nabla_a \doteq x_l\, D + x_m\, \delta + x_{\cconj{m}}\, \cconj{\delta} \,,
\end{equation}
where we have used decomposition~\eqref{eq:covder decomposition}. The action of this operator on the vector $Y^b$ yields two types of terms: derivatives of the factors $y_l$, $y_m$ and $y_{\cconj{m}}$ -- they clearly give results tangent to $\HH$; and derivatives of the vectors $l^b$, $m^b$ (and $\cconj{m}^b$) -- we shall show that they are for non-expanding horizons also tangent. Therefore, we want to demonstrate that when we apply the transport equations~\eqref{app1:transport eqs} to the latter terms, all expressions containing $n^b$ are zero. For $D l^b$ and $\delta l^b$, there are no such terms. We are left with
\begin{equation}
	D m^b \rightarrow - \kappa n^b \,, \qquad \delta m^b \rightarrow - \sigma n^b \,, \qquad \cconj{\delta} m^b \rightarrow - \scrho n^b \,.
\end{equation}
The first one is zero for $l^a$ being a geodesic while the other two vanish thanks to theorem~\ref{theorem:zero rho and sigma}.
\end{myproof}

\noindent A consequence of this theorem is the existence of the so-called \emph{rotational 1-form} defined by the following theorem.

\begin{theorem}\label{theorem:rotational form}
	For a non-expanding horizon $\HH$ with the induced covariant derivative $\DD_a$ of the type from theorem~\ref{theorem:induced covder}, there exists a 1-form $\omega_a$ such that
	\begin{equation}
		\DD_a l^b \doteq \omega_a l^b \,.
	\end{equation}
\end{theorem}
\begin{myproof}
	Let us contract $\DD_a l^b$ with a vector $X^a$, then we can use the theorem~\ref{theorem:induced covder} and decompose the covariant derivative in sense of~\eqref{eq:covder decomposition with X}, we get
	\begin{equation}
		X^a\, \DD_a l^b \doteq x_l\, D l^b + x_m\, \delta l^b + x_{\cconj{m}}\, \cconj{\delta} l^b \,.
	\end{equation}
	We recognize the transport equations~\eqref{app1:transport eqs} which are fortunately very simplified on the horizon in our tetrad, namely:
	\begin{equation}
		D l^b \doteq \(\epsilon + \cconj{\epsilon} \)\! l^b \,, \qquad \delta l^b \doteq \cconj{\pi} l^b \,.
	\end{equation}
	We can define a one form $\omega_a$ to be
	\begin{equation}\label{eq:omega def}
		\omega_a \doteq \(\epsilon + \cconj{\epsilon} \)\! n_a - \pi m_a - \cconj{\pi} \cconj{m}_a
	\end{equation}
	to get
	\begin{equation}
		X^a \omega_a \doteq x_l \!\(\epsilon + \cconj{\epsilon} \) + x_m \pi + x_{\cconj{m}} \cconj{\pi} \,.
	\end{equation}
	Since the vector $X^a$ is arbitrary,
	\begin{equation}
		\DD_a l^b \doteq \bigl(\(\epsilon + \cconj{\epsilon} \)\! n_a - \pi m_a - \cconj{\pi} \cconj{m}_a \bigr) l^b \doteq \omega_a l^b \,.
	\end{equation}
\end{myproof}

The rotational form $\omega_a$ has several interesting properties. First, it encodes an important part of the connection $\DD_a$, since it fully encodes all possible derivatives of the vector $l^a$. It also carries information about intrinsic geometry of the spherical cuts of the horizon. Let us see how.

The rotational 1-form, as defined in theorem \ref{theorem:rotational form}, is a four-dimensional object which lives only on the horizon. But it is not a form intrinsic to the horizon. In order to get a three-dimensional 1-form, we define the \emph{pull-back} \cite{Fecko2006,Hawking1973}
\begin{align}
 \underleftarrow{\omega_a} &= \psi^\star \omega_a \,,
\end{align}
where $\psi : \HH \mapsto \MM$ is the embedding of the horizon into the space-time. Under the pull-back, we have
\begin{align}
 \underleftarrow{l_a} &= 0 \,,
\end{align}
because the 1-form $l_a$ annihilates all vectors tangent to $\HH$. If we pull it back on the horizon, it annihilates all vectors of $T\HH$, and, hence, as a three-dimensional object, $\psi^\star l_a$ is identically zero. It is again a manifestation of the fact that the intrinsic metric on $\HH$ is degenerate. Similarly, pull-back of $n^a$ to the horizon is identically zero, because $n^a$ is not tangential to $\HH$ and so cannot arise as a push-forward of a vector from $T\HH$. On the other hand, the 1-form $n_a$ can be pulled-back,
\begin{align}
 \underleftarrow{n_a} &= \psi^\star n_a \,,
\end{align}
and similarly for $m_a$ and $\cconj{m}_a$. Since $l^a$ and $m^a$ are tangent to $\HH$ even in the space-time, we can freely identify three-dimensional $l^a, m^a$ on the horizon and their four-dimensional counter-parts. Thus, the pull-back of $\omega_a$ is
\begin{equation}
	\underleftarrow{\omega_a} \doteq \(\epsilon + \cconj{\epsilon} \)\! \underleftarrow{n_a} - \pi \underleftarrow{m_a} - \cconj{\pi} \underleftarrow{\cconj{m}_a} \,.
\end{equation}

\begin{figure}
\begin{center}
 \includegraphics[width=0.7\textwidth]{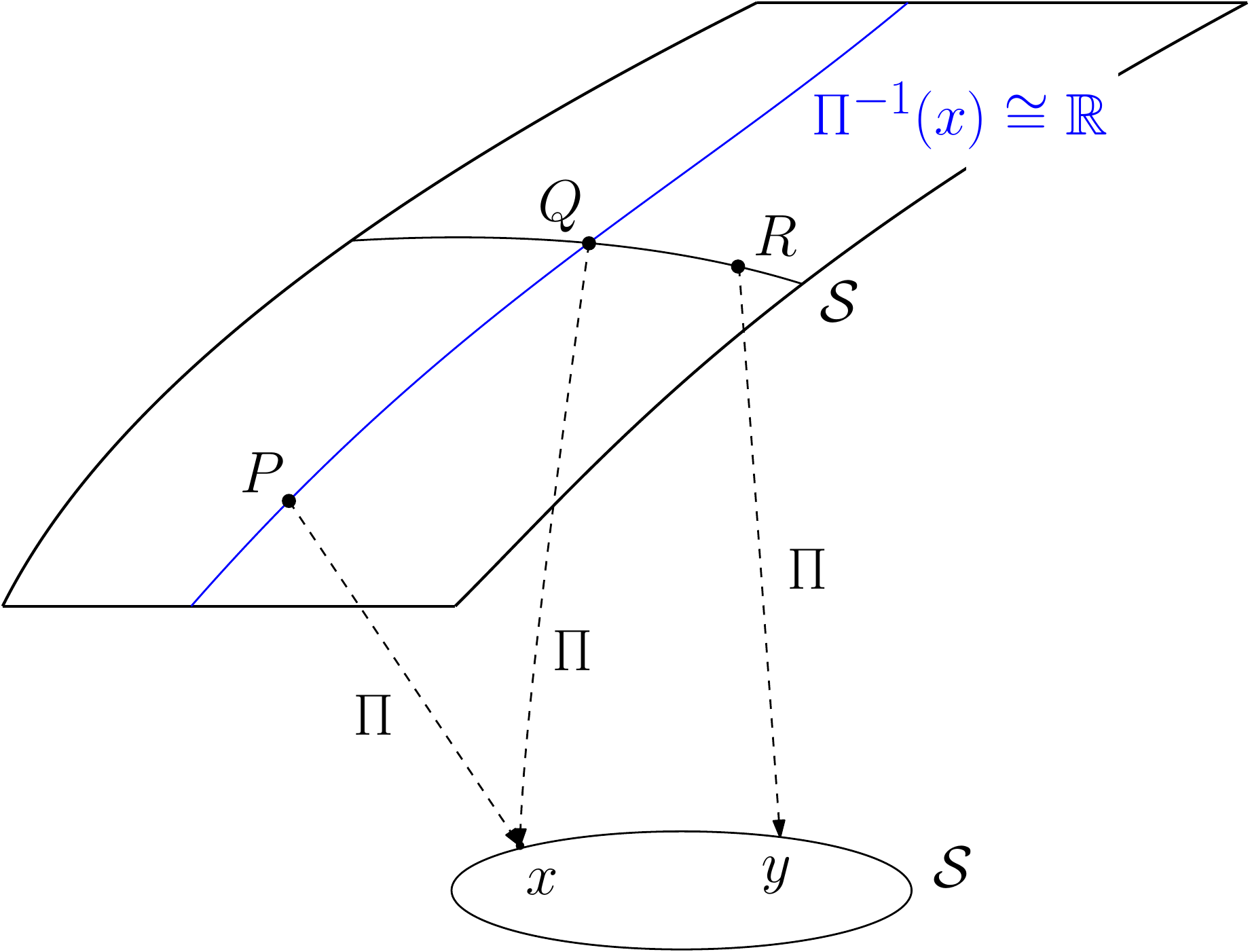}
 \caption{A non-expanding horizon as a fibre bundle $\HH$ which is diffeomorphic to $\SS \times \Real$.}
 \label{fig:fibration}
 \end{center}
\end{figure}

From a geometrical point of view, a non-expanding horizon is a fibre bundle \cite{Ashtekar2002} $\HH$ diffeomorphic to the product $\SS \times \Real$ with a canonical projection
\begin{align}
 \Pi : \HH \cong \SS \times \Real \mapsto \SS \,,
\end{align}
see figure \ref{fig:fibration}. Here, the base manifold is diffeomorphic to a sphere $\SS$ and the fibres are null geodesics generating the horizon. Local trivialization of this bundle is provided by the coordinates $x^I$ on $\SS$ and coordinate $v$ along each fibre. Thus, a point with coordinates $(v,x^I)$ is mapped onto corresponding points on the sphere with the coordinates $x^I$. For example, for the point $P,Q$ and $R$ in figure~\ref{fig:fibration}, we have
\begin{equation}
  \Pi(P) = \Pi(Q) = x \in \SS \,, \qquad 
  \Pi(R) = y \in \SS \,.
\end{equation}
The fibre at the point $x \in \SS$ is $\Pi^{-1}(x)$, i.e.\ a null geodesic in $\HH$. 

Now, on $\SS$, we have a natural metric given by the projector,
\begin{equation}
 \two q_{ab} = - m_a \cconj{m}_b - \cconj{m}_a m_b \,,
\end{equation}
where the left superscript indicates that the quantity is defined on $\SS$. This metric is non-degenerate on $\SS$, but its pull-back
\begin{equation}
 q_{ab} = \Pi^\star \two q_{ab}
\end{equation}
is exactly the projector  \eqref{eq:metric on S0} which annihilates both $l^a$ and $n^a$. Next, on $\SS$ we have a volume two form
\begin{equation}
  \two \levicivita_{ab} = \levicivita_{abcd}\sss l^c n^d \,.
\end{equation}
Newman--Penrose form of $\two\levicivita_{ab}$ can be easily found using the spinors and relation \eqref{eq:levi-civita symbol}:
\begin{align}\label{eq:epsilon 2d}
 \two\levicivita_{ab} = \levicivita_{abcd}\sss o^C \cconj{o}^{C'} \iota^D \cconj{\iota}^{D'} = 
 \im \left( o_A \iota_B \cconj{o}_{B'} \cconj{\iota}_{A'} - o_B \iota_A \cconj{o}_{A'} \cconj{\iota}_{B'} \right) = 2 \im m_{[a} \cconj{m}_{b]} \,.
\end{align}

If we think of the embedding of $\SS$ in $\HH$, the pull-back of the rotational 1-form is
\begin{equation}
 \two \omega_a = - \pi m_a -\cconj{\pi} \cconj{m}_a \,,
\end{equation}
since the pull-back of $n_a$ on $\SS$ must vanish (similarly to vanishing of $l_a$ under the pull-back from $\MM$ to $\HH$). Let us take the exterior derivative of the two-dimensional form,
\begin{equation}
 \left( \dd \two\omega \right)_{ab} = 2\nabla_{[a} \two\omega_{b]} \,.
\end{equation}
Since this is a form living on $\SS$, the volume form can have only one independent component (the space of 2-forms on vector space of dimension 2 is 1), i.e.
\begin{equation}
 2\nabla_{[a}\two\omega_{b]} = F\, \two\levicivita_{ab} \,,
\end{equation}
where the factor $F$ can be found by a contraction with $\two\levicivita^{ab}$ which is given by \eqref{eq:epsilon 2d}. Writing the result in the Newman--Penrose formalism, we get
\begin{align}
 F = \two\levicivita^{ab}\nabla_a \two\omega_b = \im \left(-\Psi_2 + \cconj{\Psi_2} \right) = 2 \Im \Psi_2 \,,
\end{align}
where we have used the Ricci identities and \eqref{eq:pi alpha beta real mu}. Thus, we have derived an important result
\begin{align}\label{eq:d omega}
 \dd \two\omega = 2 \Im \Psi_2 \two\levicivita \,.
\end{align}
Although we do not prove it here, imaginary part of $\Psi_2$ is related to the angular momentum of the horizon. In order to see this informally, consider $\Psi_2$ for the Schwarzschild metric and for the Kerr metric:
\begin{align}
 \Psi_2^{\mathrm{Schw.}} = - \frac{M}{r^3} \,,
 \qquad \qquad 
 \Psi_2^{\mathrm{Kerr}} = - \frac{M}{\left( r - \im a \cos\theta \right)^3} \,.
\end{align}
For Schwarzschild, $\Psi_2$ is real and represents non-rotating, spherically symmetric metric with zero angular momentum. On the other hand, Kerr metric describes a rotating black hole with spin $a$ which enters imaginary part of the denominator. For a full treatment in multipole moments (mass and angular momentum moments) of isolated horizons, see \cite{Ashtekar2004}.

\chapter{Weakly isolated horizons}
\label{chapter:isol}
In this chapter, we will restrict the non-expanding horizons even further. The investigated \focc{weakly isolated horizons} are those for which the zeroth law of thermodynamics holds. As in the previous chapter, we summarize results on this topic from~\cite{Krishnan2012} and partly follow its construction while completing the results with terms adequate to charge simultaneously. In~\cite{Krishnan2012}, the charge corresponding terms are presented separately and not as deeply. \snote{It's OK, but add more references in the beginning.} We have also employed~\cite{Ashtekar2002} for precise formulations of the definitions and theorems. The discussion of quasi-locality of the isolated horizons, together with application of this framework to some other space-times than we are going to investigate, can be found in~\cite{Krishnan2014}. 

\noindent Recall \cite{Wald1984} that for an event horizon we have four basic laws resembling the laws of classical thermodynamics:
\begin{enumerate}[start=0]
	\item Surface gravity $\kappa$ is constant on the horizon.\footnote{Not to be confused with the spin coefficient $\kappa$. We leave the usual letter for the surface gravity there, for we shall decorate it in just a moment.} This is an analogue of the 0th law of thermodynamics that the temperature of a body in thermal equilibrium is everywhere constant. This suggests that surface gravity plays the role of the temperature, which was fully justified by the discovery of the Hawking radiation, \cite{Hawking1974}.
	\item Under a perturbation of an event horizon, the change in the mass is
	\begin{align}\label{eq:1st law}
		\delta M &= \frac{\kappa}{8\pi}\,\delta A  + \Omega_{\mathrm{H}}\,\delta J + \Phi\,\delta Q \,,
	\end{align}
	where $A$ is the area of the horizon, $\Omega_\mathrm{H}$ its angular velocity and $J$ is the angular momentum of the black hole. In the case of charged black hole, $\Phi$ is the electric potential and $Q$ is the charge. This is analogous to the 1st law of thermodynamics, where the area $A$ plays the role of entropy, and $\Omega_{\mathrm{H}}$ and $\Phi$ play the roles of generalized forces.
	\item Under a perturbation of an event horizon, the change of the area is always non-negative,
	\begin{align}\label{eq:2nd law}
		\delta A &\geq 0 \,,
	\end{align}
	which resembles the 2nd law of thermodynamics and pushes the analogy between entropy and area even further. 
	\item It is impossible to form an extremal black hole by a finite sequence of steps. Extremal black holes are characterized by vanishing surface gravity $\kappa$, which is an analogue of the temperature. Hence, this law is an incarnation of the 3rd law of thermodynamics, that absolute zero temperature cannot be reached.
\end{enumerate}

Let us make a remark on the third law. In fact, the analogy is not perfect. One of the formulations of this law is that the entropy of a system at absolute zero temperature is zero. For black holes that would mean that extremal black hole, which has vanishing surface gravity and, hence, the temperature, should have zero entropy. This is not true, however, because the area of extremal horizon is not zero, although the temperature is, and therefore extremal black hole does not have vanishing entropy. It is interesting that it is possible to construct black holes in the (low energy limit of) string theory (``stringy black holes'') analogous to the usual Reissner-Nordstr\"om solution containing the dilaton field coupled to electromagnetic field \cite{Fabbri2005}. These metrics have the same structure like Schwarzschild black holes in their $t-r$ part, but the angular part differs and electric or magnetic charges enter just this angular part. In the extremal limit, the radius of the event horizon tends to zero and therefore the area of extremal stringy black hole is zero. The analogy with the third law of thermodynamics is then restored completely.

Although there are many concepts of the horizon, see section~\ref{sec:motivation for isol hor}, what they have in common is some sort of laws of thermodynamics. In fact, motivation for the formalism of isolated horizons in classical general relativity, as a generalization of standard event horizons, stems partially from the problems arising in loop quantum gravity \cite{Ashtekar2000} where the relation between the area and entropy is crucial. 

Non-expanding horizons defined in the previous chapter are not suitable for recovering the laws of thermodynamics. It is easy to see why.  We already know that any normal $l^a$ is necessarily a geodesic, albeit not necessarily affinely parametrized. The \qm{acceleration}, i.e.\ the deviation from affine parametrization, is called the surface gravity $\surfkappa$ and it is defined as follows.
\begin{definition}
	 Given a non-expanding horizon $\HH$, the \focc{surface gravity} $\surfkappa$ associated with a null normal $l^a$ is defined by
	 \begin{equation}
	  D l^a \doteq \surfkappa l^a \,.
	 \end{equation}
If $l^a$ is completed to a full \gls{np} tetrad, an alternative definition is
	\begin{equation}
		\surfkappa \doteq n^a D l_a \doteq \(\epsilon + \cconj{\epsilon} \) . \label{eq:def of surface gravity}
	\end{equation}
\end{definition}

An important point is that, in the present formalism, the notion of surface gravity is tied to a particular choice of the null normal $l^a$. The situation is different for an event horizon, since there exists a preferred choice of $l^a$, which coincides with the Killing vector, and the notion of surface gravity refers to this particular choice. On a general non-expanding horizon, however, no preferred choice is available a priori. 

On the other hand, the \emph{direction} of $l^a$ is unique, and the freedom in the choice of $l^a$ rests in the rescaling of $l^a$ by an arbitrary function,
\begin{align}\label{eq:rescaling}
 l^a & \mapsto f l^a ,
\end{align}
where $f$ is a function of coordinates on the horizon. Theorem \ref{theorem:rotational form} shows that the induced connection $\DD_a$ can be characterized by the rotational 1-form $\omega_a$. One can easily see that defining relation \eqref{eq:omega def} implies
\begin{align}\label{eq:rotational form transformation}
 \omega_a \mapsto \omega_a + f^{-1}\,\DD_a f
\end{align}
under rescaling. Obviously, rescaling $l^a$ with constant factor leaves the rotational form invariant. 

Since the intrinsic metric of the horizon is supposed to be stationary for a black hole in an equilibrium, it is natural to require that the connection $\DD_a$ is $v\text{-independent}$ as well. Whether $\DD_a$
is time dependent or not, however, depends on the choice of $l^a$, while $\DD_a$ is independent of such choice\footnote{Recall that the \qm{time} coordinate $v$ is chosen so that $D v = 1$, i.e.\ it reflects the parametrization of $l^a$. On the other hand, $\DD_a$ is merely a restriction of $\nabla_a$, i.e. obviously independent of the choice of $l^a$.}. It turns out that in order to ensure the zeroth law of thermodynamics it is sufficient to require that $\omega_a$ is Lie constant along $l^a$.

\begin{definition}
	\focc{Equivalence class} $[l^a]$ is a set
	\begin{equation}
		[l^a] = \{ c\sss l^a \,|\, c \in \mathbb{R} \} \,.
	\end{equation}
\end{definition}

Now we can proceed to define an appropriate notion of the horizon:
\begin{definition}
	\focc{Weakly isolated horizon} is a pair ($\mathcal{H}$, $[l^a]$) where $\mathcal{H}$ is the non-expanding horizon and $[l^a]$ is the equivalence class of normals and the condition
	\begin{equation}\label{eq:WIH def}
		[\Lie_l, \mathcal{D}_a] l^b \doteq 0
	\end{equation}
	is met. Recall that $\Lie_l$ is a Lie derivative along $l^a$ and $\mathcal{D}_a$ is the induced covariant derivative introduced in section~\ref{sec:geomtry of the horizon}. Note that it is enough to satisfy the condition for one $l^b$ and it is automatically fulfilled for the entire equivalence class.
\end{definition}

It is important to stress that there is no physical distinction between a non-expanding horizon and a weakly isolated horizon. The difference is mathematical: a non-expanding horizon was defined independently of the choice of the normal. Weakly isolated horizon is a non-expanding horizon equipped with one particular normal $l^a$ or corresponding equivalence class $[l^a]$. By \eqref{eq:rotational form transformation}, the rotational 1-form is well-defined for a weakly isolated horizon, because it is independent of the particular representative of the class $[l^a]$. However, the surface gravity $\surfkappa$ depends on the choice of such a representative, since under rescaling $l^a \mapsto c\sss l^a$ with constant $c$ the surface gravity transforms as
\begin{align}\label{eq:surf gravity rescaling}
 \surfkappa &\mapsto \kappa_{(c\sss l)} = c \kappa_l \,.
\end{align}
Formalism is built up in such a way that important results are insensitive to a different choice of normal from the equivalence class $[l^a]$. 

\begin{definition}
	If $\surfkappa = 0$ for any $l^a \in [l^a]$, the weakly isolated horizon is called \focc{extremal}.
\end{definition}
\noindent This is the only case when the whole equivalence class has the same surface gravity, as~\eqref{eq:surf gravity rescaling} implies.

\begin{theorem}
	\textsf{(Zeroth law of thermodynamics)}: Let $(\mathcal{H}, [l^a])$ be a weakly isolated horizon, then:
	\begin{enumerate}
		\item The rotational 1-form $\underleftarrow{\omega_a}$ is time independent:
			\begin{equation}\label{eq:Lie omega}
				\Lie_l \underleftarrow{\omega_a} \doteq 0 \,.
			\end{equation}
		\item The surface gravity associated with any normal $l^a \in [l^a]$ is constant on $\mathcal{H}$.
	\end{enumerate}
	And the two statements are equivalent.
\end{theorem}
The second statement is the usual formulation of the zeroth law of thermodynamics where the surface gravity is identified with temperature.
\begin{myproof}
	We have
	\begin{equation}
		\( \Lie_l \omega_a \)\! l^b = \Lie_l \big(\omega_a l^b \big) \doteq \Lie_l \mathcal{D}_a l^b = [\Lie_l, \mathcal{D}_a] l^b = 0 \label{eq:zero thermo law 1 part}
	\end{equation}
	where
	\begin{equation}
		\Lie_l l^a = [l, l]^a = 0
	\end{equation}
	have been used twice. 
	The relation~\eqref{eq:zero thermo law 1 part} holds true for every $l^b$ and; hence, $\Lie_l \omega_a \doteq 0$. Now we can use the pull-back, which commutes with the Lie derivative, to get the first statement.
	
	We are left to demonstrate that this is equivalent to the constant surface gravity on the entire horizon. We can rewrite the Lie derivative of $\omega_a$ using its definition~\eqref{eq:omega def}, and decomposition of a Lie derivative of a covector~\eqref{eq:lie derivative of covector} to get:
	\begin{equation}
		\Lie_l \omega_a \doteq - 2 \abs{\pi}^2 l_a + n_a D\! \( \epsilon + \cconj{\epsilon} \) - m_a D \pi - \cconj{m}_a D \cconj{\pi} \,.
	\end{equation}
	via
	\begin{align}
		\Lie_l n_a &\doteq l^b \nabla_b n_a + n_b \nabla_a l^b \doteq D n_a + (\epsilon \cconj{\epsilon} ) n_a - \pi m_a - \cconj{\pi} \cconj{m}_a \doteq 0 \,, \\
		\Lie_l m_a &\doteq l^b \nabla_b m_a + m_b \nabla_a l^b \doteq \cconj{\pi} l_a \,.
	\end{align}
	Recall that $\Lie_l l^a = 0$.
	
	We have to pull-back the expression, and this kills the $l_a$ term. Let us make a remark to explain why:
	\begin{itemize}
		\item[] The pull-back of an arbitrary 1-form $\underleftarrow{\alpha_a} = \phi^\star$ is defined by its action on a push-forwarded vector:
		\begin{equation}
			\(\phi^\star \alpha_a\)\! X^a = \alpha_a \phi_\star X^a .
		\end{equation}
		The vector $l^a$ is tangent to the horizon, as we already know, and is annihilated by 1-form $l_a$, which therefore is from the 1-dimensional space of forms annihilating any tangent vector to $\mathcal{H}$. The push-forwarded vector $\phi_\star X^a$ is tangent to the horizon, and therefore annihilated by $l_a$. Finally, we have $\underleftarrow{l_a} = 0$.
	\end{itemize}
	
	We are left with
	\begin{equation}
		\Lie_l \underleftarrow{\omega_a} \doteq n_a D ( \epsilon + \cconj{\epsilon} ) - m_a D \pi - \cconj{m}_a D \cconj{\pi} \overset{!}{=} 0\,.
	\end{equation}
	Since only scalars are differentiated, and all terms are linearly independent, all of them have to vanish individually. This implies, recall the definition of the surface gravity,
	\begin{equation}\label{eq:D kappa D pi}
		D \surfkappa \doteq D \pi \doteq 0 \,.
	\end{equation}
	
	Unfortunately, this is not sufficient to say that $\surfkappa$ is constant on the horizon. The direction of $m^a$ is left to allow $\surfkappa$ to change. To show that this is not the case, we have to employ equations for derivatives of the spin coefficients -- the Ricci identities. It is conspicuous that the identities with $\delta \epsilon$-like terms are the one we are looking for, they are~\eqref{np:RI:Dbeta} and~\eqref{np:RI:Dalpha}. With the aforementioned simplifications, they read\footnote{The term $-\cconj{\epsilon} \beta - \epsilon\! \(\cconj{\alpha} - \cconj{\pi}\)$ from~\eqref{np:RI:Dbeta}, as an example of a term which may not be obvious, is zeroed by both $\pi = \alpha + \cconj{\beta}$ and $\epsilon - \cconj{\epsilon} \doteq 0$.}
	\begin{equation}
		D \beta - \delta \epsilon \doteq 0 \,, \qquad D \alpha - \cconj{\delta} \epsilon \doteq 0 \,. \label{eq:Da - de}
	\end{equation}
	The surface gravity is tied to the spin coefficient $\epsilon$; hence, we would like to extract its derivative (in the direction of $m^a$) from these equations. Complex conjugation of the second one added to the first one gives (recall $\epsilon \doteq \cconj{\epsilon}$):
	\begin{equation}
		D\! \( \cconj{\alpha} + \beta \) - 2 \delta \epsilon \doteq 0 \,.
	\end{equation}
	The first term is $D \cconj{\pi}$ vanishes on $\mathcal{H}$ by \eqref{eq:D kappa D pi} and therefore
	\begin{equation}
		\delta \epsilon \doteq 0 \label{eq:delta eps} \,,
	\end{equation}
	which ensures that $\surfkappa$ is constant on $\mathcal{H}$. It is simple to check that we can revert the process, and the two statements are equivalent.
\end{myproof}

Theorem we just proved shows that in order to recover the zeroth law of black hole thermodynamics, it is sufficient to impose the condition \eqref{eq:WIH def} or, equivalently, condition \eqref{eq:Lie omega}. However, from a physical point of view it is more natural to impose a stronger condition which reflects that the intrinsic geometry of black hole in equilibrium, including connection, should be time independent.

\begin{definition}
	\focc{Isolated horizon} is a pair ($\mathcal{H}$, $[l^a]$) where $\mathcal{H}$ is the non-expanding horizon and $[l^a]$ is the equivalence class of normals for which
	\begin{equation}\label{eq:IH def}
		[\Lie_l, \mathcal{D}_a]  \doteq 0 \,. 
	\end{equation}
\end{definition}

\noindent For a weakly isolated horizon, not entire connection is time independent, only the part related to $l^a$. For an isolated horizon, full connection is required to be time independent. Hence, an isolated horizon is more restricted concept than an weakly isolated horizon, but physically well motivated. Nevertheless, in what follows we will talk mainly about \emph{weakly isolated horizons} because they are more general but still exhibit laws of black hole thermodynamics. For simplicity, though, we often use the term isolated horizon in the meaning of a weakly isolated horizon.

\section{The intrinsic geometry of the horizon}

Now we are in position to examine the intrinsic geometry of isolated horizon in more detail. We have seen that in the Newman--Penrose formalism we describe the geometry by a plenty of scalar quantities, but not all of them have physical significance. For example, by the choice of outgoing null congruence $n^a$ we were able to eliminate spin coefficients $\gamma+\bar{\gamma}$ and $\nu$, by the parallel transport of $m^a$ along $n^a$ we have achieved $\tau = 0$ and $\gamma - \bar{\gamma}=0$. Quantities which depend on the choice of tetrad or coordinates do not represent true physical degrees of freedom, instead they are part of the \emph{gauge freedom}. If they can be eliminated by a gauge transformation (e.g., rotation of the tetrad) we say they represent a \emph{pure gauge}. We have already exploited the gauge freedom because we fixed the tetrad and coordinates in a geometrical way completely, no additional gauge transformation is possible, otherwise we would break some of the established relations (for example, spin in $m^a$ would make $\epsilon$ complex). 

We can regard the space-time in the neighbourhood of an isolated horizon as a solution to a \emph{characteristic initial value problem}. In contrast to the usual Cauchy initial value problem with the initial data given on the space-like hypersurface, a characteristic problem is formulated on two intersecting null hypersurfaces. In our case, the two null hypersurfaces are the horizon $\HH$ and the transversal hypersurface $\NN_0$, where surfaces $\NN_v$ have been defined in the section \ref{sec:coordinates}. 

The Newman--Penrose formalism is particularly useful for the study of a characteristic initial value problem. By construction of the tetrad, vector $l^a$, with associated derivative $D$, is tangent to the horizon, while vector $n^a$, with associated operator $\Delta$, is tangent to $\NN_0$. So, the Newman--Penrose equations, containing $D$ and $\Delta$ derivatives describe the evolution along the horizon or along $\NN_0$. Altogether they determine the solution in the interior between the two null characteristic hypersurfaces. Equations containing just $\delta$ and $\cconj{\delta}$ derivatives are constraints which must be satisfied by the initial data given on the initial hypersurfaces. Hence, in the Newman--Penrose formalism, it is easy to split field equations into evolution equations and constraints, which is much more difficult in the \gls{adm} formalism with space-like initial hypersurface, \cite{Arnowitt2008}.

Let $\SS_0 = \HH \cap \NN_0$ be the spherical cut where the horizon and $\NN_0$ intersect. For an isolated horizon, cut $\SS_0$ plays a special role because it turns out that the initial data on $\HH$ are \emph{completely determined} by data on $\SS_0$. Thus, we are interested in the characteristic initial value problem formulated on the set $\HH \cup \SS_0 \cup \NN_0$. The values of quantities on the $\SS_0$ shall be decorated with $\szero$. For example $\alpha\szero$ is the value of $\alpha$ on $\SS_0$.

The spin coefficient $\alpha$ in the previous example was not chosen thoughtlessly. If we go back to the proof of the zeroth law of thermodynamics, and plug~\eqref{eq:delta eps} into~\eqref{eq:Da - de}, we get
\begin{equation}
	D \alpha \doteq D \beta \doteq 0 \,.
\end{equation}
This means that $\alpha$ and $\beta$ are $v$-independent on the horizon, and therefore we can prescribe for them
\begin{equation}
	\alpha \doteq \alpha\szero \,, \qquad \beta \doteq \beta\szero \,,
\end{equation}
which implies also
\begin{equation}
	\pi \doteq \pi\szero \,.
\end{equation}

And also the Weyl scalar $\Psi_2$ is sufficient to introduce only on $\SS_0$ thanks to the Bianchi identity~\eqref{np:BI:DPsi2}, which is on the horizon reduced to
\begin{equation}
	D \Psi_2 \doteq 0 \,.
\end{equation}
For completeness, this implies
\begin{equation}
	\Psi_2 \doteq \Psi_2\szero \,.
\end{equation}

In section \ref{sec:spin weights}, we introduced operators $\eth$ and $\cconj{\eth}$ and shown that they act as a spin raising/lowering operators. Some of the Newman--Penrose equations transform covariantly under the spin and hence, several terms in the equations group together to give appropriate spin weight. For this reason, operators $\eth$ and $\bar{\eth}$ appear frequently in the Newman--Penrose formalism and effectively they reduce the number of terms necessary to be written down explicitly. In order to define these operators in the present context, we denote
\begin{equation}
	a\szero \doteq \alpha\szero - \cconj{\beta}\szero \,,
\end{equation}
and $\cconj{a}\szero$ will denote its complex conjugate. In agreement with \eqref{eq:eth op} and \eqref{eq:eth op cc}, we define (with $\eta$ being a scalar, spin weight $s$ quantity)
\begin{equation}
	\eth \eta \doteq \delta \eta + s \cconj{a}\szero \eta \,, \qquad \cconj{\eth} \eta \doteq \cconj{\delta} \eta - s a\szero \eta \,.
\end{equation}
This kind of combinations appears in the Ricci identities, namely~\eqref{np:RI:Dlambda} and \eqref{np:RI:Dmu}. On the horizon (together with usage of the operators), they read
\begin{equation}
	D \lambda + \surfkappa \lambda \doteq \cconj{\eth} \pi + \pi^2 \,, \qquad D\mu + \surfkappa \mu \doteq \eth \pi + \left| \pi \right|^2 + \Psi_2 \,. \label{eq:lambda and mu eq}
\end{equation}
The spin weight of $\pi$ could have been found in table~\ref{tab:spin weights} to be $-1$.

We are not lucky enough to have as simple expression for $\lambda$ and $\mu$ on the horizon, as for, e.g., $\alpha$. However, we can solve these equations to get a formula for them. Although $D$ is a covariant derivative operator, it acts on scalar quantities in these cases; hence, it can be regarded as a partial derivative. Still, it would be complicated to project the gradient onto the $l^a$. We can get around using the notion of a quantity being on $\SS_0$, and $D$ is then only a partial derivative with respect to $v$. We can only get the solution on the horizon.

The solution of the first equation (the one for $\lambda$, the other one analogically) is
\begin{equation}
	\lambda \doteq \frac{1}{\surfkappa} \!\( \cconj{\eth} \pi\szero + \pi\szero{}^2 \) + \eu^{- \surfkappa v} C_\lambda \,.
\end{equation}
We have employed that $\pi$ is sufficiently given on $\SS_0$, and $C_\lambda$ is an integration constant which can be chosen to be such that for $v = 0$ it holds true that $\lambda = \lambda\szero$. This choice leads us to
\begin{equation}
	C_\lambda = \lambda\szero - \frac{1}{\surfkappa} \!\( \cconj{\eth} \pi\szero + \pi\szero{}^2 \)\! \,,
\end{equation}
which gives the solution in the form
\begin{equation}
	\lambda \doteq \lambda\szero \eu^{- \surfkappa v} + \frac{1}{\surfkappa} \!\( \cconj{\eth} \pi\szero + \pi\szero{}^2 \) \! \( 1 - \eu^{- \surfkappa v} \)\! \,,
\end{equation}
while for the spin coefficient $\mu$ we get
\begin{equation}
	\mu \doteq \mu\szero \eu^{- \surfkappa v} + \frac{1}{\surfkappa} \!\( \eth \pi\szero + \bigl|\pi\szero\bigr|^2 + \Psi_2\szero\) \! \( 1 - \eu^{- \surfkappa v} \)\! \,.
\end{equation}

Although it is not discussed in~\cite{Krishnan2012}, it is essential to come to realize that such a simple solution is, in fact, another consequence of the non-expanding horizon framework. In the equations to solve~\eqref{eq:lambda and mu eq}, there are $\eth$ operators (the complex conjugates are omitted from discussion as they are completely analogical) with $\delta$ operator in them which in turn have vector $m^a$ built in it. Therefore, when solving the equations, we should consider all terms with the $\eth$ operator as functions of $v$, for we a priory do not know what is the dependence of $\xi^I$ (recall~\eqref{eq:m on H}) on $v$. At this place, the frame equation~\eqref{eq:DxiI-deltaXI} comes in saying that $\xi^I \neq \xi^I(v)$, at least on the horizon.

We have already found that $\msp_0 \doteq 0$, cf.~\eqref{eq:msp0 0}, which can help us simplify, through the Maxwell equation~\eqref{eq:ap:maxwell D1}, also $\msp_1$. It turns out that (this is the reduced Maxwell equation)
\begin{equation}
	D \msp_1 \doteq 0 \,,
\end{equation}
which implies
\begin{equation}
	\msp_1 \doteq \msp_1\szero \,.
\end{equation}

The last scalar from the Maxwell spinor which is not known is $\msp_2$. We have Maxwell equation~\eqref{eq:ap:maxwell D2} for it. This equation can be solved analogically to~\eqref{eq:lambda and mu eq} to obtain
\begin{equation}
	\msp_2 \doteq \msp_2\szero \eu^{- \surfkappa v} + \frac{1}{\surfkappa} \!\( \cconj{\eth} \msp_1\szero + 2 \pi \msp_1\szero \)\! \( 1 - \eu^{- \surfkappa v} \)\! \,.
\end{equation}

Remaining unknowns which are to be determined are $\Psi_2$, $\Psi_3$ and $\Psi_4$. While the first two have the Ricci identities with derivatives tangential to the horizon, the third one has a non-tangential derivative, and would require more care, and is in fact a major difference from~\cite{Krishnan2012}, for charge brings the non-tangential term.

The aforementioned Ricci identities are~\eqref{np:RI:deltaalpha} and~\eqref{np:RI:deltalambda} of which the first one, on the horizon, is
\begin{equation}
	\delta \alpha - \cconj{\delta}\beta \doteq \cconj{a} \alpha - a \beta - \Psi_2 + \abs{\msp_1}^2 \,.
\end{equation}
Isolating the real and imaginary parts of this equation, we find
\begin{align}
	\Re  \Psi_2\szero &= {\abs{a}}^2  - \frac{1}{2} \!\( \delta a + \cconj{\delta} \cconj{a} \) + \abs{\msp_1\szero }^2 \,, \\
	\Im  \Psi_2\szero &= - \Im  \eth \pi\szero \,.
\end{align}
The second Ricci identity in our case reads:
\begin{equation}
	\delta \lambda - \cconj{\delta} \mu \doteq \pi \mu + \( \cconj{\alpha} - 3 \beta \)\! \lambda - \Psi_3 + \msp_2 \cconj{\msp_1} \,.
\end{equation}
The solution is, similarly as before,
\begin{align}
	\Psi_3 &\doteq \Bigl( \( \cconj{\eth} + \pi\szero \)\! \mu\szero - \( \eth + \cconj{\pi}\szero \)\! \lambda\szero + \msp_2\szero \sss \cconj{\msp_1}\szero \Bigr) \eu^{- \surfkappa v} \nonumber \\
	&\qquad{}+ \frac{1}{\surfkappa} \Bigl( \( \cconj{\eth} + \pi\szero \) \! \bigl( \eth \pi\szero + \bigl|\pi\szero\bigr|^2 + \Psi_2\szero \bigr) \nonumber \\
	&\qquad{}- \(\eth + \cconj{\pi}\szero \) \! \( \cconj{\eth}\pi\szero + \pi\szero{}^2 \) + \( \cconj{\eth}\msp_1\szero + 2 \pi\szero\msp_1\szero \)\! \sss \cconj{\msp_1}\szero \Bigr) \! \( 1 - \eu^{- \surfkappa v} \)\! \,.
\end{align}

We are forced to use different approach for the last Weyl scalar. It is more convenient to use, instead of the Ricci identity, the Bianchi identity, namely~\eqref{np:BI:DPsi4}, where is the non-tangential term $\Delta \ricsp_{20}$. The term can be decomposed using the Leibniz rule as
\begin{equation}
	\Delta \ricsp_{20} = \msp_2 \, \Delta \msp_0 + \msp_0 \, \Delta \msp_2 \doteq \msp_2 \, \Delta \msp_0 \,.
\end{equation}
Hence, we need the Maxwell equation~\eqref{eq:ap:maxwell Delta0} to compute the derivative. The equation is in our case simplified to:
\begin{equation}
	\Delta \msp_0 \doteq \delta \msp_1 \,. \label{eq:maxw phi 0 np}
\end{equation}
We can use, for we are discussing horizon, expansion:
\begin{equation}
	\msp_0 = r \msp_0^{(1)} + \mathcal{O}\!\(r^2\) 
\end{equation}
where the zeroth power is omitted, for it is zero according to aforementioned reasons. When we plug~\eqref{eq:maxw phi 0 np} into it, it yields
\begin{equation}
	\msp_0^{(1)} \doteq \eth \msp_1^{(0)} \,.
\end{equation}
Recall the zero spin weight of $\msp_1$ (cf.\ table~\ref{tab:spin weights}).
The equation to solve for $\Psi_4$, \eqref{np:BI:DPsi4}, is then\snote{Probably found the solution. Do really integration and derivative with $v$ commute with $\eth$ and $\delta$ operators?}
\begin{align}
	D \Psi_4 - \cconj{\delta} \Psi_3 + \msp_2 \, \eth \msp_1^{(0)} - \cconj{\delta} \!\( \msp_2 \cconj{\msp_1} \) &\doteq - 3 \lambda \Psi_2 + 2 \!\(\alpha\szero + 2 \pi\szero \)\! \Psi_3 - 2 \surfkappa \Psi_4  \nonumber \\
	&\qquad{}- 2 \lambda \left| \msp_1 \right|^2 + 2 \alpha\szero \msp_2 \cconj{\msp_1} \,.
\end{align}
We can write the solution as
\begin{equation}
	\Psi_4 \doteq \Psi_4\szero \eu^{-2 \surfkappa v} + \frac{A\szero}{\surfkappa} \eu^{- \surfkappa v} (1 - \eu^{- \surfkappa v}) + \frac{B\szero}{2 \surfkappa^2} \!\(1 - \eu^{- \surfkappa v}\)^2 \,. 
\end{equation}
The functions are
\begin{align}
	A &= \cconj{\eth}\Psi_3\szerot + \cconj{\eth}\msp_2\szerot \, \cconj{\msp_1}\szerot - \cconj{\eth} \cconj{\msp_1}\szerot \, \msp_2\szerot - \eth \msp_1\szerot \, \msp_2\szerot + \cconj{\msp_1}\szerot \msp_2\szerot \pi\szerot + 5 \Psi_3\szerot \pi\szerot - 2 \msp_1\szerot \cconj{\msp_1}\szerot \lambda\szerot - 3 \Psi_2\szerot \lambda\szerot \,, \\
	B &= \cconj{\eth}\cconj{\eth} \Psi_2\szerot + \cconj{\eth}\cconj{\eth}\msp_1\szerot \, \cconj{\msp_1}\szerot + 8\, \cconj{\eth} \Psi_2\szerot \, \pi\szerot - 2\, \cconj{\eth}\cconj{\msp_1}\szerot \, \msp_1\szerot \pi\szerot - 2\, \eth \msp_1\szerot \, \msp_1\szerot \pi\szerot + 12 \Psi_2\szerot \pi\szerot^2 \nonumber \\
	&\qquad{}- \cconj{\eth} \msp_1\szerot \!\( \cconj{\eth} \cconj{\msp_1}\szerot + \eth \msp_1\szerot - 3 \cconj{\msp_1}\szerot \pi\szerot\)\! \,.
\end{align}
All quantities in these functions are taken on the cut $\SS_0$ and the $\eth$ operators act on the nearest term only.

It is desirable to have well-defined solution in the limit of extremal horizon, $\surfkappa \to 0$. One can check that for the solutions we found, this limit is indeed well-defined and coincides with the solution which we would obtain by setting $\surfkappa = 0$ from the beginning.



\subsection{Physical meaning of the Weyl and Maxwell scalars}
We already know that some of the scalars vanish. In the case of the spin coefficients we have been able to state that it, e.g., means that $l^a$ is tangent to a geodesic. We are left to interpret the \qm{non-spin} scalars. Vanishing of $\Psi_0$ and $\Psi_1$ implies that there are no gravitational waves going through the horizon. In great analogy, the scalar $\msp_0$ is zero which means there is also no electromagnetic radiation on the way across the horizon, \cite{Chandrasekhar1983}.

\section{The extrinsic geometry of the horizon} 
The analysis of how the variables propagate off the horizon is much simpler than the intrinsic one. It is so, for we have some of the variables already zeroed, and also because the analysis is almost simultaneous for all of them.

Since we have the unknowns explored on the horizon, we simply want to compute how they are propagated from the horizon to its neighbourhood. This means we want to compute covariant derivative in direction of $n^a$, i.e.\ the $\Delta \text{-derivative}$. The only difference between the scalars is from where the equations for these derivatives come, and this splits them into three expectable sets: the spin coefficients, the Weyl scalars and the Maxwell scalars.

The most simple are the \textbf{spin coefficients}. For example, from~\eqref{np:RI:Deltamu} we get:
\begin{equation}
	\Delta \mu\hor = -\(\mu\hor\)^2 - \left|\lambda\hor\right|^2 - \big|\msp_2\hor\big|^2 \,.
\end{equation}
This equation gives us the first order of the expansion in $r$. Rewriting it together with all other equations we have:
\begingroup
\allowdisplaybreaks
\begin{subequations}
\begin{align}
	\kappa\rone &= 0 \,, \\
	\epsilon\rone + \cconj{\epsilon}\rone &= - 2\big|\pi\hor\big|^2 - 2 \Re \big(\Psi_2\hor\big) - 4\big|\msp_1\hor\big|^2 \,, \\
	\epsilon\rone - \cconj{\epsilon}\rone &= - \cconj{\pi}\hor \!\( \alpha\hor - \cconj{\beta}\hor \)\! + \pi\hor \!\(\cconj{\alpha}\hor - \beta\hor \) - \Psi_2\hor + \cconj{\Psi_2}\hor \,, \\
	\alpha\rone + \cconj{\beta}\rone &= \pi\rone = - \pi\hor\mu\hor - \cconj{\pi}\hor\lambda\hor - \Psi_3\hor - 2\msp_2\hor\cconj{\msp_1}\hor \,, \\
	\alpha\rone - \cconj{\beta}\rone &= - \mu\hor\! \(\alpha\hor - \cconj{\beta}\hor \) + \lambda\hor\! \(\cconj{\alpha}\hor - \beta\) - \Psi_3\hor - 2\msp_2\hor\cconj{\msp_1}\hor \,, \\
	\sigma\rone &= 0 \,, \\
	\mu\rone &= - \(\mu\hor\)^2 - \left|\lambda\hor\right|^2 - \big|\msp_2\hor\big|^2 \,, \\
	\rho\rone &= - \Psi_2\hor \,, \\
	\lambda\rone &= - 2\mu\hor\lambda\hor - \Psi_4\hor \,.
\end{align}
\end{subequations}
\endgroup
We could get the higher orders by subsequently substituting into the Ricci identities.

The \textbf{Weyl scalars} are obtainable from the Bianchi identities~\eqref{eq:bianchi id}. We immediately see that we are missing an equation for $\Psi_4$. Hence, we can not obtain $\Psi_4$ in terms of quantities on the horizon, and we need to introduce it at least on an arbitrary hypersurface of constant $v$ -- the most convenient choice is the hypersurface $\NN_0$. The equations for the first orders of the scalars are:
\begin{align}
	\Psi_0\rone &= 0 \,, \\
	\Psi_1\rone &= \eth \Psi_2\hor + 2 \cconj{\msp_1}\hor \, \eth \msp_1\hor \,, \\
	\Psi_2\rone &= \(\eth + \cconj{\pi}\hor\)\!\Psi_3\hor + 2 \cconj{\msp_1}\hor \!\(\eth + \cconj{\pi}\hor\)\! \msp_2\hor - 2 \cconj{\msp_2}\hor \, \cconj{\eth}\msp_1\hor \nonumber \\
	&\qquad{}- 3 \mu\hor\Psi_2\hor - 4 \mu\hor \big|\msp_1\hor\big|^2 \,, \\
	\Psi_3\rone &= \(\eth + 2 \pi\hor \)\!\Psi_4\hor - 4\mu\hor\Psi_3\hor - 2\cconj{\msp_1}\hor\msp_2\rone \nonumber \\
	&\qquad{}- 2 \cconj{\msp_2}\hor \!\(\eth - 2\pi\hor\)\! \msp_2\hor + 4\lambda\hor \msp_1\hor \cconj{\msp_2}\hor - 4\pi\hor \big| \msp_2\hor \big|^2 \,.
\end{align}
The only remaining undetermined quantities are the Maxwell scalars, which also determine the Ricci scalars. The equations for them are the Maxwell equations~\eqref{eq:ap:maxwell eq}. Similarly to the Weyl scalars, there is no equation for $\msp_2$ and; hence, it has to be given on $\NN_0$. The first order derivatives are:
\begin{align}
	\msp_0\rone &= \eth\msp_1\hor \,, \\
	\msp_1\rone &= \(\eth - \pi\hor\)\! \msp_2\hor - 2 \mu\hor \msp_1\hor \,.
\end{align}

\section{Initial value problem}
\begin{figure}
  \catcode`-=12
  \centering
  \begin{tikzpicture}
\newcommand\normalize[5][1cm]{%
  \draw[-latex] #2 -- ($#2!#1!#3$) node[#4] {#5};}


\coordinate (S0l) at (-1,2);
\coordinate (S0r) at (1.7,2.8);
\coordinate (Hl) at (-4,6.3);
\coordinate (Hr) at (-1.5,7);
\coordinate (Nl) at (2.5,6);
\coordinate (Nr) at (5,6);

\draw[name path=S0,red] (S0l) to [bend left=20] coordinate[pos=0.499] (m0d) coordinate[pos=0.5] (S0M) (S0r);

\draw[name path=horL] (S0l) to [bend right=20] (Hl);
\path[name path=horR] (S0r) to [bend right=20] (Hr);
\draw[name path=horT] (Hl) to [bend left=20] coordinate[pos=0.5] (horTM) (Hr);

\draw[name path=NL,blue] (S0l) to [bend left=20] coordinate[pos=0.6] (parMl) (Nl);
\draw[name path=NR,blue] (S0r) to [bend left=20] coordinate[pos=0.6] (parMr) (Nr);
\draw[name path=NT,blue] (Nl) to [bend left=20] coordinate[pos=0.5] (NTM) (Nr);

\draw[name path=partr, dashed,blue] (S0M) to [bend left=20] coordinate[pos=0.001] (n0d) coordinate[pos=0.6] (parM) coordinate[pos=0.601] (n1d) (NTM);

\path[name path=HorMid] (S0M) to [bend right=20] coordinate[pos=0.001] (l0d) (horTM);
\normalize[2cm]{(S0M)}{(l0d)}{above}{$l^a$};
\normalize{(S0M)}{(n0d)}{right}{$n^a$};
\normalize{(S0M)}{(m0d)}{anchor=south west}{$m^a$};

\normalize{(parM)}{(n1d)}{anchor=north west}{$n^a$};
\path[name path=HorMid] (parMl) to [bend left=20] coordinate[pos=0.499] (m1d) (parMr);
\normalize[0.5cm]{(parM)}{(m1d)}{anchor=north east}{$m^a$};
\coordinate (l0e) at ($(S0M)!2cm!(l0d)$);
\draw[-latex] (parM) -- ($ (l0e)+(parM)-(S0M) $) node[above] {$l^a$};

\path[name intersections={of = horR and NL}];
\coordinate (bod)  at (intersection-1);

\begin{scope}
\clip (bod) rectangle (S0r);
\draw[dashed] (S0r) to [bend right=20] (Hr);
\end{scope}

\begin{scope}
\clip (bod) rectangle (Hr);
\draw (S0r) to [bend right=20] (Hr);
\end{scope}

\node at (-2,6) {$\mathcal{H}: \kappa_{(l)}$};
\node[red] at (2,2) {$\mathcal{S}_0: \Psi_2, \Psi_3, \alpha, \pi, \mu, \lambda, \msp_1$};
\node[blue] at (4,4) {$\mathcal{N}_0: \Psi_4, \msp_2$};

\end{tikzpicture}
  \caption[The neighbourhood of the horizon and initial value surfaces.]{The neighbourhood\snote{comment non-presence of $\Lambda$} of the horizon and initial value surfaces. The scalars enlisted after the colons have to be introduced on the corresponding hypersurface.}
  \label{img:spacetime structure}
\end{figure}
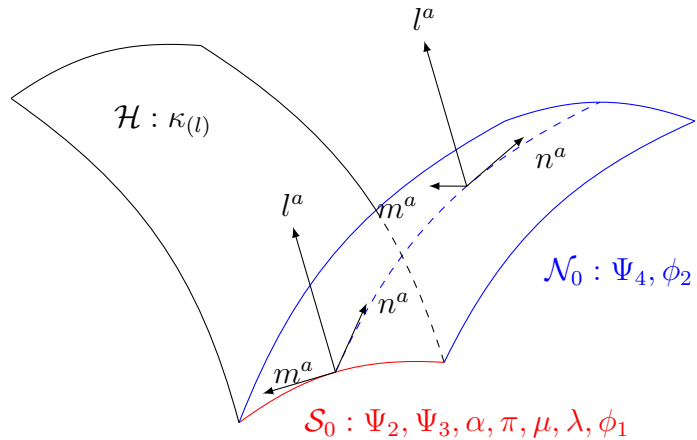

The analysis of the geometry of the space-time comes close to its end, and we shall slow down to recapitulate the results spread among many equations. The situation we have found is shown in figure~\ref{img:spacetime structure}. There are three important hypersurfaces: the horizon $\mathcal{H}$ with the chosen cut $\SS_0$, and the hypersurface of constant $v$ (in our case the one emerging from the chosen cut) $\NN_0$. On each of these hypersurfaces, there are some of the scalars to be given to have well-defined initial value problem. The scalars which are determined on $\SS_0$ are also immediately given on the entire horizon. The remaining scalars, namely: $\kappa, \tau, \sigma, \scrho, \gamma, \nu, \msp_0$ and $\Psi_1$, are zero on the horizon (some of them even everywhere).

\section{Motivation for the isolated horizons}
\label{sec:motivation for isol hor}
So far, we have been introducing a \qm{special} class of horizons. Now we would like to show, with the aid of~\cite{Ashtekar-LivRR}, that this class is, in fact, very general, and was created as a result of need for a unified framework for a large spectrum of different black hole treatments instead of having a very restrictive formalisms adapted only to individual problems, as well as inability to answer some questions in terms of then existing approaches. We also give briefly some other examples of usage of the isolated horizons than the one to follow in this work.

\subsection{Why should we introduce the isolated horizons}
Black holes are very interesting objects connecting general relativity, quantum mechanics, and statistical physics, which have ever had much attention of physicists. They have been studied from many points of view, among which are, e.g.\ numerical relativity, astrophysics, string theory. However, for a long time, the only main result of \emph{exact} general relativity for dynamical black holes was a theorem that the area of the black hole event horizon can not decrease for black holes whose matter satisfies the null energy condition. The theorem was proved by Hawking, and is similar to the second thermodynamic law, recall \eqref{eq:2nd law}. The first law \eqref{eq:1st law}, though, is only qualitative, the equation is not the fully dynamical as one would prefer. It shows an adverse attribute of the event horizons: It is often very difficult to physically interpret the results. This time, the peculiarity is that while the surface gravity $\kappa$ and the angular velocity $\Omega$ are defined at the horizon, the angular momentum $J$ and the mass $M$ are defined at infinity, and therefore it is perplexing to interpret any matter outside the horizon.

Another problem with approaches allowed by then existing frameworks were too strong limitations. For example, it was supposed to have time Killing vector field \emph{everywhere}, while it is sufficient to have it in a small neighbourhood of the black hole. The difference is that, although, it is reasonable to ensure that the black hole itself is \emph{isolated} when referring to the zeroth and first thermodynamic laws, it is not justified to restrict the whole space-time to be in an equilibrium.

\begin{figure}
	\begin{minipage}{\textwidth}
	  \catcode`-=12
	  \centering
	  \begin{tikzpicture}[scale = 1.5]
\coordinate (O) at (0,0);
\coordinate (TL) at (0,6);
\coordinate (TR) at (2,6);
\coordinate (R) at (4,4);
\coordinate (LM) at (0,4);

\draw (O) -- (TL);
\draw (O) -- (R);
\draw (R) -- coordinate[pos=0.5] (RM) (TR);
\draw[name path=singularity,decorate, decoration={zigzag}] (TL) -- coordinate[pos=0.3] (col) (TR);
\path (LM) -- coordinate[pos=0.6] (M) (TR);
\draw (M) -- (R);
\path[name path=surface] (O) to [out=80, in=269] (col);
\path (M) -- coordinate[pos=0.5] (delta) (TR);

\path[name intersections={of = singularity and surface}];
\coordinate (col2) at (intersection-1);
\begin{scope}
\clip (O) rectangle ([xshift=1] col2);
\draw (O) to [out=80, in=269] (col);
\end{scope}

\fill[blue,opacity=0.5]
(O) -- (TL) [snake=zigzag] -- (col2)
(col2) to [out=269, in=80] (O); 

\draw[red] (LM) -- (M);
\draw[red, thick] (M) -- (TR);

\node[anchor=south west] at (R) {$i^0$};
\node[anchor=south west] at (TR) {$i^+$};
\node[anchor=south west] at (RM) {$\mathcal{I}^+$};
\node[anchor=south east,red] at (delta) {$\Delta$};

\pgfmathsetmacro\fo{0.6cm}
\coordinate (S) at (2.7,3.7);
\coordinate (S1) at ([xshift=-\fo,yshift=\fo] S);
\coordinate (S2) at ([xshift=-0.7cm,yshift=0.1cm] S1);
\draw[decorate, decoration={snake}] (S) -- ([xshift=\fo,yshift=\fo] S);
\draw[-latex] ([xshift=\fo,yshift=\fo] S) -- ++(0.05,0.05);
\draw[decorate, decoration={snake}] (S1) -- ([xshift=\fo,yshift=\fo] S1);
\draw[-latex] ([xshift=\fo,yshift=\fo] S1) -- ++(0.05,0.05);
\draw[decorate, decoration={snake}] (S2) -- ([xshift=-\fo,yshift=\fo] S2);
\draw[-latex] ([xshift=-\fo,yshift=\fo] S2) -- ++(-0.05,0.05);

\node[matrix,red,anchor=east] at (LM) 
{
\node {event};\\
\node {horizon};\\
};
\node[matrix,blue,anchor=east] at ([yshift=-2cm] LM) 
{
\node {collapsing};\\
\node {matter};\\
};

\end{tikzpicture}
	  \caption[A diagram of a typical gravitational collapse.]{A diagram of a typical gravitational collapse. The thick part of the event horizon, denoted by $\Delta$, which is time distant from the collapsing matter, is \emph{isolated}.\footnote{The figure was redone according to~\cite[p. 8]{Ashtekar-LivRR}.}
	  }
	  \label{img:grav collapse}
	\end{minipage}
\end{figure}
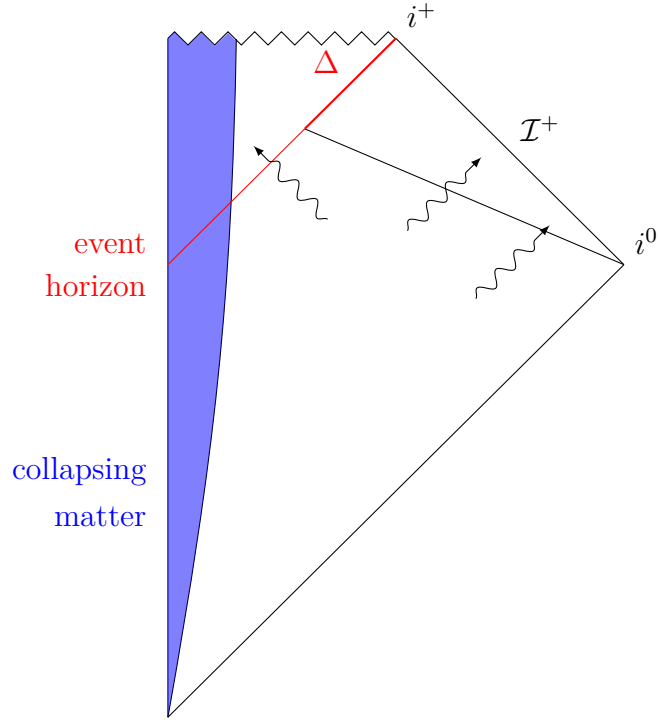

In figure~\ref{img:grav collapse}, we can find the situation of a collapsing matter where the event horizon is for late times isolated, while there can be e.g.\ radiation far outside the black hole.

\begin{figure}
	\begin{minipage}{\textwidth}
	  \centering
	  \begin{tikzpicture}[scale=1.8]
\draw (-1,0) -- (1,0);
\draw (-1,0) to [out=90,in=215] (0,2);
\draw (1,0) to [out=90,in=325] (0,2);

\fill[blue, opacity=0.5]
(-1,0) to [out=90,in=215] (0,2) to [out=325,in=90] (1,0) -- (-1,0);

\draw (0,1) to [out=130,in=270] (-0.5,2.5);
\draw (0,1) to [out=50,in=270] (0.5,2.5);

\begin{scope}
\clip (-2,1) rectangle (2,4);
\draw[thick, red] (-2,0) to [out=90,in=215] (0,3.5);
\draw[thick, red] (2,0) to [out=90,in=325] (0,3.5);
\end{scope}

\path[name path=tleva] (-2,0) to [out=90,in=215] (0,3.5);
\path[name path=tprava] (2,0) to [out=90,in=325] (0,3.5);

\path[name path=leva] (-0.5,2.5) -- (-0.5,4);
\path[name path=prava] (0.5,2.5) -- (0.5,4);

\path[name intersections={of = leva and tleva}];
\coordinate (L)  at (intersection-1);
\path[name intersections={of = prava and tprava}];
\coordinate (R)  at (intersection-1);

\draw[thick] (-0.5,2.5) -- ([yshift=-0.1cm] L);
\draw[thick] (0.5,2.5) -- ([yshift=-0.1cm] R);

\draw ([yshift=-0.1cm] L) -- ++(-0.1cm,0.15cm);
\draw[thick] ([xshift=-0.1cm,yshift=0.05cm] L) -- ++(0cm,1cm);
\draw ([yshift=-0.1cm] R) -- ++(0.1cm,0.15cm);
\draw[thick] ([xshift=0.1cm,yshift=0.05cm] R) -- ++(0cm,1cm);

\node at (0,2.8) {$\Delta_1$};
\node at (0,4) {$\Delta_2$};

\draw[latex-latex, shorten >=0.05, shorten <=0.05] (-0.5,2.6) -- ++(1,0);
\draw[latex-latex, shorten >=0.05, shorten <=0.05] (-0.6,3.8) -- ++(1.2,0);

\node at (1.6,2.6) {\textcolor{red}{$\delta M$}};
\node at (0,0.5) {$M$};
\end{tikzpicture}
	  \caption[Global nature of the event horizons.]{This figure demonstrates the global nature of the event horizons. While both parts $\Delta_1$ and $\Delta_2$ are isolated, only $\Delta_2$ is a part of the event horizon when a mass $\delta M$ falls into a collapsed star. However, if the mass shell did not fall into the black hole, $\Delta_1$ would be the event horizon.
	  \footnote{The figure was adopted from~\cite[p. 10]{Ashtekar-LivRR}.}
	  }
	  \label{img:event horizon problem}
	\end{minipage}
\end{figure}
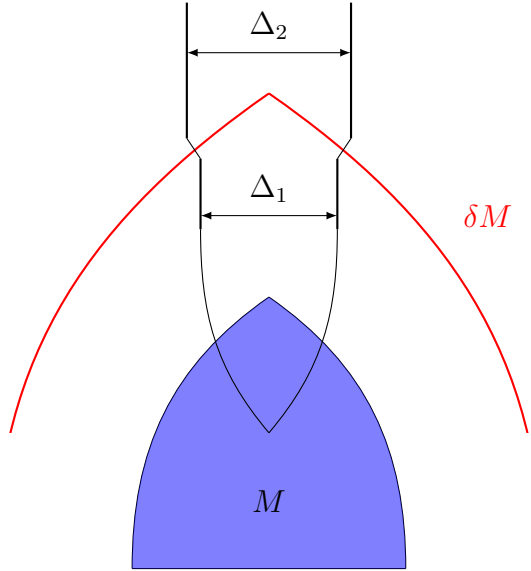

Another problem with event horizons is that they have \emph{global nature} (teleological nature). Graphical illustration of it can be found in figure~\ref{img:event horizon problem}.
The definition of the event horizon: \emph{it is the future boundary of the causal past of the future null infinity}, allows us to speak about the black hole only when the entire space-time was constructed.

Of course, event horizons were not the only framework, e.g.\ in numerical relativity apparent horizons were used, however, attempts to derive the black hole mechanics laws have been unsuccessful.

\subsection{Successes of the isolated horizon notation}
Now we would like to briefly discuss the contribution of the isolated horizons, which appeared as the resulted framework to overcome the aforementioned problems. The notion is richer than we are discussing since we are treating only black holes in equilibrium, for which the \emph{isolated horizons} were derived, while we leave so called \emph{dynamical horizons} for the interest of the reader  (they can also be found in~\cite{Ashtekar-LivRR}).

Considerations showed that the new paradigm should be quasi-local, as objects under our astrophysical observations are. They encapsulate all physical areas in which black holes appear, while being independent of the choice of the Cauchy slice (in contrast with the apparent horizons). Namely:
\begin{itemize}
	\item In \textbf{black hole mechanics}, it was possible to extend the zeroth and first laws to isolated horizons without any additional assumptions (except for the time independence) such as no radiation nearby. Moreover, the first law looks the same as before, cf.~\eqref{eq:1st law}, while all quantities are considered on the horizon, and the law ensures that the time evolution is Hamiltonian.
	\item \textbf{Quantum gravity} successfully used the isolated horizons to show proportionality of the area of the horizon to entropy, and is applicable also to physical objects which are distorted and are not extremal.
	\item This approach appears to be more robust in \textbf{numerical relativity}, while being invariant, and having no need for a priory considerations about the resulting horizon. 
	\item Also for \textbf{gravitational waves}, the isolated horizons can be very useful, as it led to notion of \emph{horizon multipole moments} which represent sources and; hence, are more relevant in equations of motion than older Hansen multipoles.
\end{itemize}
\chapter{Non-twisting null tetrad in Kerr--Newman space-time}
\label{chapter:kerr}
Now, our task is to find a non-twisting null tetrad in the Kerr--Newman space-time. We have already been discussing the twist of the horizon and its connection to the imaginary part of the spin coefficient $\scrho$, however, this was a characteristic of the generators of the horizon. This time, we want to construct a non-twisting congruence going off the horizon -- in the direction of $n^a$. We could go through a similar analysis as in the case of $l^a$, instead, we use a \qm{symmetry} of table~\ref{tab > tetrad formalism > spin coef > spin coef} of the spin coefficients. It turns out that the corresponding spin coefficient from which we take imaginary part for the twist in the case of $n^a$ is $\mu$.

\section{Kerr--Newman space-time}
We start with a recapitulation of the Kerr--Newman space-time. For more information about this topic refer to~\cite{Visser2007} or~\cite{Wald1984}. We use Bondi-like (ingoing null) coordinates $(r, v , \theta, \phi)$ in which the metric (the line element) is:
\begin{align}\label{eq:KN metric}
 \d s^2 &= \left( 1-\frac{2Mr-Q^2}{\abs{\rho}^2} \right)\d v^2 - 2\,\d v\,\d r + \frac{2a}{\abs{\rho}^2}\left( 2Mr-Q^2 \right)\sin^2\theta\,\d v \,\d \phi \, + \nonumber \\
 &\qquad{}+ 2a\sin^2\theta\,\d r\,\d \phi  - \abs{\rho}^2\,\d\theta^2 + \frac{\sin^2\theta}{\abs{\rho}^2}\left( \tilde{\Delta}a^2\sin^2\theta - \left( a^2+r^2 \right)^2 \right)\d\phi^2
\end{align}
where $M$ is mass, $Q$ is charge, $a$ is spin of the black hole (angular momentum per unit mass) and we have used functions defined as:
\begin{equation}
	\rho = r + \im a \cos \theta \,, \qquad \tilde{\Delta} = a^2 + r^2 - 2 M r + Q^2 \,.
\end{equation}
We identify the inner and outer horizons with hypersurfaces of constant $r$ with values:
\begin{equation}
	r_\pm = M \pm \sqrt{M - a^2 - Q^2}
\end{equation}
where $\tilde{\Delta} = 0$. Functions evaluated at the horizons shall be denoted with the appropriate sign ($+$~for the outer).

The space-time is stationary and axisymmetric which yields two Killing vectors: time-translational and azimuthal. These two symmetries are very important for we can conclude not to be bothered by coordinates $v$ and $\phi$ in our considerations as it will be done.

Another important characteristic of the Kerr--Newman space-time is that it has two degenerated null principal directions and, therefore, is type D space-time from the Petrov classification, also denoted as type $(2,2)$.

\subsection{Kinnersley tetrad}
Although we could start from scratch, we are fortunate to have a null tetrad (accomplishing the \gls{np} scalar products) in hand. The tetrad, named after William Morris Kinnersley, reads, \cite{Kinnersley1969a}: 
\begin{align}
l\_K  &=  \partial_v + \frac{\tilde{\Delta}}{2\!\left(a^2+r^2 \right)}\,\partial_r + \frac{a}{a^2+r^2}\,\partial_\phi \,, \nonumber\\
n\_K &= - \frac{a^2+r^2}{\abs{\rho}^2}\,\partial_r \,, \label{eq:kinnersley}\\
m\_K &= \frac{1}{\sqrt{2}\,\rho}\!\left( \im a\sin\theta\,\partial_v + \partial_\theta + \frac{\im}{\sin\theta}\,\partial_\phi \right)\! \,. \nonumber
\end{align}
It is no surprise that the last vector is complex conjugation of $m^a$. The tetrad is adapted to the principal null direction of the Weyl tensor, \cite{Stephani2009}.

In this tetrad we can compute the spin coefficients. We have the metric in hand and can use the Levi-Civita connection together with the tetrad vectors definition. As an example we outline the computation of the spin coefficient $\kappa$:
\begin{align}
	\kappa = m^a D l_a = m^a l^b \nabla_b l_a = m^a l^b (\partial_b l_a + \Gamma^c{}_{ba} l_c)
\end{align}
where $\Gamma$ are the usual Christoffel symbols. The covariant version of the tetrad is simply given by contraction with the metric. For the tedious computation itself we can with advantage use \texttt{Mathematica} software. In appendix~\ref{appendix:mathematica} a function \matcom{ComputeSpinCoeffsNP} is defined for this purpose. We get the spin coefficients as follows:
\begin{align}
\kappa &= \sigma = \nu = \lambda = 0 \,, & \gamma &=  - \frac{a \!\left( a+\im r\cos\theta \right)}{ \rho {\cconj{\rho}}^{2} } \,, \nonumber\\
 \scrho &= - \frac{\tilde{\Delta}}{2 \cconj{\rho} \!\left( a^2+r^2 \right)} \,,&  
 \tau & = -\frac{\im a \sin\theta}{\sqrt{2}\,\abs{\rho}^2} \,, \nonumber\\
 \epsilon &= - \frac{M a^2 + r Q^2-M r^2}{2\!\(a^2+r^2\)^2} \,, &
 \mu &= - \frac{a^2+r^2}{\rho {\cconj{\rho}}^{ 2}} \,, \nonumber\\
 \pi &= \alpha + \cconj{\beta} = \frac{\im a\sin\theta}{\sqrt{2}\, {\cconj{\rho}}^{2}} \,, &
 \alpha - \cconj{\beta} &= \frac{\im a - r\cos\theta}{\sqrt{2}\, {\cconj{\rho}}^{2} \sin\theta} \,,\nonumber \\
 \alpha &= \frac{2\im a- \rho \cos\theta}{2\sqrt{2}\, {\cconj{\rho}}^{2}\sin\theta} \,, &
 \beta &= \frac{\cot\theta}{2\sqrt{2}\, \rho} \,.
 \label{eq:spin coefficients 1}
\end{align}

The Kinnersley tetrad was not constructed by the process we have described and, therefore, it is very convenient to look at the spin coefficients, which have geometrical meaning, to find in what the two tetrad differs. However, the properties of the tetrad are encoded also in the Weyl and Ricci tensors. Hence, we find also them to be
\begin{equation}
	\Psi_2 = - \frac{M}{{\cconj{\rho}}^3} + \frac{Q^2}{{\cconj{\rho}}^3 \rho} \label{eq:kinnersley psi2}
\end{equation}
while all other are zero where can just use its definition through the Riemann tensor. Similarly the Ricci scalars are all zero except one:
\begin{equation}
	\ricsp_{11} = \frac{Q^2}{2\abs{\rho}^4} \,.
\end{equation}
And, therefore, the only non-zero Maxwell spinor is, recall~\eqref{eq:electrovac einstein}:
\begin{equation}
	\msp_1 = \frac{Q}{\sqrt{2}\, {\cconj{\rho}}^2} \,.
\end{equation}
In appendix~\ref{appendix:mathematica}, we define functions for computation of the curvature tensors (section~\ref{sec:curvature tensors}) and then the functions \matcom{ComputeWeylsNP} and \matcom{ComputeRicciNP} which can be used for this task.

Now, it is clear from the already done discussion that vector field $l^a$ is tangent to geodesics in this tetrad, because the spin coefficient $\kappa = 0$ and the congruence of the vectors is shear free, for $\sigma = 0$ while its expansion and twist vanish only on the horizon. This is because $\tilde{\Delta}_+ = 0$.

\subsection{Transformations of the spin coefficients}
If we compare the two tetrads, the constructed (which we do not know yet how does it look like) and the Kinnersley tetrad, using the values of the scalars, we immediately see they are different. The main difference is that the Kinnersley tetrad do not have purely real spin coefficient $\mu$ which we want as commented before while the aforementioned construction gave us a non-twisting congruence tangent to the vector field $n^a$ as a by-product, cf.~\eqref{eq:pi alpha beta real mu}.

To get a tetrad with properties of the constructed one, we have to transform the Kinnersley tetrad. Since we have to conserve the normalization of the tetrad (for we are looking for \gls{np} tetrad), the proper transformation is the Lorentz one. We have already been discussing, in section~\ref{sec: unknown parameters}\snote{check if any change in chapter 1 occurs}, how it can be divided into four different simpler transformations, it comes in handy now. We shall go through these four transformations, and show how the individual scalars transforms to be able to find such parameters for them to be applied on the Kinnersley tetrad to get the desirable one.

\subsubsection{Boost}
If we apply the boost transformation~\eqref{eq:boost}, the spin coefficients transform as:
\begin{align}
\begin{aligned}
	\kappa &\mapsto A^2 \kappa \,, \\
	\tau &\mapsto \tau \,, \\
	\sigma &\mapsto A \sigma \,, \\
	\scrho &\mapsto A \scrho \,, 
\end{aligned}
\qquad
\begin{aligned}
	\epsilon &\mapsto A \epsilon + \tfrac{1}{2} D A \,, \\
	\gamma &\mapsto A^{-1} \gamma + \tfrac{1}{2} A^{-2} \Delta A \,, \\
	\beta &\mapsto \beta + \tfrac{1}{2} A^{-1} \delta A \,, \\
	\alpha &\mapsto \alpha + \tfrac{1}{2} A^{-1} \cconj{\delta} A \,,
\end{aligned}
\qquad
\begin{aligned}
	\pi &\mapsto \pi \,, \\
	\nu &\mapsto A^{-2} \nu \,, \\
	\mu &\mapsto A^{-1} \mu \,, \\
	\lambda &\mapsto A^{-1} \lambda \,.
\end{aligned}\label{np:boost spins}
\end{align}
where we can with convenience use \texttt{Mathematica} package imposed in~\cite{Scholtz2012} for treating \gls{np} formalism where it is enough to define substitution rules for the basis vectors. It shall be used also for computation of all the other transformations of the spin coefficients and we display an example of how it can be done in appendix~\ref{app:example transf}.

The Weyl scalars and the electromagnetic scalars transform as
\begin{align}
\Psi_m &\mapsto A^{2-m} \Psi_m \,, & m&=0,1,2,3,4, \label{np:boost weyls}\\
\msp_m &\mapsto A^{1-m} \msp_m \,, & m&=0,1,2. \label{np:boost EM}
\end{align}
This can be seen from the definitions~\eqref{eq:weyl scalars} and~\eqref{eq:ricci scalars}, for $\Psi_{ABCD}$ and $\ricsp_{ABaA'B'}$ being general spinors without any weight.

\subsubsection{Spin}
The spin transformation translates the spin coefficients as follows:
\begin{align}
\begin{aligned}
	\kappa & \mapsto \eu^{\im \chi} \kappa \,, \\
	\tau &\mapsto \eu^{\im \chi}\tau \,, \\
	\sigma &\mapsto \eu^{2 \im \chi} \sigma \,, \\	
	\scrho &\mapsto \scrho \,,	 	
\end{aligned}
\qquad
\begin{aligned}
	\epsilon & \mapsto \epsilon + \tfrac{\im}{2} D\chi \,, \\
	\gamma & \mapsto \gamma + \tfrac{\im}{2} \Delta \chi \,, \\
	\beta & \mapsto \eu^{\im \chi} \!\left( \beta + \tfrac{\im}{2} \delta\chi \right)\! \,, \\
	\alpha &\mapsto \eu^{-\im \chi}\!\left( \alpha + \tfrac{\im}{2} \cconj{\delta}\chi \right)\! \,,	
\end{aligned}
\qquad
\begin{aligned}
	\pi &\mapsto \eu^{-\im \chi}\pi \,, \\
	\nu &\mapsto \eu^{-\im \chi}\nu \,, \\
	\mu &\mapsto \mu \,, \\
	\lambda &\mapsto \eu^{-2 \im \chi} \lambda \,.
\end{aligned}
\label{np:spin spins}
\end{align}
while the Weyl scalars and the Maxwell scalars are after the transformation:
\begin{align}
\Psi_m &\mapsto \eu^{(2-m)\im \chi} \Psi_m \,, & m&=0,1,2,3,4, \label{np:spin weyls}\\
\msp_m &\mapsto \eu^{(1-m)\im \chi} \msp_m \,, & m&=0,1,2. \label{np:spin EM}
\end{align}
with the same reasoning as in the case of the boost.

\subsubsection{Rotation about $\bm{l^a}$}
Under the rotation about $l^a$ with a parameter $c$, the principle of transformation of the Weyl and Ricci scalars is all the same, although, the formulas look more intricately:
\begin{align}
&\begin{aligned}
 \Psi_0 &\mapsto \Psi_0 \,, \\
 \Psi_1 &\mapsto \Psi_1 + c \Psi_0 \,, \\
 \Psi_2 &\mapsto \Psi_2 + 2 c \Psi_1 + c^2 \Psi_0 \,, \\
 \Psi_3 &\mapsto \Psi_3 + 3 c \Psi_2 + 3 c^2 \Psi_1 + c^3 \Psi_0 \,, \\
 \Psi_4 &\mapsto \Psi_4 + 4 c \Psi_3 + 6 c^2 \Psi_2 + 4 c^3 \Psi_1 + c^4 \Psi_0 \,, 
\end{aligned}\label{np:rotation l weyls}
 \\[1.5ex]
&\begin{aligned}
 \msp_0 &\mapsto \msp_0 \,, \\
 \msp_1 &\mapsto \msp_1 + c \msp_0 \,, \\
 \msp_2 &\mapsto \msp_2 + c \msp_1 + c^2 \msp_0 \,.
\end{aligned}\label{np:rotation l EM}
\end{align}
The spin coefficients undergo the change:
\begin{align}
\begin{aligned}
 {\kappa} &\mapsto \kappa \,, \\
 {\tau} &\mapsto \tau + c \sigma + \cconj{c} \scrho + \kappa  \abs{c}^2 \,, \\
 {\sigma} &\mapsto \sigma + \kappa \cconj{c} \,, \\
 {\scrho} &\mapsto \scrho + \kappa c \,, \\[1.5ex]
 {\epsilon} &\mapsto \epsilon + c \kappa \,, \\
 {\gamma} &\mapsto \gamma + c(\beta+\tau)+\alpha \cconj{c} + \sigma c^2 + \left( \epsilon+\scrho \right) \! \abs{c}^2 + \kappa c^2 \cconj{c} \,, \\
 {\beta} &\mapsto \beta + c \sigma + \epsilon \cconj{c} + \kappa  \abs{c}^2 \,, \\
 {\alpha} &\mapsto \alpha + c\!\left( \epsilon+\scrho \right)\!+\kappa c^2 \,, \\[1.5ex]
 {\pi} &\mapsto \pi + 2 c \epsilon + c^2 \kappa + Dc \,, \\
 {\nu} &\mapsto \nu + c( 2\gamma+\mu ) + \cconj{c} \lambda + c^2( 2 \beta+\tau ) + c^3 \sigma + \abs{c}^2 ( \pi + 2 \alpha )  \\
 &\qquad{}+ c^2 \cconj{c}\left( 2\epsilon+\scrho \right) + c^3 \cconj{c} \kappa + \abs{c}^2 \sss  Dc + \Delta c + c \, \delta c + \cconj{c} \, \cconj{\delta} c \,, \\
 {\mu} &\mapsto \mu + 2 c \beta + \cconj{c} \pi + c^2 \sigma + 2  \abs{c}^2 \epsilon + c^2 \cconj{c} \kappa + \cconj{c}\, Dc + \delta c \,, \\
 {\lambda} &\mapsto \lambda + c( \pi + 2 \alpha )+c^2( \scrho + 2 \epsilon ) + \kappa c^3 + c \, Dc +\cconj{\delta} c \,.
\end{aligned}\label{np:rotation l spins}
\end{align}

\subsubsection{Rotation about $\bm{n^a}$}
To distinguish the rotation about $n^a$ from the previous one more visibly we choose the parameter to be $d$, then the transformation of the spin coefficients is
\begin{equation}
\begin{aligned}
 \kappa & \mapsto \kappa + d\!\left( 2\epsilon+\scrho \right) + \cconj{d} \sigma + d^2 ( \pi+2\alpha ) + d^3 \lambda + \abs{d}^2 ( \tau+2\beta ) \\
 &\qquad{}+ d^2 \cconj{d} ( 2\gamma+\mu ) + d^3 \cconj{d} \nu - \abs{d}^2 \sss \Delta d - D d - d \, \cconj{\delta}d - \cconj{d} \, \delta d \,, \\
 \tau &\mapsto \tau + 2 d \gamma + d^2 \nu-\Delta d \,, \\
 \sigma &\mapsto \sigma + d ( \tau+2\beta )+d^2 ( \mu+2\gamma )+d^3 \nu - d \, \Delta d-\delta d \,, \\
 \scrho &\mapsto \scrho + 2 d \alpha + \cconj{d} \tau + d^2 \lambda + 2 \abs{d}^2 \gamma + d^2 \cconj{d} \nu - \cconj{d} \, \Delta d - \cconj{\delta}d \,, \\[1.5ex]
 \epsilon &\mapsto \epsilon + d ( \alpha+\pi )+\beta \cconj{d} + \lambda d^2 + \left( \mu+\gamma \right) \abs{d}^2+\nu d^2 \cconj{d} \,, \\
 \gamma &\mapsto \gamma + d \nu \,, \\
\beta &\mapsto \beta + d ( \gamma+\mu )+d^2 \nu \,, \\
\alpha &\mapsto \alpha + d \lambda + \cconj{d} \epsilon +  \abs{d}^2 \nu \,, \\[1.5ex]
\pi &\mapsto \pi + d \lambda + \cconj{d} \mu +  \abs{d}^2 \nu \,, \\
\nu & \mapsto \nu \,, \\
\mu &\mapsto \mu + d \nu \,, \\
\lambda &\mapsto \lambda + \nu \cconj{d} \,.
\end{aligned}
\label{np:rotation n spins}
\end{equation}
For the Weyl and electromagnetic scalars we get:
\begin{align}
&\begin{aligned}
 \Psi_0 & \mapsto \Psi_0 + 4 d \Psi_1 + 6 d^2 \Psi_2 + 4 d^3 \Psi_3 + d^4 \Psi_4 \,, \\
 \Psi_1 &\mapsto \Psi_1 + 3 d \Psi_2 + 3 d^2 \Psi_3 + d^3 \Psi_4 \,, \\
 \Psi_2 &\mapsto \Psi_2 + 2 d \Psi_3 + d^2 \Psi_4 \,, \\
 \Psi_3 &\mapsto \Psi_3 + d \Psi_4 \,, \\
 \Psi_4 &\mapsto \Psi_4 \,,
\end{aligned}\label{np:rotation n Weyls}
 \\[1.5ex]
&\begin{aligned}
 \msp_0 &\mapsto \msp_0 + 2 d \msp_1 + d^2 \msp_2 \,, \\
 \msp_1 &\mapsto \msp_1 + d \msp_2 \,, \\
 \msp_2 &\mapsto \msp_2 \,. 
\end{aligned}\label{np:rotation n EM}
\end{align}

\section{Expansion in neighbourhood of the horizon}
\label{sec:expansion}
Although it might seem that we are ready just to perform appropriate transformations on the Kinnersley tetrad, there is a major drawback of this approach, to be discussed in section~\ref{sec:diff eq for lorentz}. Instead of going directly to the transformations we use perturbations at this place.

In \cite{Krishnan2012}, there has been found a perturbative solution for the isolated horizons. However, it is not clear then, how to choose the initial values of the scalars to obtain Kerr--Newman space-time. Our approach starts with the desired space-time (which is given by the Kinnersley tetrad) and then makes the tetrad to have identical properties as the one found from the construction instead. The perturbation lies in the fact that, though, it is difficult to find the transformation in general, as mentioned in previous paragraph, we are able to do so on the horizon itself.

The difference, of course, is that many quantities are zero on the horizon. However, the most fruitful simplification, for us at this moment, is the much plainer look of the metric functions.

We start with renormalization of the Kinnersley tetrad. The vector $m\_K$ is left while $l\_K$ is multiplied by $\left( r^2 + a^2 \right)/\abs{\rho}^2$ and $n\_K$ is divided by the same factor to conserve its scalar product with $l\_K$. Let us refer to this \qm{new} tetrad with small $\text{k}$ in subscript instead of the $\text{K}$.\footnote{We could have have omit this step, however, it bring us to a more nicely normalized $n\_k$.} Simultaneously we relabel the coordinate $\phi$ to $\tilde{\phi}$.\footnote{We proceed with coordinate transformation and do not want to use tildes (or other decorations) after it.} On the horizon, where $r = r_+$, the function $\Delta$ vanishes and vector $l\_k$ becomes
\begin{equation}
	l\_k \doteq \frac{r_+^2 + a^2}{\abs{\rho_+}^2}\, \pd_v +  \frac{a}{\abs{\rho_+}^2}\, \pd_{\tilde{\phi}} \,.
\end{equation}

Recall how the vector $l^a$ of the previously constructed tetrad looks like on the horizon -- \eqref{eq:l on H}. It has only $v$-component for what we would like to make the $\partial_\phi$ part disappear. We can do it more conveniently a by coordinate transformation than by a Lorentz transformation because the vector $n\_k$ is not affected by the coordinate transformation. We transform
\begin{equation}
	\phi = \tilde{\phi} - \frac{a v}{a^2+r_+^2} \,.
\end{equation}
The resulting tetrad on the horizon reads
\begin{align}
	l\_k &\doteq \frac{a^2+r_+^2}{r_+^2 +a^2 \cos^2\theta} \, \pd_v \,, \nonumber \\
	n\_k &\doteq -  \pd_r \,, \label{eq:kinnersley before lorentz} \\
	m\_k &\doteq \frac{1}{\sqrt{2}\left( r_++\im a \cos\theta \right)} \!\left( 
	\im a \sin\theta \, \pd_v + \pd_\theta + 
	\frac{\im{} \!\(r_+^2 + a^2 \cos^2\theta\)}{\left(a^2 + r_+^2\right) \sin\theta} \, \pd_{\phi} \right)\! \,. \nonumber
\end{align}

We have the vector $l\_k$ almost in the form of~\eqref{eq:l on H}, except the $v$-component is not properly normalized. To accomplish that, we perform a boost of the form~\eqref{eq:boost} with the parameter
\begin{equation}
	 A = \frac{r_+^2 + a^2 \cos^2\theta}{a^2 + r_+^2} \,.
\end{equation}
The tetrad is changed to:
\begin{align}
	l\_B &\doteq \pd_v \,, \nonumber \\
	n\_B &\doteq -\frac{a^2 + r_+^2}{r_+^2 + a^2 \cos^2\theta} \, \pd_r \,, \\
	m\_B &\doteq \frac{1}{\sqrt{2}\left( r_++\im a \cos\theta \right)} \!\left( 
		\im a \sin\theta \, \pd_v + \pd_\theta + 
		\frac{\im{} \!\(r_+^2 + a^2 \cos^2\theta\)}{\left(a^2 + r_+^2\right) \sin\theta} \, \pd_{\phi} \right)\! \,. \nonumber
\end{align}

As we have shown, cf.~\eqref{eq:m on H}, vector fields $m^a$ and $\cconj{m}^a$ do not have $v$-com\-po\-nent. To remove it we want to perform another transformation, however, the transformation has to conserve vector $l^a$ as it has the desired form on the horizon. There are two such transformations: spin (which is unable to zero $v$-component) and the null rotation about $l^a$ \eqref{eq:rotation about l}. It is not difficult to find that the parameter $c$ has to be
\begin{equation}
	 c = \frac{\im a \sin\theta}{\sqrt{2}\left( r_+ - \im a \cos\theta \right)}
\end{equation}
in order to transform $m^a$ as desired. The entire basis, after the rotation, is
\begin{align}
	l\_R & \doteq \pd_v \,, \nonumber \\
	n\_R &\doteq -\frac{a^2 \sin^2\theta}{2 \! \left(r_+^2 + a^2 \cos^2\theta
	\right)} \, \pd_v - \frac{a^2 + r_+^2}{r_+ + a^2 \cos\theta} \, \pd_r - 
	\frac{a}{a^2 + r_+^2} \, \pd_\phi \,,\\
	m\_R &\doteq \frac{1}{\sqrt{2}\left( r_+ + \im a \cos\theta \right)} \, \pd_\theta + \frac{\im\left(r_+ - \im a \cos\theta\right)}{\sqrt{2}\left( a^2 + r_+^2 \right)\sin\theta} \, \pd_\phi \,. \nonumber
\end{align}
and has the property that $m^a$ is Lie dragged along $l^a$ which was required.

Nevertheless, we want to have the tetrad at least in some neighbourhood of the horizon. To get it, we can use a power series of $r$ as it have been already done during the construction the tetrad. This we manifestly expand around the value of $r$ on the horizon, namely
\begin{align}
 l^a &= l^a\zero + (r-r_+) l^a\one + \mathcal{O}{\left( r-r_+\right)^2} \,, &
 l^a\one=l^a\one (\theta,\phi) \,,
\end{align}
and similarly for other vectors. Recall the existence of the two Killing vectors which give us the simplified coordinate independence.

\subsection{The first order}
From the metric, we can calculate the 
connection and covariant derivatives
\begin{align}
 \Delta l^a ,\; \Delta n^a ,\; \Delta m^a \quad \text{on the horizon},
\end{align}
with yet undetermined functions $l\one^a, n\one^a, m\one^a$, and 
require that these covariant derivatives vanish on the horizon.

The condition of the parallel transport (on the horizon) gives us system of linear algebraic equations which can be solved to obtain:\footnote{We have rewritten the charge $Q$ in terms of $r_+$, $a$ and $M$ in order to simplify the equations.}
\begin{subequations}
\begin{align}
	l\one &= \frac{a^2  (M-r_+) \sin^2\theta}{2 \!\left(a^2 + r_+^2\right)^2} \, \pd_v + \frac{r_+ - M}{a^2 + r_+^2} \, \pd_r \nonumber \\
	&\qquad{}- \frac{a^2 \sin (2 \theta )}{2\!\left(a^2+r_+^2\right)\!\(r_+^2 + a^2 \cos^2\theta\) } \, \pd_\theta - \frac{a r_+}{\left(a^2 + r_+^2\right)^2} \, \pd_\phi \,,
\end{align}
\begin{align}
	n\one &= 
		\left(\frac{a^2 \sin ^2 \theta \(-a^4 (M - 33 r_+) + 32 r_+^5 + 4 a^2 r_+^2 (-M + 17 r_+)\)}{32 \left(a^2+r_+^2\right)^2 \left(a^2 \cos ^2\theta + r_+^2\right)^2} \right.
		\nonumber \\
		&\qquad{}\left. +\frac{a^2 \sin ^2 \theta \(a^2 (M-r_+) \left(a^2 \cos (4 \theta )+4 r_+^2 \cos (2 \theta )\right)\)}{32 \left(a^2+r_+^2\right)^2 \left(a^2 \cos ^2\theta + r_+^2\right)^2} \right)\!
			\, \pd_v \nonumber \\[1.5ex]
		&\qquad{}-\frac{a^2 \sin^2 \theta \left(a^2 \cos (2 \theta ) (M-r_+)+a^2 (M-5 r_+)+2 r_+^2 (M-3 r_+)\right)}{2 \left(a^2+r_+^2\right) \left(a^2 \cos^2 \theta+r_+^2\right)^2}	
			\, \pd_r \nonumber \\[1.5ex]
		&\qquad{}+\frac{a^2 \sin \theta \cos \theta}{\left(a^2+r_+^2\right) \left(a^2 \cos^2 \theta+r_+^2\right)}	
			\, \pd_\theta
		+\frac{a r_+ \left(a^2 \cos (2 \theta )+3 a^2+4 r_+^2\right)}{2 \left(a^2+r_+^2\right)^2 \left(a^2 \cos^2 \theta+r_+^2\right)}	
			\, \pd_\phi \,,
\end{align}
%
%
\begin{align}
	m\one &=
		\left(\frac{a^2 \sin \theta \cos \theta \left( 8 r_+^3 + a^2 (M + 7 r_+) \right)}{8 \sqrt{2} \left(a^2 + r_+^2\right) \left(a^2 \cos^2 \theta+r_+^2\right)} 
		-\frac{2 \im a^3 \sin \theta \cos (2 \theta ) \left(a^2 + M r_+\right) }{8 \sqrt{2} \left(a^2 + r_+^2\right) \left(a^2 \cos^2 \theta+r_+^2\right)}
		\right. \nonumber \\
		&\qquad{}\left. 
		-\frac{\im a^3 \sin \theta \left(6 a^2 - \im a \cos (3 \theta ) (M-r_+) - 2 M r_+ + 8 r_+^2 \right)}{8 \sqrt{2} \left(a^2 + r_+^2\right) \left(a^2 \cos^2 \theta+r_+^2\right)}
		\right)\!
			\, \pd_v \nonumber \\[1.5ex]
		&\qquad{}+\frac{a \sin \theta  \bigl(a \cos \theta (M-r_+) - \im \left(a^2 + r_+ (2r_+ - M)\right)\bigr)}{\sqrt{2} \left(a^2 + r_+^2\right) \left(a^2 \cos^2 \theta+r_+^2\right)}		
			\, \pd_r \nonumber \\[1.5ex]
		&\qquad{}+\frac{
		a^2 + 2r_+^2 + 2\im a r_+ \cos\theta - a^2 \cos(2\theta)
		}{2 \sqrt{2} \left(a^2+r_+^2\right) (a \cos\theta - \im r_+)^2}	
			\, \pd_\theta 
		-\frac{r_+ (a \cos \theta +\im
		   r_+)}{\sqrt{2} \sin \theta \left(a^2 + r_+^2\right)^2}	
			\, \pd_\phi \,.
\end{align}
%
\end{subequations}

Since the spin coefficients are derivatives of the tetrad and we have the tetrad up to the first order, we can compute the spin coefficient on the horizon (the zeroth order). The spin coefficients should (on the horizon) meet the conditions, which can serve as a check that the vectors are parallelly transported as they should be.
The spin coefficients are:
\begin{equation}
 \begin{aligned}
 	\scrho\hor &= \sigma\hor = \tau\hor = \kappa\hor = \gamma\hor = \nu\hor = 0 \,, \\
	\epsilon\hor &= \frac{r_+ - M}{2\!\left( a^2+r_+^2 \right)} \,, \\
	\pi\hor &= \frac{a \sin \theta \bigl(a \cos \theta (M-r_+)-\im \left(a^2+r_+ (2
	r_+-M)\right)\bigr)}{\sqrt{2} \left(a^2+r_+^2\right) (a \cos \theta +\im
	r_+)^2} \,, \\
	\alpha\hor-\cconj{\beta}\hor &=\frac{ r_+ \cos \theta -\im a}{\sqrt{2}\, (a \cos \theta +\im r_+)^2\sin\theta} \,,\\
	\lambda\hor &=\frac{a^2 \sin ^2\theta \bigl(-a \cos \theta (M-r_+)+\im \left(2 a^2+r_+
	(3 r_+-M)\right)\bigr)}{2 \left(a^2+r_+^2\right) (a \cos \theta +\im
	r_+)^3} \,,\\
	\mu\hor &= \frac{a^2 \cos (2 \theta ) (M-r_+)-a^2 (M+3 r_+)-4 r_+^3}{4   \left(a^2+r_+^2\right) \left(a^2 \cos^2 \theta+r_+^2\right)} \,.
 \end{aligned}
\end{equation}
As we have desired, the spin coefficient $\mu$ is, at least on the horizon, manifestly real (no twist is present). Also note that the spin coefficient $\epsilon$ gives through~\eqref{eq:def of surface gravity} the right surface gravity, compare \cite{Poisson2004}. 

To have the complete set of the variables we want to compute also the zeroth order values of the Weyl scalars and of the Maxwell scalars. For we have the transformation formulas~\eqref{np:rotation l weyls} and~\eqref{np:rotation l EM} where no derivatives are present, we can compute the zeroth order simply by multiplication without actual need of the first order of the tetrad. The Weyl scalars turn out to be
\begin{equation}
\begin{aligned}
	\Psi_0\hor &= \Psi_1\hor = 0 \,, \\
	\Psi_2\hor &= -\frac{a^2+\im a M \cos \theta +r_+ (r_+-M)}{(r_+-\im a \cos \theta )^3 (r_+ + \im a \cos \theta )} \,, \\
	\Psi_3\hor &=-\frac{3 \im a \sin \theta  \left(a^2+\im a M \cos \theta +r_+ (r_+-M)\right)}{\sqrt{2}\, (r_+-\im a \cos \theta )^4 (r_+ + \im a \cos \theta)} \,, \\
	\Psi_4\hor &= \frac{3 a^2 \sin ^2\theta  \left(a^2+\im a M \cos \theta + r_+ (r_+-M)\right)}{(r_+-\im a \cos \theta )^5 (r_++\im a \cos \theta )} \,.
\end{aligned}
\end{equation}
Before, in the Kinnersley tetrad, we had only one non-zero Weyl scalar: $\Psi_2$, cf.~\eqref{eq:kinnersley psi2}. This was left unchanged, however, as we see, other component arose. Finally, the components of the electromagnetic field are
\begin{equation}
\begin{aligned}
 \msp_0\hor &= 0 \,, \\
 \msp_1\hor &= \frac{Q}{\sqrt{2}\, (r_+-\im a \cos \theta )^2} \,, \\
 \msp_2\hor &=\frac{\im a Q \sin \theta }{(r_+-\im a \cos \theta)^3} \,.
\end{aligned}
\end{equation}


\subsection{Higher orders}
Of course we do not have to stop at the first order. The routine is almost all the same. We solve higher order of the linear algebraic system of equations, and substitute for the lower orders which we have to have computed before. Together with the 3rd order, derivatives of the lower ones come into play. Then we can use the resulting order of the tetrad to compute a new order of the spin coefficients. The other scalars require new treatment, for we do not know them off the horizon and the transformations are giving us nothing new for the higher orders. We have to use the metric and compute the curvature. We start with the Christoffel symbols, which give us the Riemann tensor. Then we compute, using contractions, its traceless part -- the Weyl tensor, and project it onto the tetrad. For the Maxwell scalars, we can similarly use the Ricci tensor, which we have already computed during the computation of the Weyl scalars. It is projected onto the tetrad, and, by the means of~\eqref{eq:electrovac einstein}, we get the Maxwell scalars.

Nevertheless, the higher the order the more complicated the formulas are. In fact, from the second order of the tetrad the results are pretty much useful only in a computer. Nevertheless, the second order is needed to compute the first order of the spin coefficients and, therefore, to know how they propagate off the horizon.

In appendix~\ref{appendix:mathematica}, we introduce a \texttt{Mathematica} source code for computation of, theoretically, any order of the expansion. The limitation is the computational time and, probably  more importantly, memory consumption. We have computed the tetrad up to the third order while we are displaying, in appendix~\ref{app:series}, only the second order of it (only the tetrad, the scalars are omitted), just to demonstrate the \qm{jump} in the intricacy.

\section{Discussion of further difficulties}
\label{sec:diff eq for lorentz}
We would like to discuss the main difficulties of finding the tetrad non-per\-tur\-ba\-ti\-ve\-ly in the neighbourhood of the horizon and why we have to employ other methods together with the already used. In the perturbative process we have used the simplicity of the tetrad on the horizon in our favour. We have been able to transform the vector $l^a$ into the very restricted form while, as discussed, there was almost no freedom left for the vector $m^a$.

In the general case, the tetrad looks quite difficult, recall~\eqref{eq:tetrad gen coor l}, \eqref{eq:tetrad gen coor n} and \eqref{eq:tetrad gen coor m}, while the equations for the metric functions \eqref{eq:eqs for metric functions} are not useful at first place, for they depend on the tetrad itself.

Therefore, it is a better idea not to treat the tetrad and look at the transformations of the spin coefficients instead. They encode all the important properties -- the parallel transport and congruences being geodesics. However, their transformation rules are complicated and unlike in the case of the tetrad itself, we are not guided to what transformation we should do.

For this reason, we turned our attention to description of the congruence of the geodesics by the Carter constants~\cite{Carter1968}. This simplifies equations for the transformations, however, there remains a very troublesome complication. It is clear that we have to solve differential equations when trying to find the parameters of transformations from how spin coefficients change. The problem is that, although we have been able to determine the three integrals of motion with only the Carter one left, the formula for non-twisting $n^a$ exhibits functions which can not be integrated to elementary functions. This formulas, elliptic functions with very complicated arguments (pages long in some cases), appear in the differential equations to solve and complicate the solution too much.

It turned out that this type of terms appeared in any type of approach of ours. For example, when we want to start with the integral curves for the congruence of $n^a$ instead of the transformations, these terms appear as early, in fact, in equations already given by Carter, \cite{Carter1968}:
\begin{subequations}
\begin{align}
	v &= \alpha^0 + \int \frac{a}{\sqrt{\Theta}} \!\left( L + a E \sin^2\theta \right)\! \,\d \theta + \int \frac{a^2+r^2}{\tilde{\Delta}} \!\left( 1 - \frac{\PP}{\sqrt{\RR}} \right)\! \,\d r \,, \\
	\phi &= \alpha^1 + \int \frac{a}{\tilde{\Delta}} \!\left( 1- \frac{\PP}{\sqrt{\RR}} \right)\! \,\d r + \int \frac{a E + L \sin^{-2}\theta}{\sqrt{\Theta}}\,\d \theta \,, \\
	\mu \lambda &= - \alpha^2 - \mu \int \frac{r^2 \,\d r}{\sqrt{\RR}} - \mu \int \frac{a^2 \cos^2\theta\,\d \theta}{\sqrt{\Theta}} \,, 
	%
\end{align}
and most importantly
\begin{align}
	\frac{1}{2}\int \frac{\d \theta}{\sqrt{\Theta}} &= \alpha^3 + \frac{1}{2}\int \frac{\d r}{\sqrt{\RR}} \,.
\end{align}
\end{subequations}
The functions $\sqrt{\Theta}$ and $\sqrt{\RR}$ are these which are complicated, and this resulted in us unable to find explicit formula for dependence of $\theta$ on $r$ along a geodesic, and therefore also to solve the rest of equations.

Nevertheless, we have been able to find the non-perturbative tetrad at least formally (it is not written explicitly) with the aid of other constructions such as Killing--Yano tensor. 
The results
together with the procedure and commentaries are to be presented in the upcoming paper.

%

\chapter*{Conclusion}
\addcontentsline{toc}{chapter}{Conclusion}

In this thesis, we presented a review of basic results on black hole horizons and motivated the formalism of isolated horizons. We gave detailed explanation of the Newman--Penrose formalism and related two-spinor formalism. With this tool, we were able to define and analyse non-expanding and isolated horizons, following the works by Ashtekar et al., which are referenced in the text. Then we explained the construction of Krishnan \cite{Krishnan2012} which was a main pillar of our work. Generalizing the procedure of \cite{Fletcher2003a}, we have constructed a null tetrad satisfying the criteria imposed in \cite{Krishnan2012}, which are in detail explained in the text. In particular, the tetrad is non-twisting and parallelly propagated along the null vector $n^a$ which is transversal to the horizon.

To summarize, we have provided an analysis of the Kerr--Newman metric in the framework of isolated horizons. Technically more challenging exact solution is to be given in the paper of ours. With the developed formalism, let us sketch some possible applications. First, the analysis of the Meissner effect undertaken in \cite{Guerlebeck2016} can be pushed further by analysing the physical properties of deformations of the Kerr metric considered therein. As a part of this programme, it would be useful to formulate appropriate boundary conditions for electromagnetic fields in the neighbourhood of an isolated horizon. 

For example, especially in the context of the Meissner effect, fields which are asymptotically aligned with the axis of symmetry of the Kerr metric are considered. How can one impose such a condition in the present formalism? More generally, how to interpret different boundary conditions for fields? What happens to these conditions when deformations of the Kerr geometry are allowed? We stress once again that we are not talking about \emph{perturbations} of the Kerr metric in a usual sense, because the back-reaction effects are not neglected.

Another interesting question would be whether it is possible to get appropriate model for the accretion disk surrounding the black hole; again, we do not mean just the test matter on the Kerr background, but Kerr geometry deformed by the presence of an accretion disk. In order to do that, one will need Bondi-like expansions for the solution where the sources are not just electromagnetic fields, like in \cite{Krishnan2012}, but, say, fluid or dust.

These interesting questions will be analysed in the future work.

\appendix
\allowdisplaybreaks
\chapter{Newman--Penrose formalism}
\label{app:np formalism}
The purpose of this appendix is to reveal mostly well-known (sets of) equations in the form of the Newman--Penrose formalism.

Firstly, we review the basis:
\begin{equation}
	l^a n_a = 1 \,, \qquad m^a \cconj{m}_a = -1 \,, \qquad l^a l_a = l^a m_a = n^a n_a = m^a m_a = 0 \,.
\end{equation}
And the metric
\begin{equation}
	g_{ab} = l_a n_b + n_a l_b - m_a \cconj{m}_b - \cconj{m}_a m_b \equiv 2 l_{(a} n_{b)} - 2 m_{(a} \cconj{m}_{b)}
\end{equation}
which gives for the covariant derivative
\begin{equation}\label{eq:covder decomposition}
	\nabla_a = g_a^b \nabla_b = l_a \Delta + n_a D - m_a \cconj{\delta} - \cconj{m}_a \delta \,.
\end{equation}

\section{Transport equations}
\label{app:sec:transport eq}
Directional derivatives of the tetrad vectors are called \emph{transport equations} and are as follows
\begin{subequations}
\begin{align}
 Dl^a &= \left( \epsilon+\cconj{\epsilon} \right)\! l^a - \cconj{\kappa}m^a - 
\kappa \cconj{m}^a \,, \label{np:transport eqs-D la}\\
\Delta l^a &= \left( \gamma+\cconj{\gamma} \right)\! l^a-\cconj{\tau}m^a - 
\tau \cconj{m}^a \,, \label{np:transport eqs-Delta la}\\
\delta l^a &= \left( \cconj{\alpha}+\beta \right)\! l^a-\cconj{\scrho}m^a - 
\sigma \cconj{m}^a \,, \label{np:transport eqs-delta la}\\
Dn^a &= -\!\left( \epsilon+\cconj{\epsilon} \right)\! l^a +\pi m^a 
+\cconj{\pi} \cconj{m}^a \,,\label{np:transport eqs-D na}\\
\Delta n^a &= -\!\left( \gamma+\cconj{\gamma} \right)\! n^a +\nu m^a + 
\cconj{\nu} \cconj{m}^a \,, \label{np:transport eqs-Delta na}\\
\delta n^a &= -\!\left(\cconj{\alpha}+\beta\right)\! n^a+\mu m^a + \cconj{\lambda} \cconj{m}^a \,, 
\label{np:transport eqs-delta na}\\
D m^a &= \cconj{\pi} l^a - \kappa n^a + \left( \epsilon-\cconj{\epsilon} \right)\! m^a \,, 
\label{np:transport eqs-D ma} \\
\Delta m^a &= \cconj{\nu} l^a -\tau n^a + \left( \gamma-\cconj{\gamma} 
\right)\! m^a \,, \label{np:transport eqs-Delta ma}\\
\delta m^a &= \cconj{\lambda} l^a-\sigma n^a+\left( \beta-\cconj{\alpha} 
\right)\! m^a \,, \label{np:transport eqs-delta ma} \\
\cconj{\delta}m^a &= \cconj{\mu} l^a -\scrho n^a+\left( \alpha-\cconj{\beta} 
\right)\! m^a \,.\label{np:transport eqs-deltabar ma}
\end{align}
\label{app1:transport eqs}
\end{subequations}
Note that $m^a$ is complex and therefore we need both $\delta$ and $\cconj{\delta}$ directions.

\section{Ricci identities}
\label{sec:ricci identities}
Derivation of the vectorial form of them can be found in~\cite{Chandrasekhar1983} while the spinorial approach to this topic is present in~\cite{Penrose1987}.

We list foregoing equations without torsion for we are using Levi-Civita covariant derivative. Then the Ricci identities, in the most common way, are written as
\begin{equation}\label{eq:adx:ricci id}
	- R^i{}_{jkl} Z_i = \nabla_l\nabla_k Z_j - \nabla_k \nabla_l Z_j \,.
\end{equation}

In the spinor form, the Ricci identities can be written as follows, \cite{Scholtz2012}:
\begin{align}
\nabla_{A^\prime (A}\nabla_{B)}^{A^\prime}\xi_C &= 
\Psi_{ABCD}\xi^D-2\Lambda\xi_{(A}\epsilon_{B)C} \,, \label{np:Ricci spinor}\\
\nabla_{A (A^\prime}\nabla_{B^\prime)}^A\xi_C &= \Phi_{C D A^\prime 
B^\prime}\xi^D \,.
\end{align}

The equation\eqref{eq:adx:ricci id} can be simply rewritten as
\begin{equation}\label{np adx:ricci}
	- R_{ijkl} e_{\hat{a}}{}^i = \nabla_l \nabla_k e_{\hat{a} j} - \nabla_k \nabla_l e_{\hat{a} j} \,.
\end{equation}
We have chosen the basis covector of the tetrad for the completely general $Z_j$ because any other covector is linear combination of the basis for what it is enough to ensure validity for the basis.

By means of the projection of~\eqref{np adx:ricci} onto the tetrad basis, we get
\begin{align}
	-\!\tensor{R}{_{\hat{a}\hat{b}\hat{c}\hat{d}}} = -2\tensor{\gamma}{_{\hat{a}\hat{b}[\hat{c},\hat{d}]}} + 2\tensor{\gamma}{_{\hat{b}\hat{a}\hat{f}}} \tensor{\gamma}{_{[\hat{c}}^{\hat{f}}_{\hat{d}]}} + 2\tensor{\gamma}{_{\hat{f}\hat{a}[\hat{c}|}} \tensor{\gamma}{_{\hat{b}}^{\hat{f}}_{|\hat{d}]}} \,.
\end{align}
Comma is used for partial derivative. We have used the definition of the Ricci rotation coefficients~\eqref{eq:tetrad formalism:def spin coef} twice to rewrite the covariant derivative.
When written out explicitly they read (the components of the Riemann tensor from which each of the equations comes from, if needed, can be found in~\cite{Chandrasekhar1983})
\begin{subequations}
\begin{align}
  D\tau-\Delta\kappa &= 
(\tau+\cconj{\pi})\scrho+(\cconj{\tau}+\pi)\sigma+(\epsilon-\cconj{\epsilon})\tau 
-(3\gamma+\cconj{\gamma})\kappa+\Psi_1+\Phi_{01} \,, \label{np:RI:Dtau}\\
  D\gamma-\Delta\epsilon &= (\tau+\cconj{\pi})\alpha + (\cconj{\tau}+\pi)\beta - 
(\epsilon+\cconj{\epsilon})\gamma \nonumber \\
&\qquad{}- (\gamma + \cconj{\gamma})\epsilon + \tau \pi - \nu \kappa + \Psi_2 - \Lambda + \Phi_{11} \,, \label{np:RI:Dgamma}\\
  D\nu-\Delta\pi &= 
(\pi+\cconj{\tau})\mu+(\cconj{\pi}+\tau)\lambda+(\gamma-\cconj{\gamma})\pi -(3\epsilon+\cconj{\epsilon})\nu+\Psi_3+\Phi_{21} \,, \label{np:RI:Dnu}\\
  D\sigma-\delta\kappa &= (\scrho+\cconj{\scrho}+3\epsilon-\cconj{\epsilon})\sigma - 
(\tau-\cconj{\pi}+\cconj{\alpha}+3\beta)\kappa+\Psi_0 \,,  \label{np:RI:Dsigma}\\
  D\beta-\delta\epsilon &= (\alpha+\pi)\sigma + 
(\cconj{\scrho}-\cconj{\epsilon})\beta-(\mu+\gamma)\kappa-(\cconj{\alpha}-\cconj{\pi})\epsilon + 
\Psi_1 \,, \label{np:RI:Dbeta}\\
  D\mu-\delta\pi &= (\cconj{\scrho}-\epsilon-\cconj{\epsilon})\mu+\sigma\lambda+ 
(\cconj{\pi}-\cconj{\alpha}+\beta)\pi - \nu \kappa + \Psi_2 + 2 
\Lambda \,, \label{np:RI:Dmu}\\
  D \scrho - \cconj{\delta} \kappa &=\scrho ^2+\left(\epsilon +\cconj{\epsilon
   }\right) \scrho -\kappa  \left(3 \alpha +\cconj{\beta }-\pi \right)-\tau
   \cconj{\kappa }+\sigma  \cconj{\sigma }+\Phi_{00} \,,  \label{np:RI:Drho}\\
  D\alpha-\cconj{\delta}\epsilon &= (\scrho 
+\cconj{\epsilon}-2\epsilon)\alpha+\beta\cconj{\sigma}-\cconj{\beta}\epsilon - \kappa \lambda - 
\cconj{\kappa}\gamma + (\epsilon+\scrho)\pi + \Phi_{10} \,, \label{np:RI:Dalpha}\\
  D\lambda-\cconj{\delta}\pi &= (\scrho - 3\epsilon+\cconj{\epsilon})\lambda + 
\cconj{\sigma}\mu + (\pi+\alpha-\cconj{\beta})\pi - 
\nu\cconj{\kappa}+\Phi_{20} \,, \label{np:RI:Dlambda}\\
  \Delta\sigma-\delta\tau& = -(\mu-3\gamma+\cconj{\gamma})\sigma - 
\cconj{\lambda}\scrho - (\tau + \beta - \cconj{\alpha})\tau + \kappa 
\cconj{\nu}-\Phi_{02} \,, \label{np:RI:Deltasigma}\\
  \Delta\beta-\delta\gamma &= (\cconj{\alpha}+\beta-\tau)\gamma - \mu \tau + 
\sigma \nu + \epsilon \cconj{\nu} + (\gamma-\cconj{\gamma}-\mu)\beta - 
\alpha\cconj{\lambda}-\Phi_{12} \,, \label{np:RI:Deltabeta}\\
  \Delta\mu-\delta\nu &= 
-(\mu+\gamma+\cconj{\gamma})\mu-\lambda\cconj{\lambda}+\cconj{\nu}\pi+(\cconj{\alpha}
+3\beta-\tau)\nu-\Phi_{22} \,, \label{np:RI:Deltamu}\\
  \Delta\scrho-\cconj{\delta}\tau &= (\gamma+\cconj{\gamma}-\cconj{\mu})\scrho - \sigma 
\lambda + (\cconj{\beta}-\alpha-\cconj{\tau})\tau + \nu \kappa - \Psi_2 - 2 
\Lambda \,, \label{np:RI:Deltarho}\\
  \Delta\alpha-\cconj{\delta}\gamma &= (\scrho+\epsilon)\nu - (\tau+\beta)\lambda + 
(\cconj{\gamma}-\cconj{\mu})\alpha + (\cconj{\beta}-\cconj{\tau})\gamma - 
\Psi_3 \,, \label{np:RI:Deltaalpha}\\
  \Delta\lambda-\cconj{\delta}\nu &= 
-(\mu+\cconj{\mu}+3\gamma-\cconj{\gamma})\lambda+(3\alpha+\cconj{\beta}+\pi-\cconj{\tau}
)\nu-\Psi_4 \,, \label{np:RI:Deltalambda}\\
  \delta\scrho-\cconj{\delta}\sigma &= (\cconj{\alpha}+\beta)\scrho - 
(3\alpha-\cconj{\beta})\sigma+(\scrho-\cconj{\scrho})\tau+(\mu-\cconj{\mu})\kappa -\Psi_1 
+ \Phi_{01} \,, \label{np:RI:deltarho}\\
  \delta\alpha-\cconj{\delta}\beta &= \mu\scrho-\lambda\sigma + 
\alpha\cconj{\alpha}+\beta\cconj{\beta}-2\alpha\beta \nonumber \\
		&\qquad{}+ (\scrho-\cconj{\scrho})\gamma + 
(\mu-\cconj{\mu})\epsilon  -  \Psi_2 + \Lambda + \Phi_{11} \,, \label{np:RI:deltaalpha}\\
  \delta\lambda-\cconj{\delta}\mu &= (\scrho-\cconj{\scrho})\nu + (\mu-\cconj{\mu})\pi 
+ (\alpha+\cconj{\beta})\mu+(\cconj{\alpha}-3\beta)\lambda-\Psi_3 + 
\Phi_{21} \,. \label{np:RI:deltalambda}
\end{align}\label{eq:ricci id}
\end{subequations}

\section{Bianchi identities}
\label{sec:bianchi identities}
The common way to write the Bianchi identities is
\begin{equation}\label{eq:adx: bianchi common}
	\tensor{\nabla}{_{[e}}\tensor{R}{_{ab]cd}} = 0 \,.
\end{equation}
As discussed in section~\ref{sec:ricci identities}, torsion is not included.
The Bianchi identities have the spinor form~\cite{Scholtz2012}
\begin{align}
\nabla^D_{B^\prime} \Psi_{ABCD}&= \nabla_A^{A^\prime}\Phi_{BC A^\prime 
B^\prime} + \epsilon_{C(A}\,\nabla_{B)B^\prime}\Lambda - \frac{3}{2}\,
\epsilon_{AB}\,\nabla_{CB^\prime}\Lambda.
\label{np:Bianchi spinor}
\end{align}

Projecting~\eqref{eq:adx: bianchi common} onto the basis leads through similar process as in the case of the Ricci identities (section~\ref{sec:ricci identities}) to the Bianchi 
identities in the \gls{np} formalism. Let us rewrite one of the three terms arising from the antisymmetrization to see how the term will look like and then apply the antisymmetrization to it (it would be difficult to make the use of abbreviated notation for the antisymmetrization clear with both types of indices, therefore we \qm{remember} the antisymmetrization for \qm{later usage})
\begin{align}
	\nabla_e R_{abcd} &= e^{\hat{e}}{}_e \nabla_{\hat{e}} \(R_{\hat{a}\hat{b}\hat{c}\hat{d}} e^{\hat{a}}{}_a e^{\hat{b}}{}_b e^{\hat{c}}{}_c e^{\hat{d}}{}_d \) \nonumber \\
	&= \( \nabla_{\hat{e}} R_{\hat{a}\hat{b}\hat{c}\hat{d}} \) e^{\hat{e}}{}_e e^{\hat{a}}{}_a e^{\hat{b}}{}_b e^{\hat{c}}{}_c e^{\hat{d}}{}_d + e^{\hat{e}}{}_e R_{\hat{a}\hat{b}\hat{c}\hat{d}} \sum_{\varsigma} \( \! \(\nabla_{\hat{e}} \varsigma_\dagger e^{\hat{a}}{}_a \) \varsigma e^{\hat{b}}{}_b \varsigma e^{\hat{c}}{}_c \varsigma e^{\hat{d}}{}_d \) \nonumber \\
	&= \( \nabla_{\hat{e}} R_{\hat{a}\hat{b}\hat{c}\hat{d}} \) e^{\hat{e}}{}_e e^{\hat{a}}{}_a e^{\hat{b}}{}_b e^{\hat{c}}{}_c e^{\hat{d}}{}_d + e^{\hat{e}}{}_e R_{\hat{a}\hat{b}\hat{c}\hat{d}} \sum_{\varsigma} \( \tensor{\gamma}{_{\hat{e}}^{\varsigma_\dagger \hat{a}}_{\hat{f}}} e^{\varsigma_\ast \hat{f}}{}_a e^{\varsigma \hat{b}}{}_b e^{\varsigma \hat{c}}{}_c e^{\varsigma \hat{d}}{}_d \)
\end{align}
where the summation over $\varsigma$ means that are added all terms where each term (one at a time) marked with $\varsigma$ interchanges with the term marked with $\varsigma_\dagger$. The term preceded with $\varsigma_\ast$, if present, goes to the place of the interchanged one while the term decorated with $\varsigma_\dagger$ fills the remaining position. The three terms (from antisymmetrization) together after the projection onto the tetrad are
\begin{equation}
	\nabla \tensor[_{[\hat{e}}]{R}{_{\hat{a}\hat{b}]\hat{c}\hat{d}}} = - \mathcal{A}_{\hat{e}\hat{a}\hat{b}}\!\( \sum_{\varsigma} \( \tensor{R}{_{\varsigma_\ast \hat{f}\, \varsigma \hat{b}\, \varsigma \hat{c}\, \varsigma \hat{d}}} \, \tensor{\gamma}{_{\hat{e}}^{\hat{f}}_{\varsigma_\dagger \hat{a}}} \) \! \) \!\,.
\end{equation}
The symbol $\mathcal{A}_{\hat{e}\hat{a}\hat{b}}\!\(X\)$ is an antisymmetrization of $X$ in indices $\hat{e}\hat{a}\hat{b}$.
When the summations are rewritten with aid of the special symbols for the spin coefficients we get
\begin{subequations}
\begin{align}
		&{}D\Psi_1-\cconj{\delta}\Psi_0-D\Phi_{01}+\delta\Phi_{00} = 
			(\pi - 4 \alpha) \Psi_0 + 2(2\scrho+\varepsilon)\Psi_1 - 3\kappa\Psi_2 + 2\kappa\Phi_{11} \nonumber\\ 
				&\qquad - (\cconj{\pi} - 2\cconj{\alpha} - 2\beta) \Phi_{00} - 2\sigma\Phi_{10} - 2(\cconj{\scrho}+\varepsilon)\Phi_{01} + \cconj{\kappa}\Phi_{02} \,, \label{np:BI:DPsi1} \\[1.5ex]
		&{}D\Psi_2-\cconj{\delta}\Psi_1 + \Delta\Phi_{00} - \cconj{\delta}\Phi_{01} + 2D\Lambda = - \lambda\Psi_0 + 2 (\pi-\alpha)\Psi_1 + 3\scrho \Psi_2 - 2\kappa\Psi_3  \nonumber\\
			&\qquad + 2\scrho\Phi_{11} + \cconj{\sigma}\Phi_{02} + (2\gamma + 2\cconj{\gamma} - \cconj{\mu})\Phi_{00} - 2(\alpha + \cconj{\tau})\Phi_{01} - 2\tau\Phi_{10} \,,\label{np:BI:DPsi2} \\[1.5ex]
		&{}D\Psi_3-\cconj{\delta}\Psi_2-D\Phi_{21}+\delta\Phi_{20}-2\cconj{\delta}\Lambda = -2\lambda \Psi_1+3\pi\Psi_2 + 2 (\scrho-\varepsilon)\Psi_3 - \kappa\Psi_4  \nonumber\\
			&\qquad + 2\mu\Phi_{10} - 2\pi\Phi_{11} - (2\beta+\cconj{\pi} - 2\cconj{\alpha})\Phi_{20} - 2(\cconj{\scrho} - \varepsilon)\Phi_{21} + \cconj{\kappa}\Phi_{22} \,, \label{np:BI:DPsi3} \\[1.5ex]
		&{}D\Psi_4-\cconj{\delta}\Psi_3 + \Delta\Phi_{20} - \cconj{\delta}\Phi_{21} = - 3\lambda\Psi_2 + 2(\alpha + 2\pi)\Psi_3 + (\scrho - 4\varepsilon)\Psi_4 + 2\nu\Phi_{10} \nonumber\\
			&\qquad - 2\lambda\Phi_{11} - (2\gamma - 2\cconj{\gamma} + \cconj{\mu})\Phi_{20}- 2(\cconj{\tau} - \alpha)\Phi_{21} + \cconj{\sigma}\Phi_{22} \,, \label{np:BI:DPsi4} \\[1.5ex]
		&{}\Delta\Psi_0 - \delta\Psi_1 + D\Phi_{02} - \delta\Phi_{01} = (4\gamma-\mu)\Psi_0 -2(2\tau+\beta)\Psi_1+ 3\sigma\Psi_2 \nonumber\\
			&\qquad +(\cconj{\scrho} +2\varepsilon -2\cconj{\varepsilon})\Phi_{02} + 2\sigma\Phi_{11} - 2\kappa\Phi_{12} -\cconj{\lambda}\Phi_{00} +2(\cconj{\pi} -\beta)\Phi_{01} \,, \label{np:BI:DeltaPsi0} \\[1.5ex]
		&{}\Delta\Psi_1 -\delta\Psi_2 -\Delta\Phi_{01} +\cconj{\delta}\Phi_{02} -2\delta\Lambda =\nu\Psi_0 +2(\gamma-\mu)\Psi_1 -3\tau\Psi_2 +2\sigma\Psi_3 \nonumber\\
			&\qquad -\cconj{\nu}\Phi_{00} + 2(\cconj{\mu}-\gamma)\Phi_{01} +(2\alpha+\cconj{\tau} -2\cconj{\beta})\Phi_{02} +2\tau\Phi_{11} -2\scrho\Phi_{12} \,, \label{np:BI:DeltaPsi1} \\[1.5ex]
		&{}\Delta\Psi_2 - \delta\Psi_3 +D\Phi_{22} - \delta\Phi_{21} + 2\Delta\Lambda = 2\nu\Psi_1-3\mu\Psi_2 + 2(\beta-\tau)\Psi_3 + \sigma\Psi_4 \nonumber\\
			&\qquad - 2\mu\Phi_{11} -\cconj{\lambda}\Phi_{20} + 2\pi\Phi_{12}+ 2(\beta+\cconj{\pi})\Phi_{21} + (\cconj{\scrho} -2\varepsilon -2\cconj{\varepsilon})\Phi_{22} \,, \label{np:BI:DeltaPsi2} \\[1.5ex]
		&{}\Delta\Psi_3-\delta\Psi_4\hfill - \Delta\Phi_{21}+\cconj{\delta}\Phi_{22} = 3\nu\Psi_2-2(\gamma+2\mu)\Psi_3 +(4\beta-\tau)\Psi_4-2\nu\Phi_{11} \nonumber\\
			&\qquad - \cconj{\nu}\Phi_{20} + 2\lambda\Phi_{12} +2(\gamma+\cconj{\mu})\Phi_{21} +(\cconj{\tau} -2\cconj{\beta}-2\alpha)\Phi_{22} \,, \label{np:BI:DeltaPsi3} \\[1.5ex]
		&{}D\Phi_{11}-\delta\Phi_{10}+\Delta\Phi_{00}-\cconj{\delta}\Phi_{01}+3D\Lambda = (2\gamma+2\cconj{\gamma}-\mu-\cconj{\mu})\Phi_{00} \nonumber\\
			&\qquad + (\pi-2\alpha-2\cconj{\tau})\Phi_{01} + (\cconj{\pi}-2\cconj{\alpha} -2\tau)\Phi_{10} +2(\scrho+\cconj{\scrho})\Phi_{11}+\cconj{\sigma}\Phi_{02} \nonumber\\
			&\qquad + \sigma\Phi_{20} -\cconj{\kappa}\Phi_{12} -\kappa\Phi_{21} \,, \label{np:BI:DPhi11} \\[1.5ex]
		&{}D\Phi_{12}-\delta\Phi_{11} +\Delta\Phi_{01}-\cconj{\delta}\Phi_{02} +3\delta\Lambda = (2\gamma-\mu -2\cconj{\mu})\Phi_{01} +\cconj{\nu}\Phi_{00} -\cconj{\lambda}\Phi_{10} \nonumber\\
			&\qquad + 2(\cconj{\pi}-\tau)\Phi_{11} +(\pi+2\cconj{\beta} -2\alpha-\cconj{\tau})\Phi_{02} +(2\scrho+\cconj{\scrho} -2\cconj{\varepsilon})\Phi_{12} \nonumber \\
			&\qquad + \sigma\Phi_{21} -\kappa\Phi_{22} \,, \label{np:BI:DPhi12} \\[1.5ex]
		&{}D\Phi_{22}-\delta\Phi_{21}+\Delta\Phi_{11}-\cconj{\delta}\Phi_{12}+3\Delta\Lambda = \nu \Phi_{01}+\cconj{\nu}\Phi_{10}-2(\mu+\cconj{\mu})\Phi_{11}-\lambda\Phi_{02} \nonumber\\
			&\qquad - \cconj{\lambda}\Phi_{20} + (2\pi-\cconj{\tau} +2\cconj{\beta})\Phi_{12} +(2\beta-\tau+2\cconj{\pi})\Phi_{21} \nonumber \\
			&\qquad + (\scrho+\cconj{\scrho}-2\varepsilon -2\cconj{\varepsilon})\Phi_{22} \,. \label{np:BI:DPhi22}
\end{align}\label{eq:bianchi id}
\end{subequations}
Again, the corresponding components of the Riemann tensor can be found in~\cite{Chandrasekhar1983}.

\section{Maxwell equations}
For convenience, we enlist also already discussed and displayed Maxwell equations which in the spinor form read
\begin{equation}
	\nabla^{AA'} \oldphi_{AB} = 0 \,
\end{equation}
while the projections onto the tetrad are
\begin{subequations}
\begin{align}
	D \msp_1 - \cconj{\delta} \msp_0 &= \(\pi - 2\alpha\)\!\msp_0 + 2\scrho\msp_1 - \kappa\msp_2 \,, \label{eq:ap:maxwell D1} \\
	D\msp_2 - \cconj{\delta}\msp_1 &= - \lambda \msp_0 + 2 \pi \msp_1 + \(\scrho - 2 \epsilon\)\! \msp_2 \,, \label{eq:ap:maxwell D2} \\
	\Delta\msp_0 - \delta \msp_1 &= \(2\gamma - \mu\) - 2 \tau \msp_1 + \sigma \msp_2 \,, \label{eq:ap:maxwell Delta0} \\
	\Delta\msp_1 - \delta\msp_2 &= \nu \msp_0 - 2 \mu \msp_1 + \(2\beta - \tau\)\! \msp_2 \,. \label{eq:ap:maxwell Delta1}
\end{align}\label{eq:ap:maxwell eq}
\end{subequations}

\chapter{Series}
\label{app:series}
\begin{subequations}
	\begin{align}
		n_{(2)}^v &= a^2 \sin ^2(\theta ) \left(64 \left(a^2+r_+^2\right)^4 \left(a^2 \cos (2 \theta )+a^2+2 r_+^2\right)^3\right)^{-1} \times \nonumber \\
		&\qquad{}\times \bigg(6 a^{10} \cos (6 \theta )+212 a^{10}+a^8 M^2 \cos (8 \theta )+3 a^8 M^2 \nonumber \\
		&\qquad{}\qquad{}-16 a^8 M r_+ \cos (6 \theta )-2 a^8 M r_+ \cos (8 \theta )+58 a^8 M r_+ \nonumber \\
		&\qquad{}\qquad{}+28 a^8 r_+^2 \cos (6 \theta ) +a^8 r_+^2 \cos (8 \theta )+59 a^8 r_+^2+8 a^6 M^2 r_+^2 \cos (6 \theta ) \nonumber \\
		&\qquad{}\qquad{}+16 a^6 M^2 r_+^2 -32 a^6 M r_+^3 \cos (6 \theta )+320 a^6 M r_+^3+30 a^6 r_+^4 \cos (6 \theta ) \nonumber \\
		&\qquad{}\qquad{}-1980 a^6 r_+^4 +48 a^4 M^2 r_+^4+576 a^4 M r_+^5-4192 a^4 r_+^6+384 a^2 M r_+^7 \nonumber \\
		&\qquad{}\qquad{}-3168 a^2 r_+^8 +4 a^4 \cos (4 \theta ) \Big(11 a^6-a^4 \left(M^2+14 M r_+-49 r_+^2\right) \nonumber \\
		&\qquad{}\qquad{}+a^2 r_+^2 \left(-4 M^2-48 M r_++87 r_+^2\right)+4 r_+^4 \left(M^2-12 M r_+ +14 r_+^2\right)\Big) \nonumber \\
		&\qquad{}\qquad{}+2 a^2 \cos (2 \theta ) \Big(125 a^8+2 a^6 r_+ (4 M+249 r_+) \nonumber \\
		&\qquad{}\qquad{}+a^4 r_+^2 \left(-4 M^2-48 M r_++801 r_+^2\right)-32 a^2 r_+^4 \left(M^2+6 M r_+-22 r_+^2\right) \nonumber \\
		&\qquad{}\qquad{}+16 r_+^7 (19 r_+-12 M)\Big)-768 r_+^{10} \bigg) \,, \\
		n_{(2)}^r &= a^2 \sin ^2(\theta ) \left(8 \left(a^2+r_+^2\right)^3 \left(a^2 \cos (2 \theta )+a^2+2 r_+^2\right)^3\right)^{-1} \times \nonumber \\ 
		&\qquad{}\times \bigg(44 a^8-a^6 M^2 \cos (6 \theta )+2 a^6 M^2+2 a^6 M r_+ \cos (6 \theta )+108 a^6 M r_+ \nonumber \\
		&\qquad{}\qquad{}-a^6 r_+^2 \cos (6 \theta )-150 a^6 r_+^2+8 a^4 M^2 r_+^2+416 a^4 M r_+^3-796 a^4 r_+^4 \nonumber \\
		&\qquad{}\qquad{}+16 a^2 M^2 r_+^4+544 a^2 M r_+^5-1008 a^2 r_+^6 \nonumber \\
		&\qquad{}\qquad{}+2 a^4 \cos (4 \theta ) \Big(2 a^4-a^2 \left(M^2-10 M r_++5 r_+^2\right) \nonumber \\ 
		&\qquad{}\qquad{}-2 r_+^2 \left(2 M^2-8 M r_++5 r_+^2\right)\Big) +a^2 \cos (2 \theta ) \Big(48 a^6 \nonumber \\
		&\qquad{}\qquad{}+a^4 \left(M^2+126 M r_++33 r_+^2\right)+16 a^2 r_+^3 (20 M-9 r_+) \nonumber \\
		&\qquad{}\qquad{}-16 r_+^4 \left(M^2-14 M r_++9 r_+^2\right)\Big)+256 M r_+^7-416 r_+^8\bigg) \,, \\
		n_{(2)}^\theta &= -a^2 \sin (2 \theta ) \left(8 \left(a^2+r_+^2\right)^3 \left(a^2 \cos (2 \theta )+a^2+2 r_+^2\right)^2\right)^{-1} \times \nonumber \\
		&\qquad{}\times \Big(-a^4 \cos (4 \theta ) (M-r_+)+a^4 M+39 a^4 r_++4 a^2 M r_+^2 \nonumber \\
		&\qquad{}\qquad{}+4 a^2 r_+ \cos (2 \theta ) \left(2 a^2+r_+ (3 r_+-M)\right)+84 a^2 r_+^3+48 r_+^5\Big) \,, \\
		n_{(2)}^\phi &= a\left(16 \left(a^2+r_+^2\right)^4 \left(a^2 \cos (2 \theta )+a^2+2 r_+^2\right)^2\right)^{-1} \times \nonumber \\
		&\qquad{}\times \bigg(28 a^8+a^6 M r_+ \cos (6 \theta )+2 a^6 M r_+-a^6 r_+^2 \cos (6 \theta )-42 a^6 r_+^2 \nonumber \\
		&\qquad{}\qquad{}+12 a^4 M r_+^3-320 a^4 r_+^4-432 a^2 r_+^6 \nonumber \\
		&\qquad{}\qquad{}+2 a^4 \cos (4 \theta ) \left(2 a^4-a^2 r_+ (M+3 r_+)+2 r_+^3 (M-4 r_+)\right) \nonumber \\
		&\qquad{}\qquad{}+a^2 \cos (2 \theta ) \Big(32 a^6+a^4 r_+ (49 r_+-M) \nonumber \\
		&\qquad{}\qquad{}-16 a^2 r_+^3 (M+3 r_+)-80 r_+^6\Big)-192 r_+^8\bigg) \,.
	\end{align}
	\end{subequations}
	\begin{subequations}
	\begin{align}
		l_{(2)}^v &= - a^2 \sin ^2(\theta ) \left(8 \left(a^2+r_+^2\right)^4 \left(a^2 \cos (2 \theta )+a^2+2 r_+^2\right)\right)^{-1} \times \nonumber \\
		&\qquad{}\times \bigg(8 a^6-a^4 M^2+a^4 \cos (4 \theta ) (M-r_+)^2-18 a^4 M r_++31 a^4 r_+^2 \nonumber \\
		&\qquad{}\qquad{}-4 a^2 M^2 r_+^2  +4 a^2 \cos (2 \theta ) \Big(a^4 + a^2 r_+ (5 r_+-3 M) \nonumber \\
		&\qquad{}\qquad{} +r_+^2 \left(M^2-5 M r_++5 r_+^2\right) \Big)-44 a^2 M r_+^3+48 a^2 r_+^4 \nonumber \\
		&\qquad{}\qquad{}-32 M r_+^5+28 r_+^6\bigg) \,, \\
		l_{(2)}^r &= \left(8 \left(a^2+r_+^2\right)^3 \left(a^2 \cos (2 \theta )+a^2+2 r_+^2\right)\right)^{-1} \times \nonumber \\
		&\qquad{}\times \Big(a^4 M^2-a^4 \cos (4 \theta ) (M-r_+)^2+22 a^4 M r_+-15 a^4 r_+^2+4 a^2 M^2 r_+^2 \nonumber \\
		&\qquad{}\qquad{}+4 a^2 \cos (2 \theta ) \left(2 a^4+2 a^2 r_+ (M+r_+)-r_+^2 \left(M^2-4 M r_++r_+^2\right)\right) \nonumber \\
		&\qquad\qquad{}+48 a^2 M r_+^3-36 a^2 r_+^4 +32 M r_+^5-24 r_+^6\Big) \,, \\
		l_{(2)}^\theta &= a^2 \sin (2 \theta ) \left(a^2 \cos (2 \theta )+a^2+2 r_+^2\right) \left(64 \left(a^2+r_+^2\right)^3 \left(a^2 \cos ^2(\theta )+r_+^2\right)^3\right)^{-1} \times \nonumber \\
		&\qquad\times \Big(-a^4 \cos (4 \theta ) (M-r_+)-3 a^4 M+39 a^4 r_+ \nonumber \\
		&\qquad\qquad{}-4 a^2 \cos (2 \theta ) \left(a^2 (M-2 r_+)+r_+^2 (2 M-3 r_+)\right) \nonumber \\
		&\qquad\qquad{}-8 a^2 M r_+^2+84 a^2 r_+^3-8 M r_+^4+48 r_+^5\Big) \,, \\
		l_{(2)}^\phi &= -a\left(4 \left(a^2+r_+^2\right)^4\right)^{-1} \times \nonumber \\
		&\qquad\times\Big(2 a^4+a^2 r_+ \cos (2 \theta ) (M-r_+) \nonumber \\
		&\qquad\qquad{}+a^2 r_+ (r_+-3 M)-2 r_+^3 (M+r_+)\Big) \,.
	\end{align}
	\end{subequations}
	\begin{subequations}
	\begin{align}
		m_{(2)}^v &= -a^2 \sin (\theta ) \left(128 \sqrt{2} \left(a^2+r_+^2\right)^4 \left(a^2 \cos ^2(\theta )+r_+^2\right)^2\right)^{-1} \times \nonumber \\
		&\qquad\times \bigg(\cos (\theta ) \Big(-8 a^8+a^6 \left(3 M^2+66 M r_++155 r_+^2\right) \nonumber \\
		&\qquad\qquad{}+8 a^4 r_+^2 \left(M^2+21 M r_++69 r_+^2\right)+16 a^2 r_+^5 (7 M+40 r_+)+256 r_+^8\Big) \nonumber \\
		&\qquad\qquad{}+a \Big[4 a^7 \cos (5 \theta )+8 \ii a^6 M \cos (4 \theta )+2 \ii a^6 M \cos (6 \theta )-8 \ii a^6 M \nonumber \\
		&\qquad\qquad{}-4 \ii a^6 r_+ \cos (4 \theta )-2 \ii a^6 r_+ \cos (6 \theta )-188 \ii a^6 r_+-a^5 M^2 \cos (5 \theta ) \nonumber \\
		&\qquad\qquad{}+a^5 M^2 \cos (7 \theta )-18 a^5 M r_+ \cos (5 \theta )-2 a^5 M r_+ \cos (7 \theta ) \nonumber \\
		&\qquad\qquad{}+27 a^5 r_+^2 \cos (5 \theta ) +a^5 r_+^2 \cos (7 \theta )-4 \ii a^4 M^2 r_+ \cos (4 \theta ) \nonumber \\
		&\qquad\qquad{}+2 \ii a^4 M^2 r_+ \cos (6 \theta )+4 \ii a^4 M^2 r_+ -4 \ii a^4 M r_+^2 \cos (4 \theta ) \nonumber \\
		&\qquad\qquad{}-2 \ii a^4 M r_+^2 \cos (6 \theta )+52 \ii a^4 M r_+^2+16 \ii a^4 r_+^3 \cos (4 \theta ) -656 \ii a^4 r_+^3  \nonumber \\
		&\qquad\qquad{}+4 a^3 M^2 r_+^2 \cos (5 \theta )-28 a^3 M r_+^3 \cos (5 \theta )+28 a^3 r_+^4 \cos (5 \theta ) \nonumber \\
		&\qquad\qquad{}+8 \ii a^2 M^2 r_+^3 \cos (4 \theta )+24 \ii a^2 M^2 r_+^3-36 \ii a^2 M r_+^4 \cos (4 \theta ) \nonumber \\
		&\qquad\qquad{}+196 \ii a^2 M r_+^4 +32 \ii a^2 r_+^5 \cos (4 \theta )-832 \ii a^2 r_+^5+a \cos (3 \theta ) \times \nonumber \\
		&\qquad\qquad{}\times\Big(4 a^6 +a^4 \left(-3 M^2-46 M r_++73 r_+^2\right)+4 a^2 r_+^2 \big(-3 M^2-35 M r_+ \nonumber \\
		&\qquad\qquad{} +47 r_+^2\big)+16 r_+^5 (8 r_+-7 M)\Big)-2 \ii \cos (2 \theta ) \Big(a^6 (M+31 r_+) \nonumber \\
		&\qquad\qquad{}+a^4 r_+ \left(M^2+23 M r_++64 r_+^2\right)+16 a^2 r_+^3 \left(M^2+5 M r_+-r_+^2\right) \nonumber \\
		&\qquad\qquad{}+8 r_+^6 (11 M-8 r_+)\Big)+176 \ii M r_+^6-384 \ii r_+^7\Big]\bigg) \,, \\
		m_{(2)}^r &= a \sin (\theta )\left(32 \sqrt{2} \left(a^2+r_+^2\right)^3 \left(a^2 \cos ^2(\theta )+r_+^2\right)^2\right)^{-1} \times \nonumber \\
		&\qquad\times{} \bigg(-8 a^7 \cos (3 \theta )+2 \ii a^6 M \cos (4 \theta )+6 \ii a^6 M-2 \ii a^6 r_+ \cos (4 \theta ) \nonumber \\
		&\qquad\qquad{}+34 \ii a^6 r_++a^5 M^2 \cos (3 \theta )+a^5 M^2 \cos (5 \theta )-18 a^5 M r_+ \cos (3 \theta ) \nonumber \\
		&\qquad\qquad{}-2 a^5 M r_+ \cos (5 \theta )+a^5 r_+^2 \cos (3 \theta )+a^5 r_+^2 \cos (5 \theta )+2 i a^4 M^2 r_+ \cos (4 \theta ) \nonumber \\
		&\qquad\qquad{}-2 \ii a^4 M^2 r_+-2 \ii a^4 M r_+^2 \cos (4 \theta )-70 \ii a^4 M r_+^2+168 \ii a^4 r_+^3 \nonumber \\
		&\qquad\qquad{}+4 a^3 M^2 r_+^2 \cos (3 \theta )-24 a^3 M r_+^3 \cos (3 \theta )+12 a^3 r_+^4 \cos (3 \theta ) \nonumber \\
		&\qquad\qquad{}-8 \ii a^2 M^2 r_+^3-176 \ii a^2 M r_+^4+256 \ii a^2 r_+^5+8 \ii a^2 \cos (2 \theta ) \left(a^4 (M-4 r_+) \right.\nonumber \\
		&\qquad\qquad{}\left.-a^2 r_+^2 (M+5 r_+)+M r_+^3 (M-4 r_+)\right)-2 a \cos (\theta ) \Big(12 a^6+a^4 \left(M^2 \right.\nonumber \\
		&\qquad\qquad{}\left.+54 M r_+-15 r_+^2\right)+2 a^2 r_+^2 \left(M^2+58 M r_+-37 r_+^2\right) \nonumber \\
		&\qquad\qquad{}+16 r_+^5 (4 M-3 r_+)\Big)-112 \ii M r_+^6+128 \ii r_+^7\bigg) \,. \\
		m_{(2)}^\theta &= \left(64 \sqrt{2} \left(a^2+r_+^2\right)^3 (a \cos (\theta )-\ii r_+)^3 (a \cos (\theta )+\ii r_+)\right)^{-1} \times \nonumber \\
		&\qquad\times \Big(-10 \ii a^7 \cos (3 \theta )-2 \ii a^7 \cos (5 \theta )-2 a^6 M \cos (4 \theta )-a^6 M \cos (6 \theta ) \nonumber \\
		&\qquad\qquad{}+2 a^6 M-2 a^6 r_+ \cos (4 \theta )+a^6 r_+ \cos (6 \theta )-14 a^6 r_+ \nonumber \\
		&\qquad\qquad{}+2 \ii a^5 M r_+ \cos (3 \theta ) +2 \ii a^5 M r_+ \cos (5 \theta )-20 \ii a^5 r_+^2 \cos (3 \theta ) \nonumber \\
		&\qquad\qquad{}-4 \ii a^5 r_+^2 \cos (5 \theta )-4 a^4 M r_+^2 \cos (4 \theta )+4 a^4 M r_+^2-96 a^4 r_+^3 \nonumber \\
		&\qquad\qquad{}+8 \ii a^3 M r_+^3 \cos (3 \theta )-16 \ii a^3 r_+^4 \cos (3 \theta )-144 a^2 r_+^5 \nonumber \\
		&\qquad\qquad{}+a^2 \cos (2 \theta ) \left(a^4 (M+47 r_+)+96 a^2 r_+^3+48 r_+^5\right) \nonumber \\
		&\qquad\qquad{}+4 \ii a \cos (\theta ) \left(11 a^6+a^4 r_+ (6 r_+-M) \right.\nonumber \\
		&\qquad\qquad{}\left.-2 a^2 r_+^3 (M+10 r_+)-16 r_+^6\right)-64 r_+^7 \Big) \,, \\
		m_{(2)}^\phi &= \csc (\theta ) \left(32 \sqrt{2} \left(a^2+r_+^2\right)^4 \left(a^2 \cos ^2(\theta )+r_+^2\right)\right)^{-1} \times \nonumber \\
		&\qquad\times \Big(-4 a^7 \cos (3 \theta )-6 \ii a^6 r_+ \cos (4 \theta )+14 \ii a^6 r_++7 a^5 M r_+ \cos (3 \theta ) \nonumber \\
		&\qquad\qquad{}-a^5 M r_+ \cos (5 \theta )+13 a^5 r_+^2 \cos (3 \theta )+a^5 r_+^2 \cos (5 \theta ) \nonumber \\
		&\qquad\qquad{}-2 \ii a^4 M r_+^2 \cos (4 \theta )-14 \ii a^4 M r_+^2-4 \ii a^4 r_+^3 \cos (4 \theta )+36 \ii a^4 r_+^3 \nonumber \\
		&\qquad\qquad{}+4 a^3 M r_+^3 \cos (3 \theta )+20 a^3 r_+^4 \cos (3 \theta )-8 \ii a^2 M r_+^4+48 \ii a^2 r_+^5 \nonumber \\
		&\qquad\qquad{}-8 \ii a^2 r_+ \cos (2 \theta ) \left(3 a^4-2 a^2 r_+ (M-2 r_+)-M r_+^3\right) \nonumber \\
		&\qquad\qquad{}-2 a \cos (\theta ) \left(6 a^6+a^4 r_+ (3 M+7 r_+) \right. \nonumber \\
		&\qquad\qquad{}\left.+2 a^2 r_+^3 (M-7 r_+)-16 r_+^6\right)+32 \ii r_+^7\Big) \,.
	\end{align}
	\end{subequations}
\chapter{Mathematica source code}
\label{appendix:mathematica}
\newcommand{\wmcnl}{%
\vspace{0.5cm}%
\\%
}
In this appendix, we present a few \texttt{Mathematica} source codes which were used during our work with non-twisting tetrads in the Kerr--Newman space-time. The version of \texttt{Mathematica} used was $10.4.1.0$, and it was the Student Edition one kindly provided by Charles University.

\section{Computation of series of the tetrad at horizon}
The source code to be presented and described in this section is, with a few modifications in favour of usability instead of self-explanatority, also included on an attached \textsc{cd-rom} together with the computed series of the tetrad up to the third order.

\subsection{Transformation of the Kinnersley tetrad}
The first code to be displayed is an implementation of the Lorentz transformations of the tetrad and the coordinate change. At the end, there is also listed a function \matcom{CheckNPTetrad}, which were used after each transformation to verify that the resulting tetrad still has the proper Newman--Penrose normalizations.
\wmcnl
\noindent\includegraphics{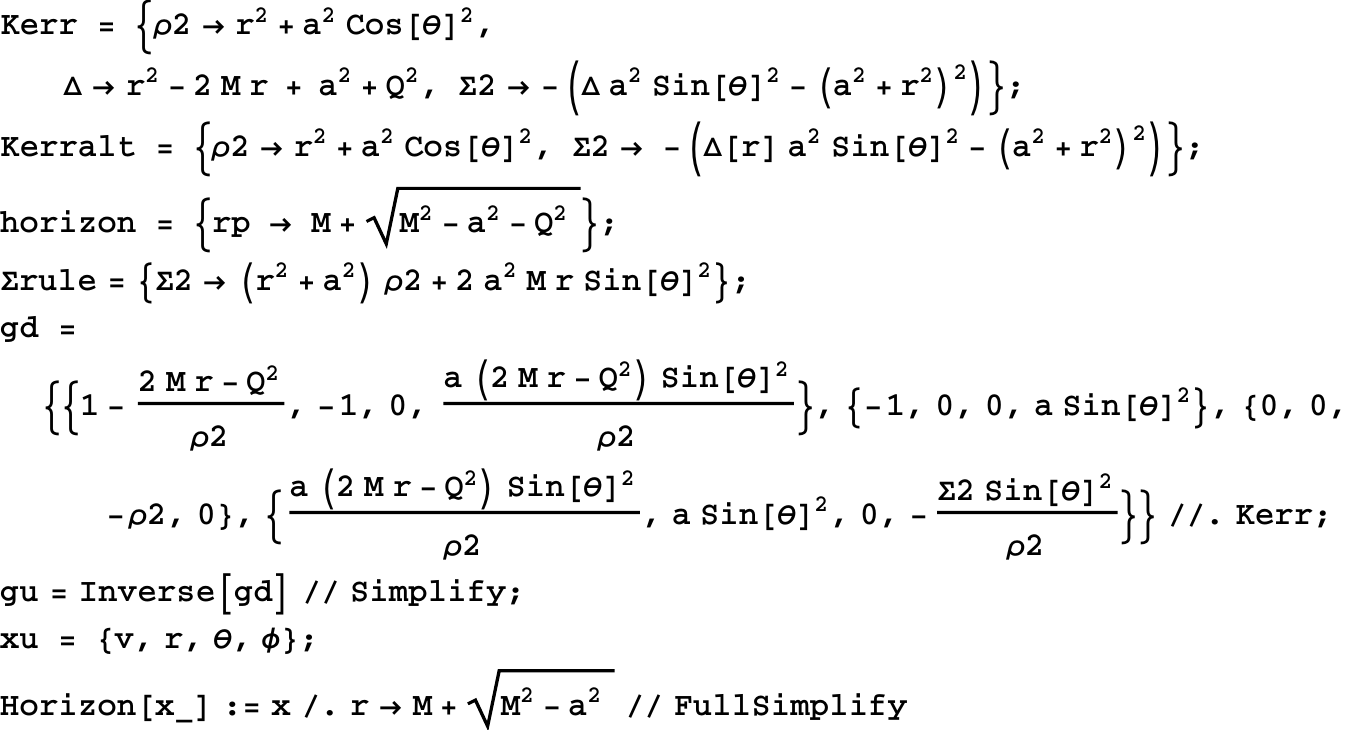} 

\subsubsection{Kinnersley tetrad}
\includegraphics{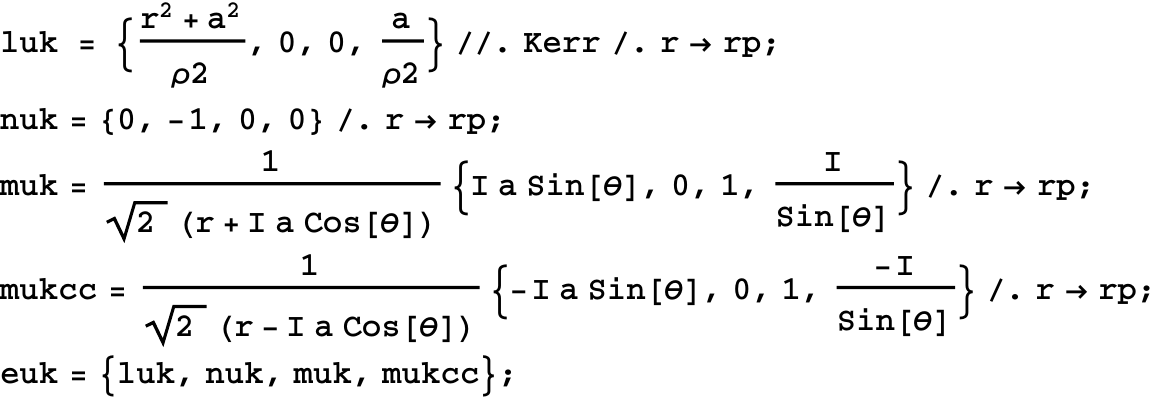} 

\subsubsection{Coordinate transformation}
\includegraphics{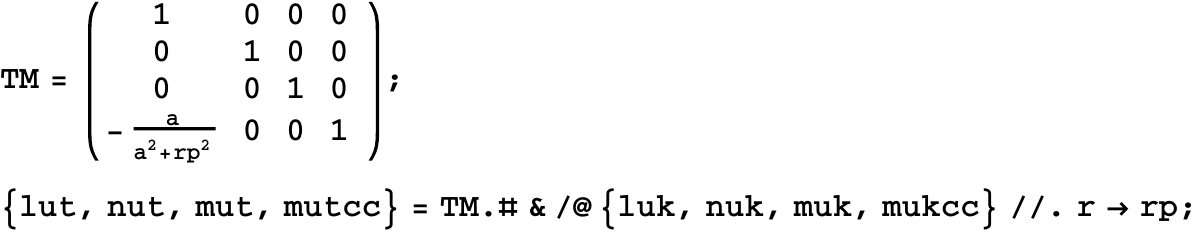}\wmcnl
\includegraphics{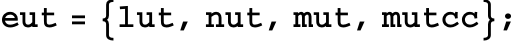}\wmcnl
\includegraphics{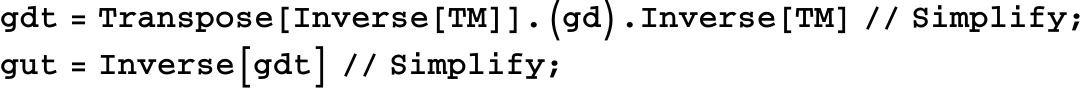} 

\subsubsection{Boost}
\includegraphics{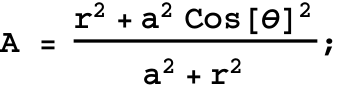}\wmcnl
\includegraphics{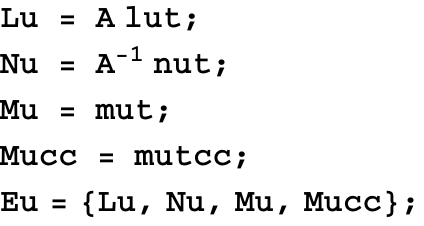} 

\subsubsection{Null rotation about $\bm{l^a}$}
\includegraphics{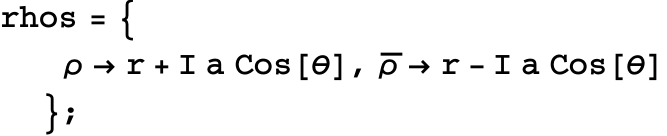}\wmcnl
\includegraphics{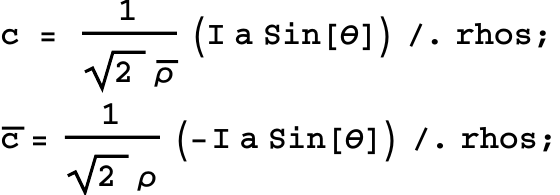}\wmcnl
\includegraphics{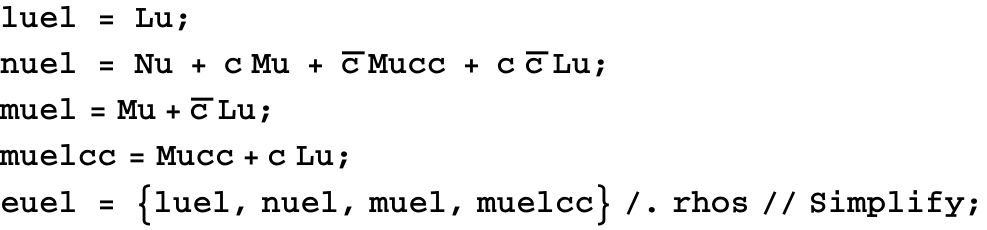}\wmcnl

\noindent\includegraphics{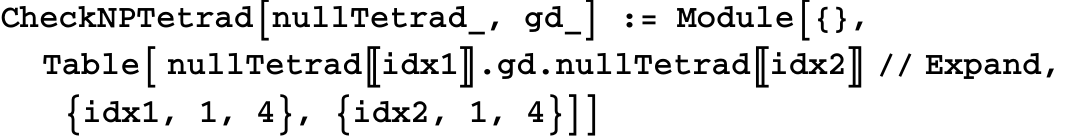}

\subsection{Expansion of the tetrad}
The main computational part consists of functions successively adding new orders of the solution. These are functions \matcom{Compute\-Order\-Of\-Vector}, \matcom{Compute\-Order\-Of\-Der}, and \matcom{Compute\-Order\-Of\-Tetrad}. None of these commands should be run twice (with the same parameters) or in wrong order of powers of the solution in the construction of ours -- they are rewriting (adding to it) the solution, so we work with one list of rules giving the solution. If run twice, any of the commands, therefore, uses the already found solution to \qm{compute it again}. This leads to information loss. Since we hope for a charitable user, we have not taken care of this. Moreover, the last one (\matcom{ComputOrderOfTetrad}) is the only needed by the end user.

Moreover, there is a function \matcom{Test\-Tetrad\-Up\-To\-Order}, which verifies that the computed series of the tetrad is up to the specified order really a proper \gls{np} tetrad. 

To decide whether the tetrad accomplishes the conditions of the parallel transport, we need the spin coefficients which are provided by functions \matcom{Compute\-Spins\-Up\-To\-Order} and \matcom{Compute\-Spin\-Coeffs\-NP}, which is used in the first one. Similarly, Weyl scalars and the Ricci scalars are given by functions \matcom{Compute\-Weyls\-Up\-To\-Order} and \matcom{Compute\-Riccis\-Up\-To\-Order}, in which a few other functions providing the curvature are employed.

Because computation of covariant derivative of the general series of the tetrad is time-consuming, there is a switch \matcom{MaxOrder} which determines the maximal order we are going to compute and has to be greater of equal to the order set in the function of type \matcom{Compute\-Order\-Of\-<Something>} or \matcom{Compute\-<Something>\-Up\-To\-Order}

The source code is:\wmcnl
\includegraphics{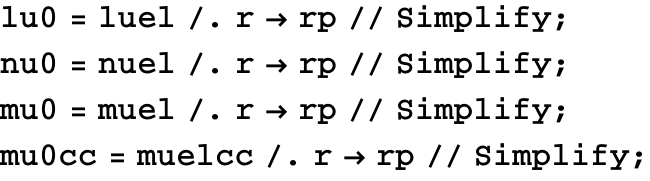}\wmcnl
\includegraphics{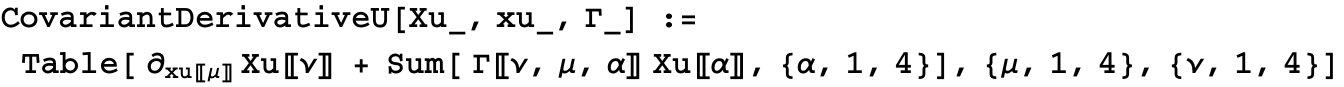}\wmcnl
\includegraphics{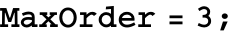}\wmcnl
\includegraphics{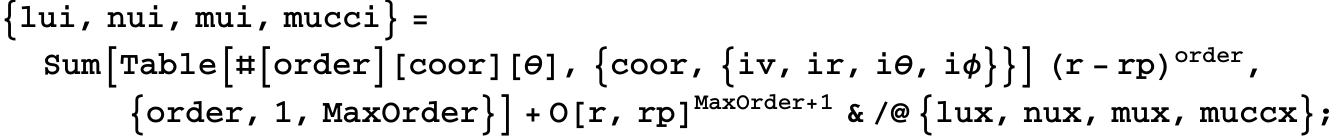}\wmcnl
\includegraphics{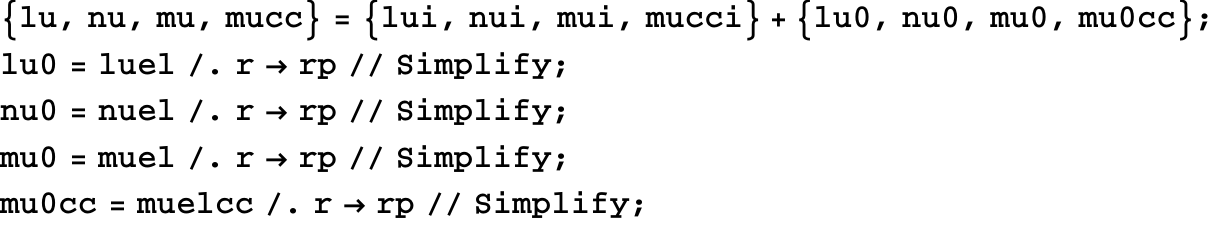}\wmcnl
\includegraphics{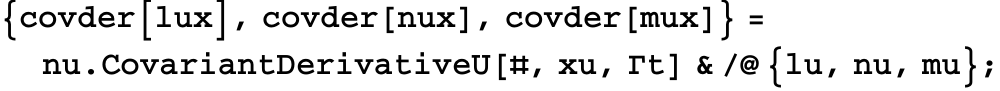}\wmcnl
\includegraphics{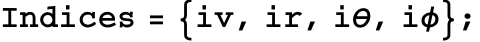}\wmcnl
\includegraphics{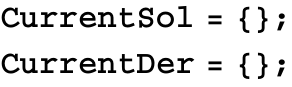} 

\subsubsection{Compute tetrad expansion}
\includegraphics{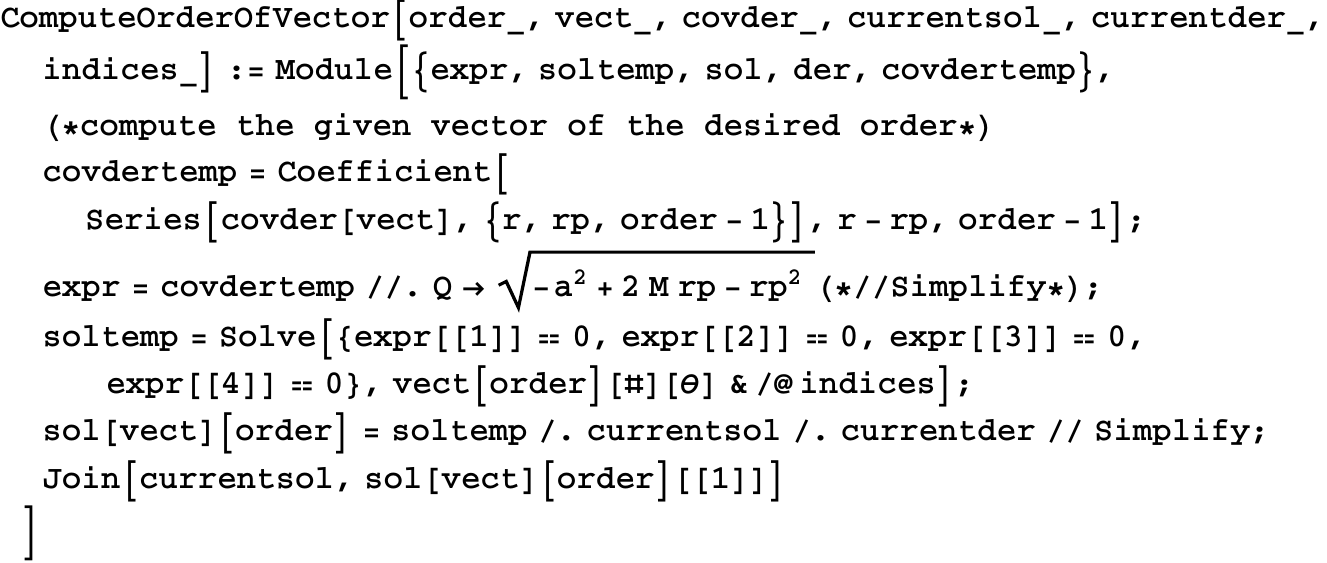}\wmcnl 
\includegraphics{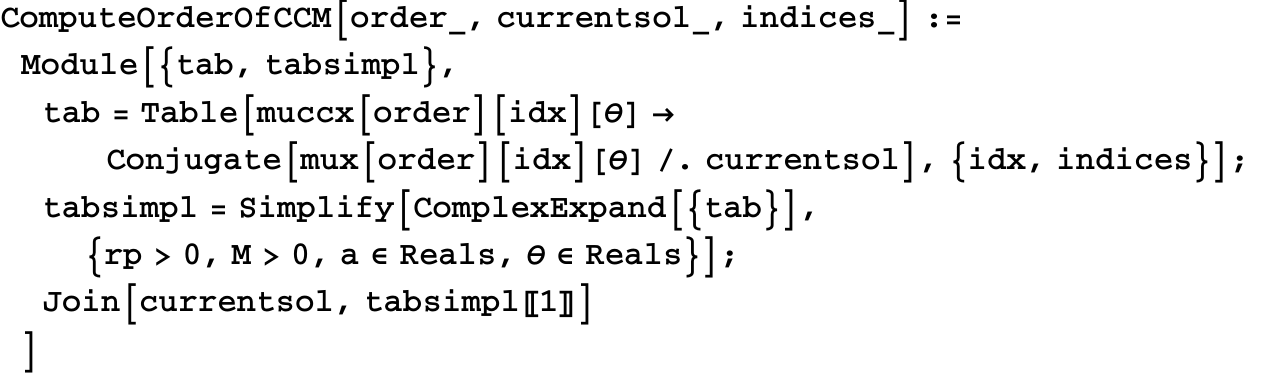}\wmcnl
\includegraphics{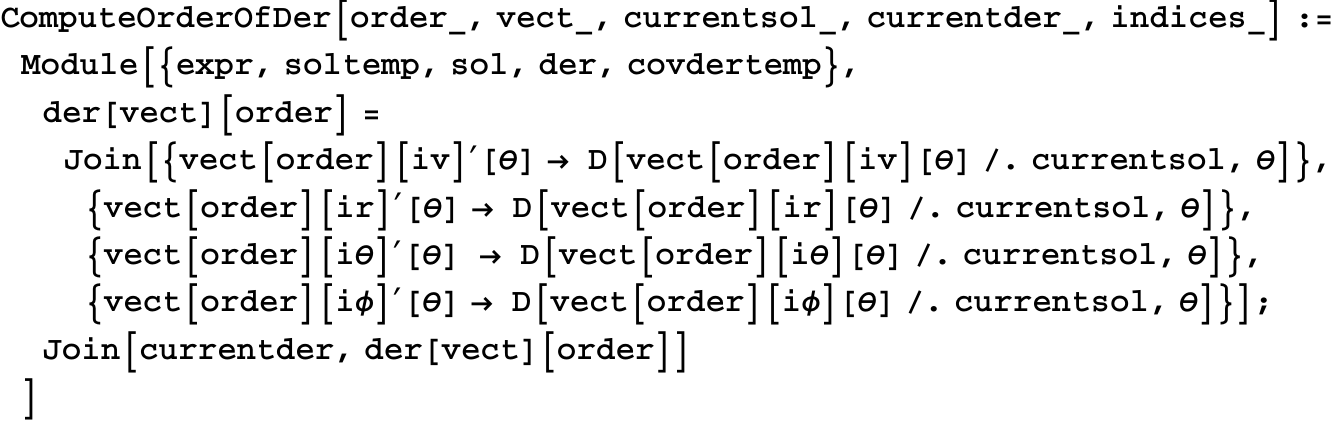}\wmcnl
\includegraphics{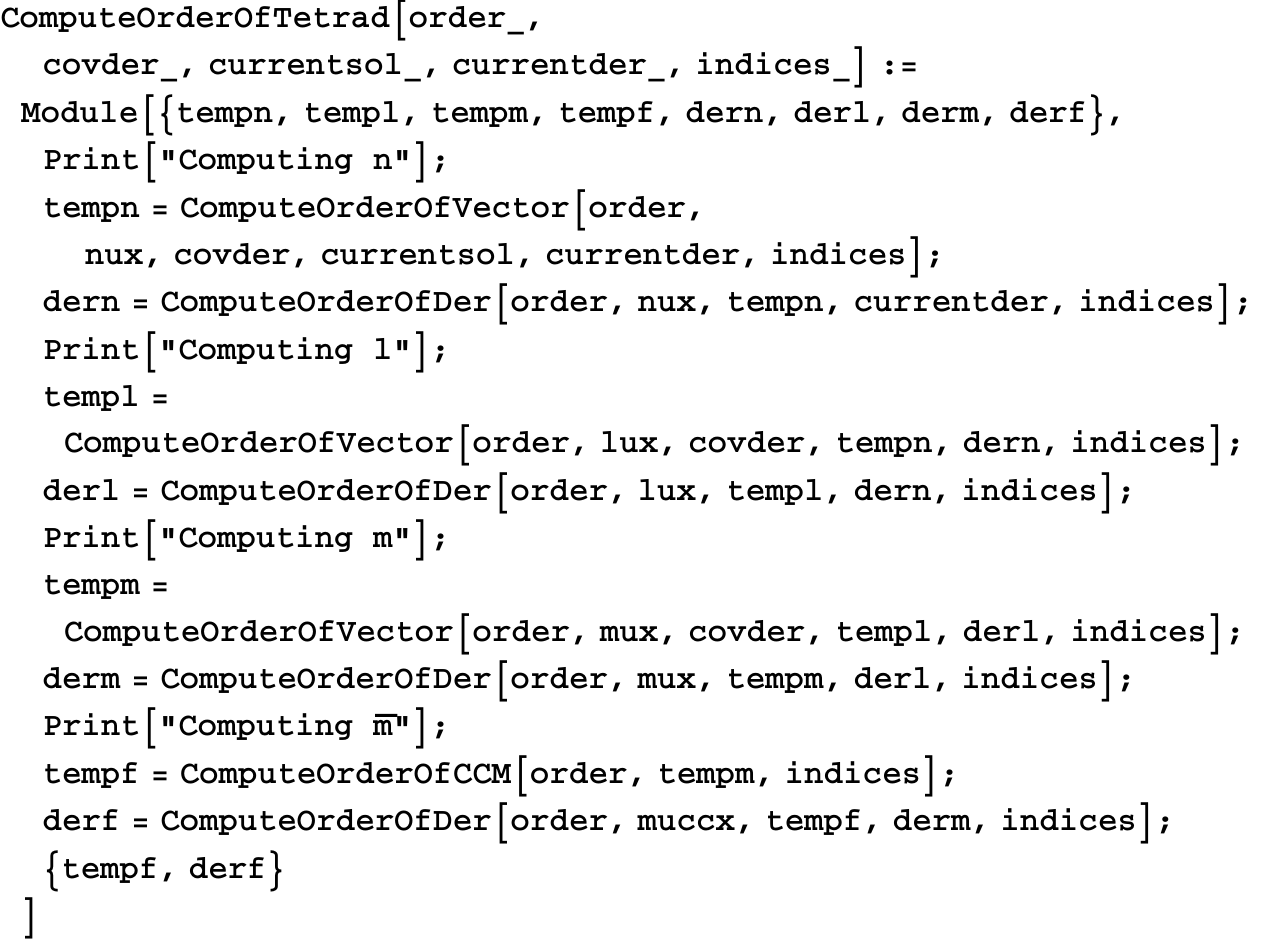} 

\subsubsection{Test of the tetrad being Newman--Penrose}
\includegraphics{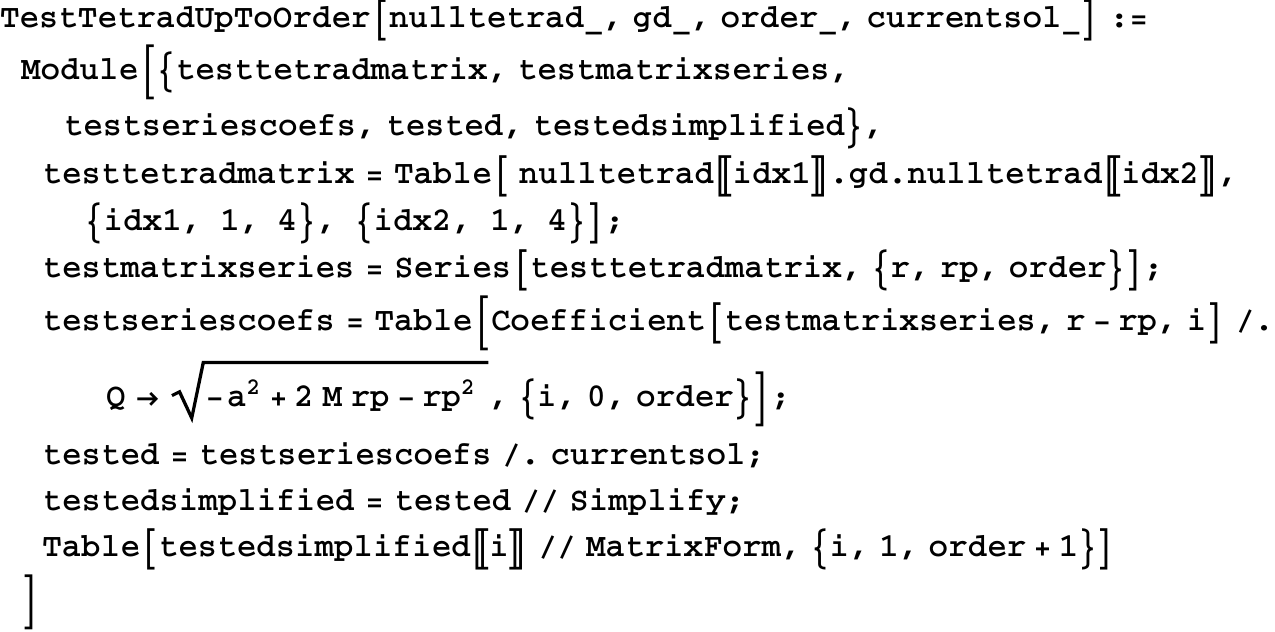} 

The scalars are remaining to be computed. Since we need derivatives for the spin coefficients, order $n+1$ of the tetrad is required to get order $n$ of them.

\subsubsection{Spin coefficients}
\includegraphics{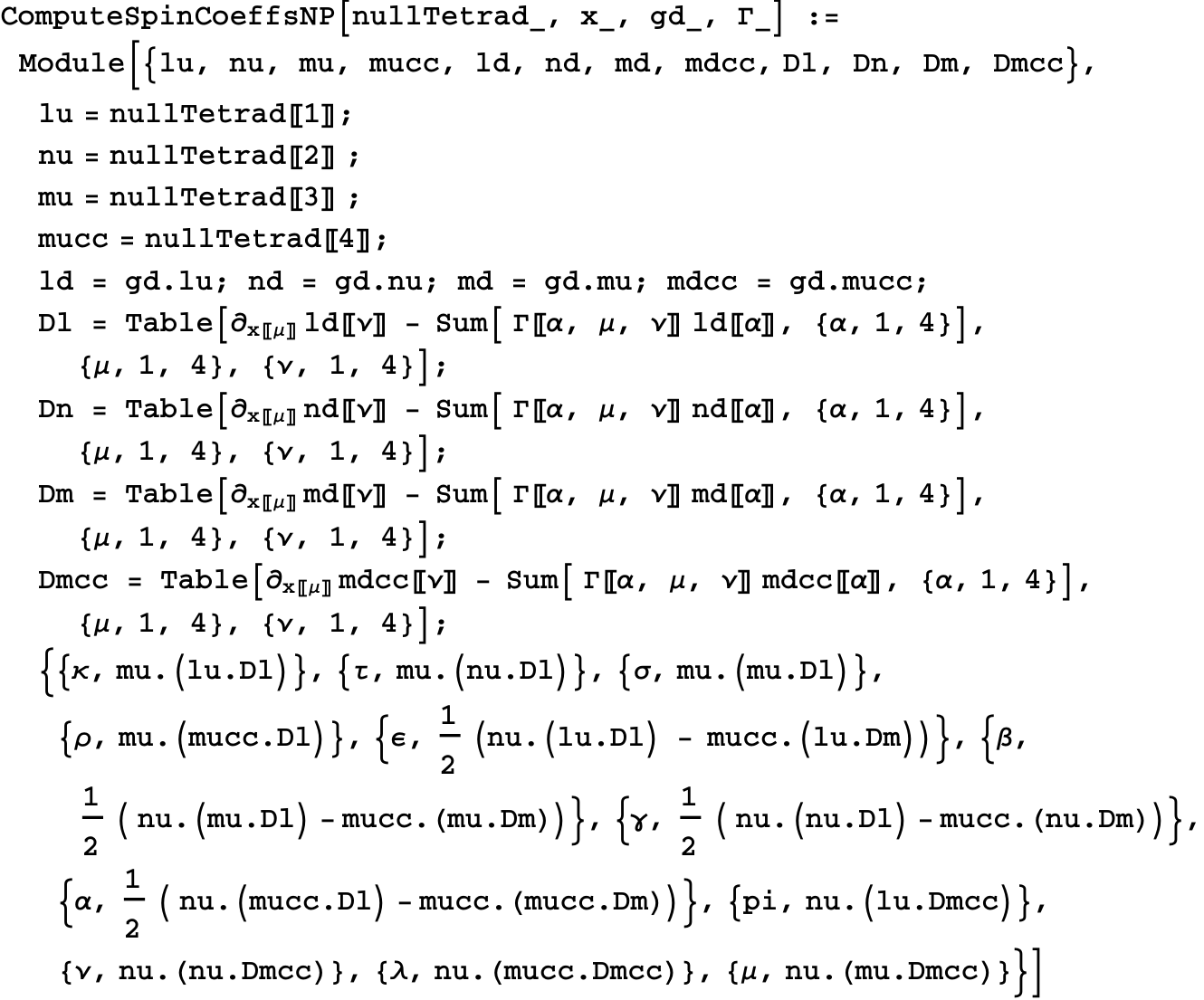}\wmcnl
\includegraphics{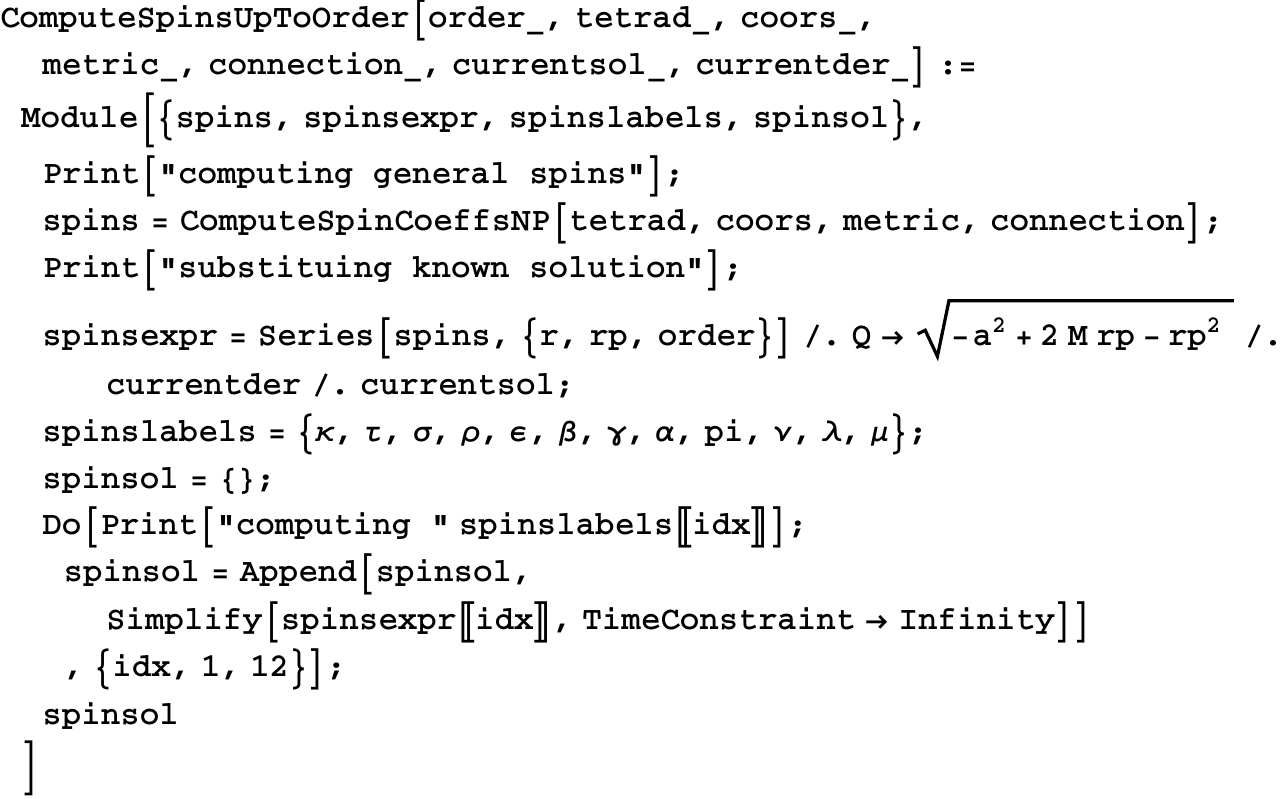} 
\vfill

\subsubsection{Curvature tensors}\label{sec:curvature tensors}
\includegraphics{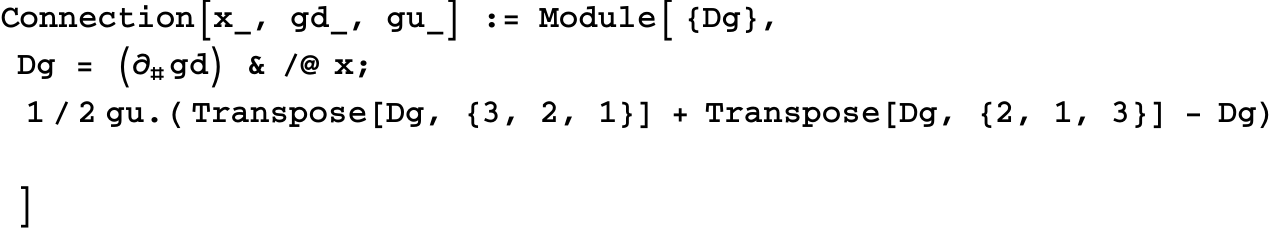}\wmcnl 
\includegraphics{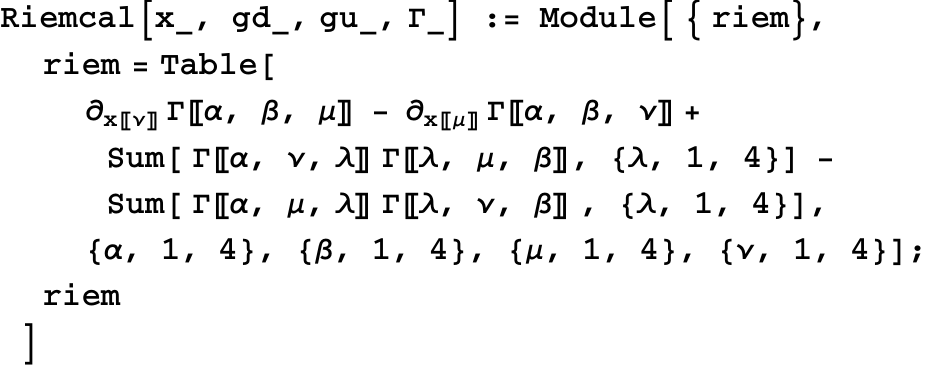}\wmcnl
\includegraphics{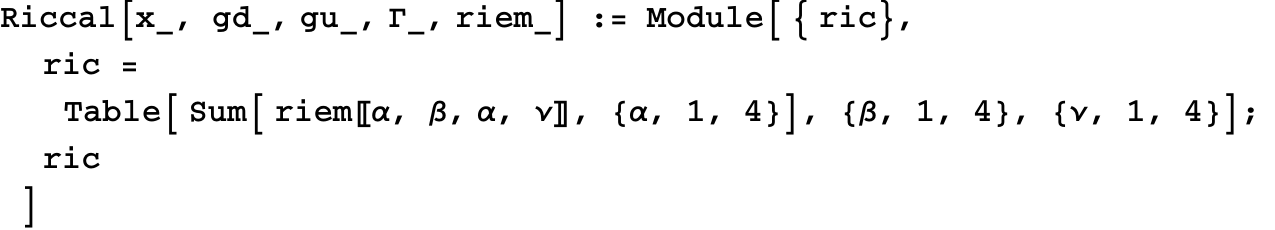}\wmcnl
\includegraphics{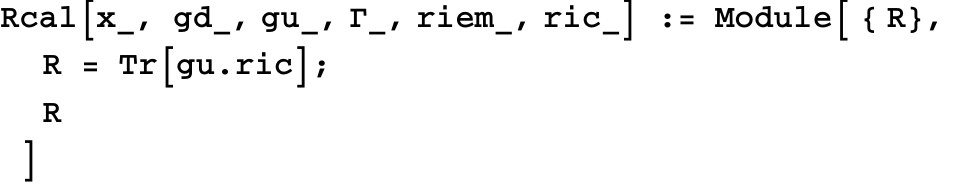}\wmcnl
\includegraphics{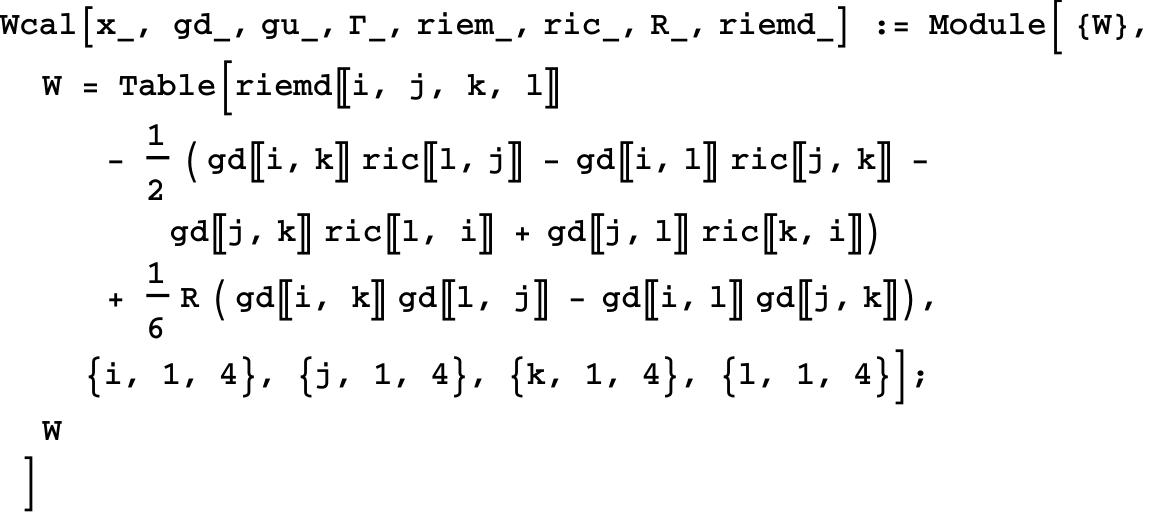}\wmcnl
\includegraphics{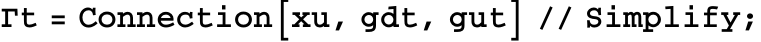}\wmcnl
\includegraphics{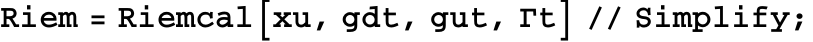}\wmcnl
\includegraphics{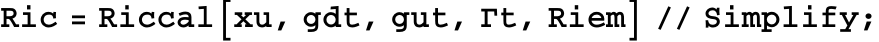}\wmcnl
\includegraphics{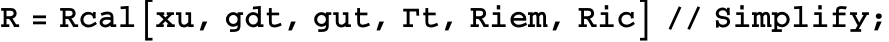}\wmcnl
\includegraphics{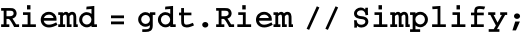}\wmcnl
\includegraphics{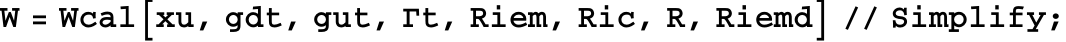} 

\subsubsection{Weyl scalars}\label{sec:weyl scalars}
\includegraphics{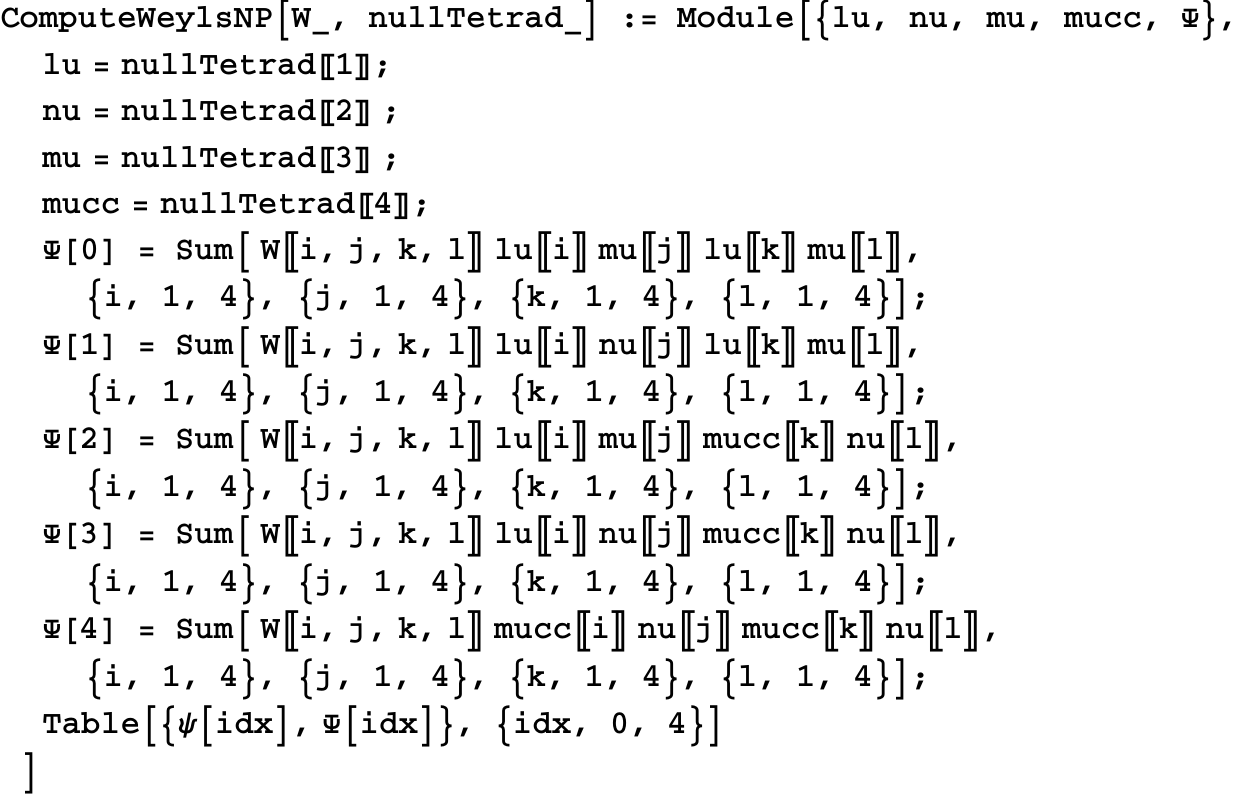}\wmcnl
\includegraphics{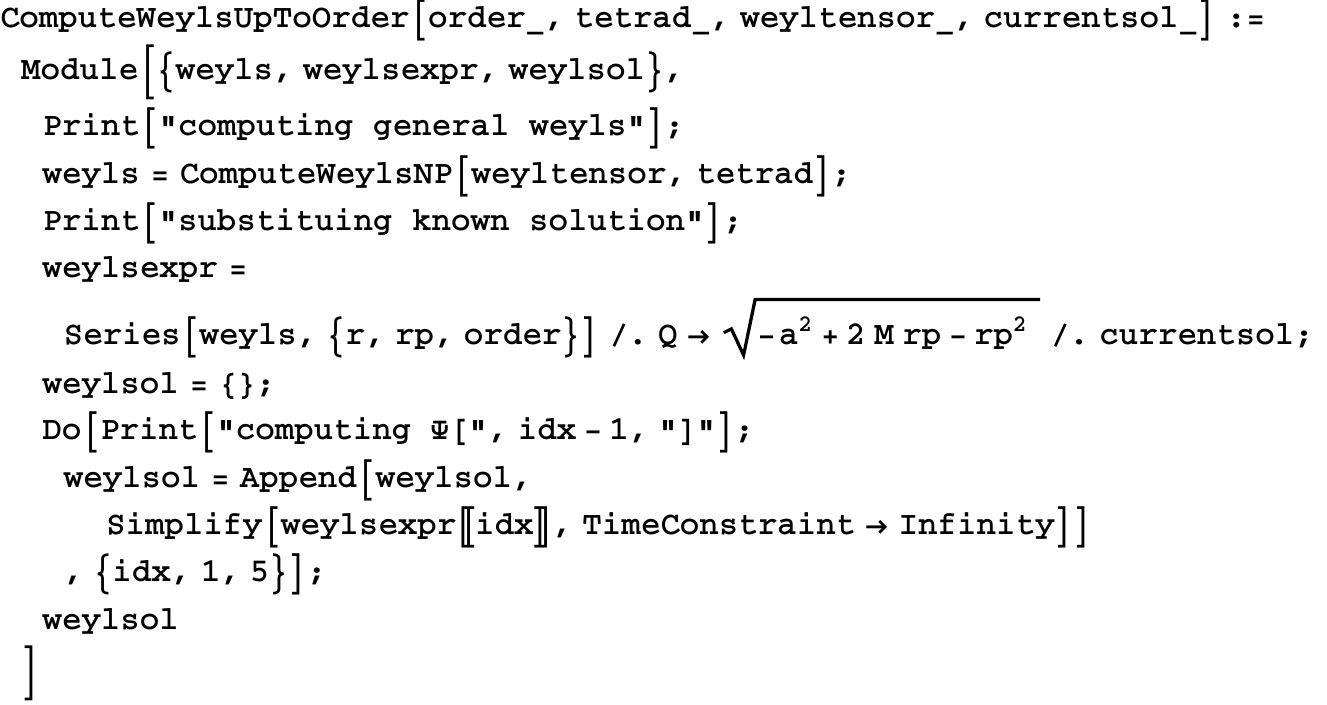} 
\vfill

\subsubsection{Ricci scalars}\label{sec:ricci scalars}
\includegraphics{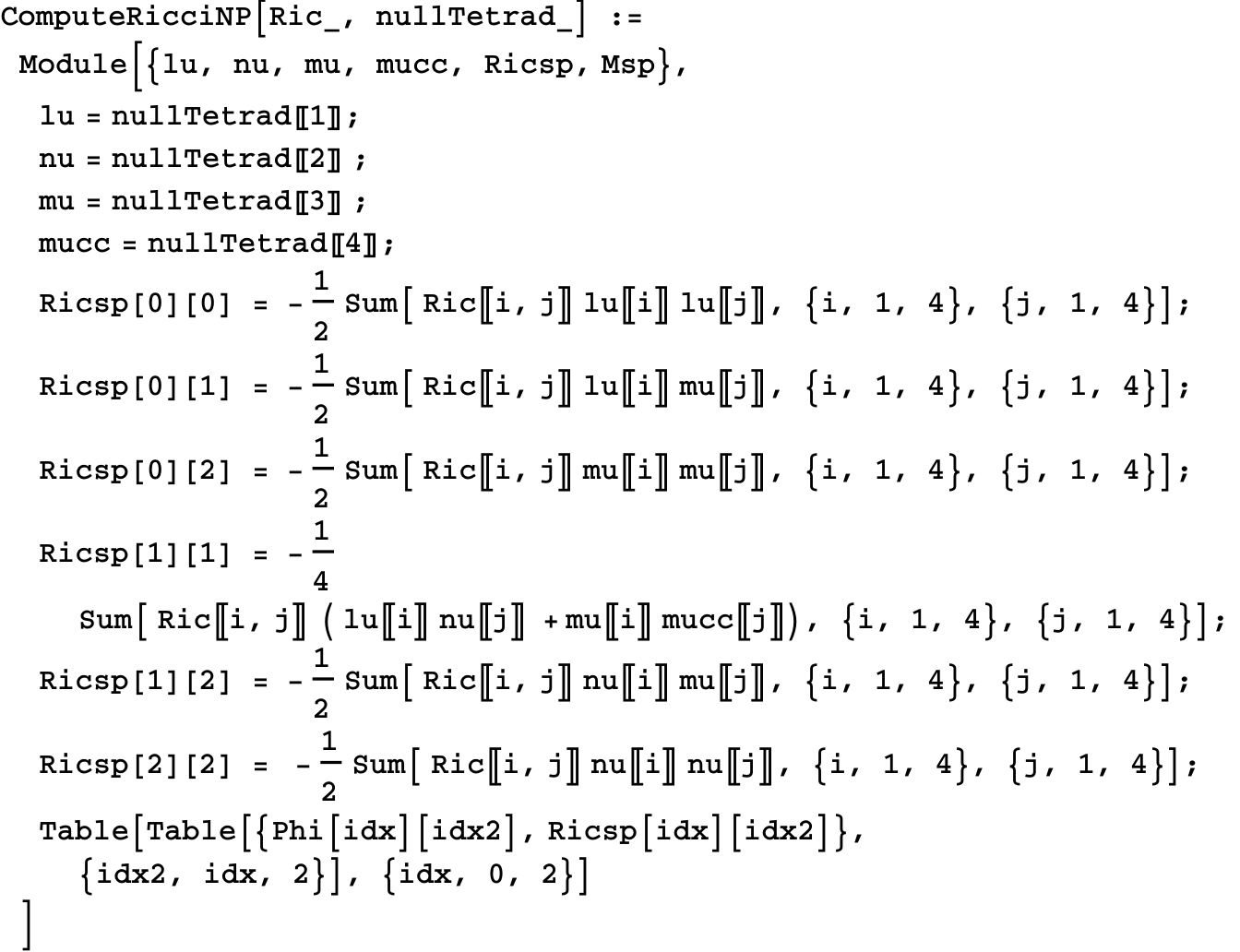}\wmcnl
\includegraphics{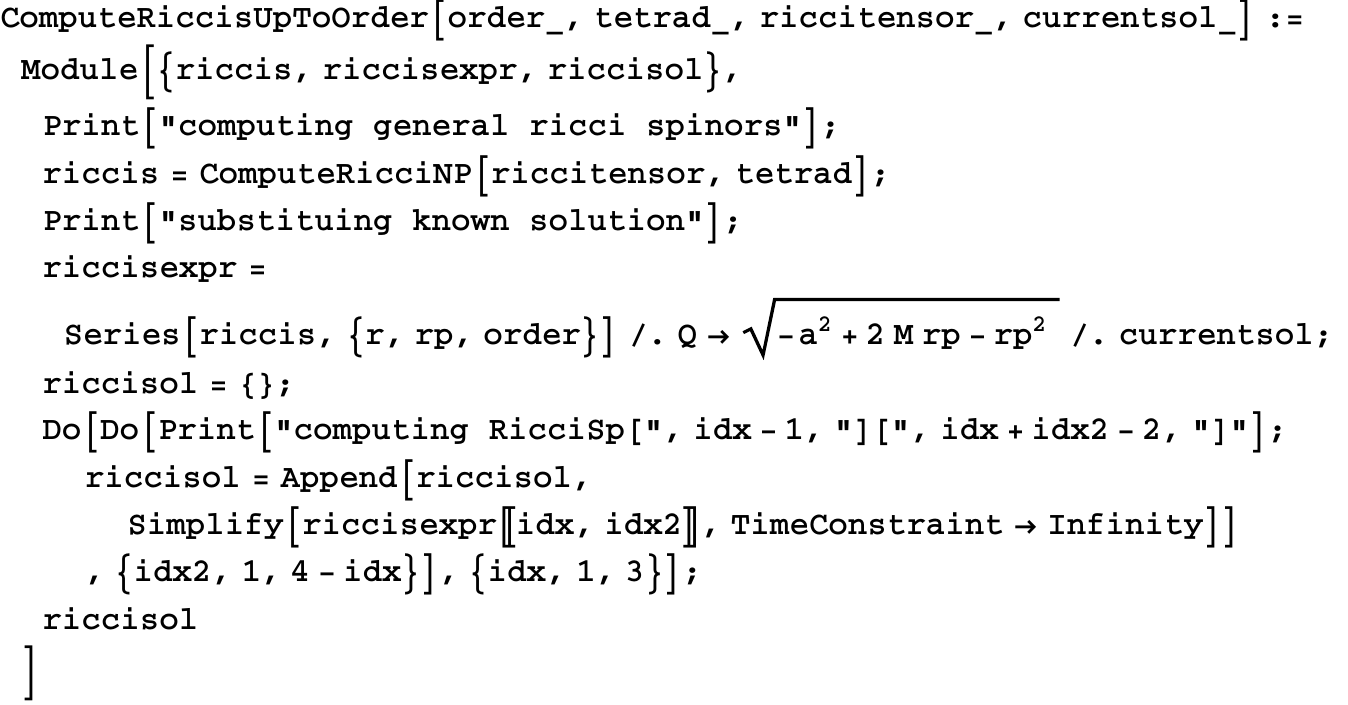} 

\section{Example of computation of the spin coefficient transformation}
\label{app:example transf}
We also present an example of how to get the transformation rules for the spin coefficients. We show only the boost transformation, for the other be only change of the list or rules at the beginning.

The other scalars are even simpler to get since there are no derivatives. We do not show any code for them.\wmcnl
\includegraphics{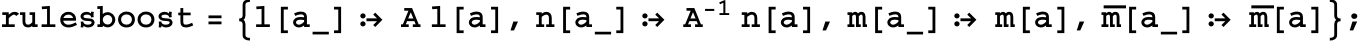}\wmcnl
\includegraphics{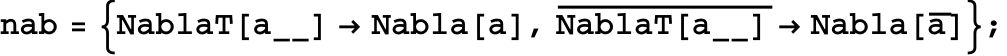}\wmcnl
\includegraphics{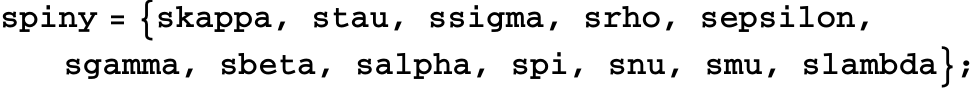}\wmcnl
\includegraphics{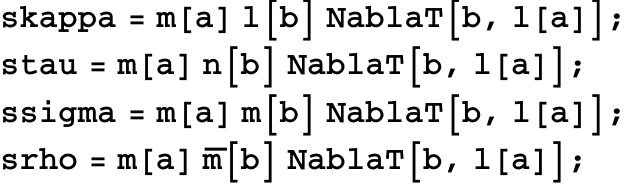}\wmcnl
\includegraphics{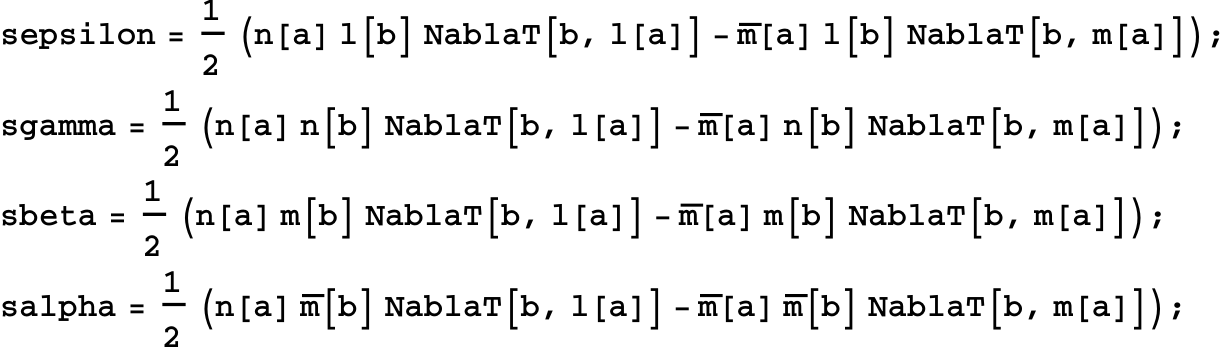}\wmcnl
\includegraphics{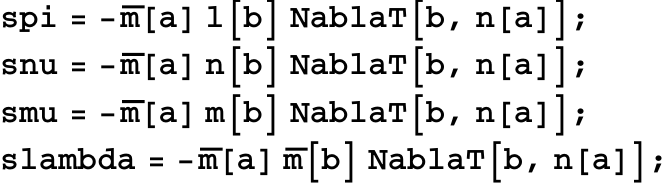}\wmcnl
\includegraphics{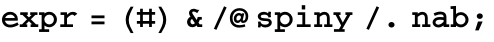}\wmcnl
\includegraphics{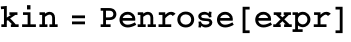}\wmcnl
\includegraphics{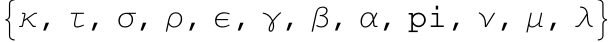}\wmcnl
\includegraphics{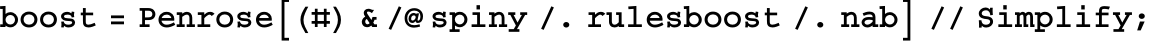}\wmcnl
\includegraphics{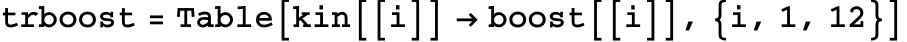}\wmcnl
\includegraphics{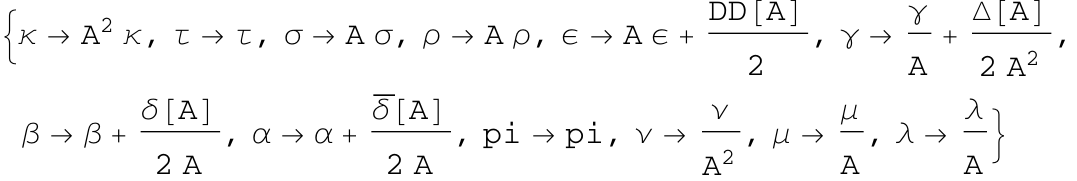}\wmcnl


\begingroup
\sloppy
\phantomsection
\addcontentsline{toc}{chapter}{Bibliography}
\printbibliography\newpage
\endgroup

\ifthenelse{\equal{\draftoption}{draft}}
{
}{

\pagebreak

\listoffigures

\listoftables

\printglossary[type=\acronymtype,title={List of Abbreviations},toctitle={List of Abbreviations}]


\openright
}
\end{document}